\documentclass[traditabstract]{aa}
\usepackage{amssymb}
\usepackage{latexsym}
\usepackage{graphicx} 
\usepackage{rotating}
\usepackage{verbatim} 
\def\f#1   {Fig.~\ref{#1}}
\def\s#1   {Sec.~\ref{#1}}
\def\tab#1   {Table~\ref{#1}}
\def\t#1   {Table~\ref{#1}}

\def\comm#1   {{\tt (COMMENT: #1) }}
\def\kms{~km~s$^{\mathrm{-1}}$}

\def\sqdeg            {$\Box^{\circ}$}

\def\smo               {Smol\v{c}i\'{c}}

\def\mmsample {1.1mm-selected sample}
\def\submmsample {870$\mu$m-selected sample}

\begin{document}
   \title{Millimeter imaging of submillimeter galaxies in the COSMOS field: Redshift distribution$^*$}

   \subtitle{ }

   \author{ V.~Smol\v{c}i\'{c}$^{1,2,3,4}$, M.~Aravena$^5$, 
 F.~Navarrete$^{3,6}$, E.~Schinnerer$^7$, D.~A.~Riechers$^8$, 
 F.~Bertoldi$^{3}$, C.~Feruglio$^9$, A.~Finoguenov$^{10}$, M.~Salvato$^{10}$, , M.~Sargent$^{11}$, 
 H.~J.~McCracken$^{12}$, M.~Albrecht$^{3}$, A.~Karim$^{7}$, P.~Capak$^8$, C.~L.~Carilli$^{13}$, 
 N.~Cappelluti$^{14}$, M.~Elvis$^{15}$, O.~Ilbert$^{16}$, J.~Kartaltepe$^{17}$, S.~Lilly$^{18}$, D.~Sanders$^{19}$, 
 K.~Sheth$^{20}$, N.~Z.~Scoville$^8$, Y.~Taniguchi$^{21}$
          }

\authorrunning{V.\ Smol\v{c}i\'{c} et al.}
\titlerunning{mm-imaging of COSMOS SMGs: Redshift distribution}

   \institute{ 
$^*$Based on observations carried out with the IRAM Plateau de Bure Interferometer. IRAM is supported by INSU/CNRS (France), 
MPG (Germany) and IGN (Spain).\\
$^{1}$ESO ALMA COFUND Fellow\\
$^{2}$European Southern Observatory, Karl-Schwarzschild-Strasse 2, 
D-85748 Garching, Germany\\
$^{3}$Argelander-Institute for Astronomy, Auf dem H\"{u}gel 71, Bonn, D-53121, Germany\\
$^{4}$University of Zagreb, Physics Department, Bijeni\v{c}ka cesta 32, 10002 Zagreb, Croatia\\
$^{5}$European Southern Observatory, Alonso de C\'{o}rdoba 3107, Vitacura, Casilla 19001, Santiago 19, Chile\\
$^6$Max-Planck-Institut f\"{u}r Radioastronomie, Auf dem H\"{u}gel 69, 53121 Bonn, Germany\\
$^7$Max Planck Institute for Astronomy, K\"{o}nigstuhl 17, 69117 Heidelberg, Germany\\
$^8$California Institute of Technology, MC 249-17, 1200 East California Boulevard, Pasadena, CA 91125\\
$^9$IRAM, 300 rue de la piscine, F-38406 Saint-Martin d'H$\grave{e}$res, France\\
$^{10}$Max-Planck-Institut f\"ur Extraterrestrische Physik, Giessenbachstra\ss e, 85748 Garching, Germany\\
$^{11}$Laboratoire AIM-Paris-Saclay, CEA/DSM/Irfu CNRS Universite Paris Diderot, CE-Saclay, pt courrier 131, F-91191 Gif-sur-Yvette, France\\
$^{12}$Institut d'Astrophysique de Paris, UMR7095 CNRS, Universite Pierre et Marie Curie, 98 bis Boulevard Arago, 75014 Paris, France\\
$^{13}$National Radio Astronomy Observatory, P.O. Box 0, Socorro, NM 87801-0387\\
$^{14}$INAF-Osservatorio Astronomico di Bologna, Via Ranzani 1, 40127 Bologna, Italy\\
$^{15}$Harvard-Smithsonian Centre for Astrophysics, 60 Garden Street, Cambridge, MA 02138, USA\\
$^{16}$Laboratoire d'Astrophysique de Marseille, Universit\'e de
Provence, CNRS, BP 8, Traverse du Siphon, 13376 Marseille Cedex 12, France\\
$^{17}$National Optical Astronomy Observatory, 950 North Cherry Avenue, Tucson, AZ 85719, USA\\
$^{18}$Institute for Astronomy, ETH Z\"{u}rich, Wolfgang-Pauli-strasse 27, 8093 Z\"{u}rich, Switzerland\\
$^{19}$Institute for Astronomy, University of Hawaii, 2680 Woodlawn Drive, Honolulu, HI, 96822, USA\\
$^{20}$National Radio Astronomy Observatory, 520 Edgemont Road, Charlottesville, VA 22903, USA\\
$^{21}$Research Center for Space and Cosmic Evolution, Ehime University, Bunkyo-cho, Matsuyama 790-8577, Japan
             }

   \date{Received xxx; accepted yyy}

% \abstract{}{}{}{}{} 
% 5 {} token are mandatory
 
  \abstract
  { We present new IRAM Plateau de Bure interferometer (PdBI) 1.3\,mm continuum observations at $\sim$1.5''
  resolution of 28 submillimeter
  galaxies (SMGs), previously discovered with the 870\,$\mu$m bolometer LABOCA at the APEX telescope
  from the central 0.7\,deg$^2$ of the COSMOS field.
 Nineteen out of the 28 LABOCA sources were detected with PdBI
  at a $\gtrsim3\sigma$ level of $\approx$ 1.4~mJy/beam.
   A combined analysis of this new sample with existing
  interferometrically identified SMGs in the COSMOS field yields the following results: 
  i) $\gtrsim15\%$, and possibly up to $\sim40\%$ of
  single-dish detected SMGs consist of multiple sources, 
   ii) statistical analysis of multi-wavelength counterparts to single-dish SMGs shows that only $\sim50\%$ have real radio or IR counterparts,
  iii) $\sim18$\% of interferometric SMGs have either no multi-wavelength counterpart or only a radio-counterpart, and 
  iv) $\sim50-70\%$ of $z\gtrsim3$ SMGs have no radio counterparts (down to an rms of 7-12~$\mu$Jy at 1.4~GHz). Using the exact interferometric positions to
  identify the multi-wavelength counterparts allows us to
  determine accurate photometric redshifts for these sources. The
  redshift distributions of the combined and the individual 1.1~mm
  and 870\,$\mu$m selected samples shows a higher mean and a broader width
  than those derived in previous studies. This study finds that on average brighter and/or mm- selected SMGs are located at higher
  redshifts, consistent with previous studies.  The mean redshift for the
  1.1~mm selected sample ($\bar{z}=3.1\pm0.4$) is tentatively higher than that for the 870\,$\mu$m selected sample ($\bar{z}=2.6\pm0.4$). Based on our nearly complete sample of AzTEC 1.1~mm SMGs in a 0.15~deg$^2$ area, we infer a higher surface density of $z\gtrsim4$ SMGs than predicted by current cosmological models.   In summary, our findings imply that 
  interferometric identifications at (sub-)millimeter wavelengths are crucial to build statistically
  complete and unbiased samples of SMGs. }
   { }
   { }
   { }
  { }

   \keywords{
               }

\maketitle

%\begin{abstract}
 % \end{abstract}

%\begin{keywords}
%galaxies: active,
%-- cosmology: observations -- radio continuum: galaxies 
%\end{keywords}

\section {Introduction}
\label{sec:intro}

\subsection{Submillimeter galaxies}
Submillimeter galaxies (SMGs; $\mathrm{S_{850\mu m}\gtrsim5}$~mJy)
are ultra-luminous, dusty, starburst galaxies with extreme star
formation rates of order $10^3$~M$_\odot$~yr$^{-1}$
(e.g. Blain et al.\ 2002).  They
trace a phase of the most intense stellar mass build-up 
in galaxies and contribute significantly to the volume-averaged
cosmic star formation rate density at $z=2-3$ ($\sim20\%$; 
Michalowski et al.\ 2010). Evidence is emerging that SMGs represent
the progenitors of massive elliptical galaxies (e.g.\ Cimatti et al.\ 2008; van Dokkum et al.\ 2008; Michalowski et al.\
2010), and their enhanced star formation properties may be intimately
related to the evolution of quasi stellar objects (Sanders et al.\
1996; Hopkins et al.\ 2006; Hayward et al. 2011).

Spectroscopic and photometric studies of SMGs locate them predominantly at
redshifts 2-3 (e.g.\ Chapman et al.\ 2005; Wardlow et al.\ 2011), and
only a few $z>4$ SMGs have recently been detected.  Identifying the highest redshift 
SMGs requires time-consuming, systematic follow-up observations to
properly identify them against strong lower-redshift selection
biases. To date $\sim10$ $z>4$ SMGs have been confirmed (Daddi et
al.\ 2009a, 2009b; Capak et al.\ 2008, 2011; Schinnerer et al.\ 2008; Coppin et al.\ 2009,
Knudsen et al.\ 2010; Carilli et al.\ 2010, 2011; Riechers et al.\ 2010; \smo\ et al.\
2011; Cox et al.\ 2011; Combes et al.\ 2012). 

Although the number of high redshift ($z > 4$) SMGs remains small it appears that 
their abundance is so high that it is only marginally
consistent with current galaxy formation models (Baugh et al.\ 2005; Coppin et
al.\ 2009; \smo\ et al.\ 2011). It has also been suggested that high-redshift SMGs may be qualitatively
different from those at intermediate redshift (Wall et al.\ 2008).  
However, such conclusions are premature given the significant uncertainties in identification of SMGs with optical/IR sources, and therefore the appropriate measurement of their redshift. 

\subsection{Identifying multi-wavelength counterparts to SMGs}

SMGs are typically first detected with single-dish mm or sub-mm
telescopes which have a relatively large ($10"-35"$) beam size which may include tens of galaxies in the visible or NIR.
Numerous methods have been applied to pinpoint the proper counterparts, such as UV/IR/radio star formation indicators or an association with AGN (Ivison et al.\ 2005 2007; Bertoldi et al.\ 2007; Biggs et al.\ 2010). These techniques are problematic because the identification is probabilistic and thus introduces the possibility of sample incompletness, contamination, and/or bias.
Moreover SMGs may be tightly clustered and thus blended typical 
in single dish observations (e.g., Younger
et al.\ 2007, 2009). 
Interferometric observations of SMGs at intermediate-resolution ($\sim2"$) shows that they often do not coincide with any  galaxy at any wavelength (Younger et al.\ 2009).  This may be due to extreme dust extinction or due to a very high-redshift for the galaxy.  To assess the overall properties of SMGs including their redshift distribution, it is therefore crucial to follow up single-dish detections with high resolution interferometric imaging. Before the improved sensitivities provided by the upgraded IRAM PdBI or ALMA, interferometric follow-up at millimeter or submillimeter wavelengths was slow and expensive.  Only about 50 SMGs have been properly identified in various survey fields
(Downes et al.\ 1999, Frayer et al.\
  2000; Dannerbauer et al.\ 2002; Downes \& Solomon 2003; Genzel et
  al.\ 2003; Kneib et al.\ 2005; Greve et al.\ 2005; Tacconi et al.\
  2006; Sheth et al.\ 2004, Iono et al.\ 2006; Younger et al.\
  2007, 2009; Aravena et al.\ 2010a; Ikarashi et al.\
2011; Tamura et al.\ 2010; Hatsukade et al.\ 2010; Wang et al.\ 2011;
Chen et al.\ 2011; Neri et al.\ 2003; Chapman et al.\ 2008; \smo\ et al.\ 2012). 
The largest statistically significant, signal-to-noise- and flux-limited sample of interferometrically
identified SMGs contains 17 sources drawn from the AzTEC/JCMT
survey of 0.15\sqdeg\ within the COSMOS field (Younger et al.\ 2007,
2009). Here we present PdBI observations towards 28 SMGs drawn from the LABOCA-COSMOS
0.7\sqdeg\ survey (Navarrete et al., in prep.), which constitutes the largest 
interferometric follow-up of SMGs drawn from bolometer imaging surveys to date.

\subsection{Determining the redshift of SMGs}

The proper identification with an optical counterpart may allow a determination of the SMG redshift through deep optical/NIR spectroscopy. Given the ambiguity of identifications through probability considerations and the optical faintness of the counterparts, and the absence of lines in particular redshift ranges,  this has been a very difficult task.
The largest SMG sample with spectroscopic
redshifts to date was established by Chapman et al.\ (2005), who followed-up SMG counterparts identified
through deep, intermediate ($\lesssim2"$) resolution radio observations, getting redshifts for 76 of 150 targets.

Where spectroscopic redshifts cannot be measured for large samples of SMGs, deep panchromatic surveys such as COSMOS or GOODS can measure photometric redshifts, which are based on $\chi^2$ minimization fits of multi-band photometry to spectral models (e.g.\ Ilbert et al.\ 2009). With an optimized choice
of spectral models and dense multi-wavelength photometric coverage
 photometric redshifts can reach accuracies of a few
percent (e.g.\ Ilbert et al.\ 2009). 
Although it was not obvious whether common photometric redshift templates could be applied to SMG counterparts, 
recent studies confirm that photometric redshifts can be estimated for SMGs, 
both on statistical and a case by case basis (e.g.\ Daddi et al.\ 2009a; Wardlow et al.\ 2010, 2011; Yun et al.\ 2012; \smo\ et al.\ 2012).
Here we further test the photometric redshift estimates for SMGs using the largest
"training set" of SMGs with secure spectroscopic redshifts to date from COSMOS.

% This allows us to derive the proper redshift distribution of an identification-unbiased, S/N-limited and nearly flux-limited
% sample of SMGs.

In \s{sec:data} \ we describe the data used for our analysis. In \s{sec:obs} \  we present the PdBI observations towards 28 SMGs drawn from the LABOCA-COSMOS survey. In \s{sec:statsamples} \ we define two samples of SMGs with mm-interferometric detections in the COSMOS field. Using these in \s{sec:counterparts} \ we investigate blending of SMGs, and usually applied statistical counterpart association methods to single-dish identified SMGs. In \s{sec:redsft} \ we calibrate photometric redshifts for SMGs.  In \s{sec:zdistrib} \ we derive redshift distributions for our statistical samples of SMGs with unambiguously determined counterparts. We discuss and summarize our results in \s{sec:discussion} \ and \s{sec:summary} .  We adopt $H_0=70$~\kms ~Mpc$^{-1}$, $\Omega_M=0.3$, $\Omega_\Lambda=0.7$, 
and use a dust emissivity index of $\beta=1$, 
and a Chabrier (2003) initial-mass function if not stated otherwise.

%%%%%%%%%%%%%%%%%%%%%%%%%%%%%%%%%%%%%%%%%%%%%

\begin{table*}
\caption{Summary of interferometrically observed COSMOS SMGs besides our work}
\label{tab:interf}
\begin{tabular}{|l|c|ccc|ccc|}
\hline
    Source       &  reference & 
    \multicolumn{3}{|c|}{LABOCA}   & \multicolumn{3}{|c|}{redshift}  \\
          &  & 
    source & separation & $F_\mathrm{870\mu m}$  &  spectroscopic & photometric$^+$ & mm-to-radio\\
          &  &   & [$"$] &  [mJy]   &  & & \\
    \hline 
    \vspace{0.5mm}
      AzTEC-1     & (1), (2), (3) & COSLA-60 & & $12.6\pm3.6$ & 4.64 & $4.26^{+0.17}_{-0.20}$ & -- \\ \vspace{0.5mm}
      AzTEC-2     & (1), (4) & COSLA-4 & & $14.4\pm3.0$ & 1.125  & -- & -- \\ \vspace{0.5mm}
      AzTEC-3     & (1), (5) & -- & -- & -- & 5.299  & $5.20^{+0.09}_{-0.21}$ & -- \\ \vspace{0.5mm}
      AzTEC-4     & (1) &  -- & -- & -- & -- & $4.70^{+ 0.43}_{- 1.11}$ & -- \\ \vspace{0.5mm}
      AzTEC-5     & (1), (6) &  -- & -- & -- &3.971 & $3.05^{+0.33}_{-0.28}$ & -- \\ \vspace{0.5mm}
      AzTEC-6     & (1) &  -- & -- & -- &0.802 & $0.82^{+0.13}_{-0.10}$ & -- \\ \vspace{0.5mm}
      AzTEC-7     & (1) &  -- &  -- & -- & -- & $2.30^{+ 0.10}_{- 0.10}$ & -- \\ \vspace{0.5mm}
      AzTEC-8     & (7), (8)&  COSLA-73 & -- & $12.3\pm3.6$  \ & $3.179$ & $3.17^{+0.29}_{-0.22}$ & -- \\ %$3.17^{+ 0.29}_{- 0.22}$ \\ \vspace{0.5mm}
      AzTEC-9     & (7) &  COSLA-3 & -- & $16.4\pm3.3$ &1.357 & $1.07^{+0.11}_{-0.10}$ & -- \\ \vspace{0.5mm}
      AzTEC-10    & (7) &  -- &  -- & -- & --& $2.79^{+1.86}_{-1.29}$ & -- \\ \vspace{0.5mm}
      AzTEC-11    & (7) &  -- &  -- & -- & 1.599 & $1.93^{+0.13}_{-0.18}$ & -- \\ \vspace{0.5mm}
      ~~~~AzTEC-11-N    & (7) &  -- &  -- & -- & -- & $1.51^{+0.41}_{-0.92}$ & -- \\ \vspace{0.5mm}
      ~~~~AzTEC-11-S    & (7) &  -- &  -- & -- & -- & --& $>2.58$ \\ \vspace{0.5mm}
      AzTEC-12    & (7) &  -- &  -- & -- &--&$2.54^{+ 0.13}_{- 0.33}$ & --\\ \vspace{0.5mm}
      AzTEC-13    & (7) &  COSLA-158 & -- & $11.8\pm3.9$   & --&--&$>3.59$\\ \vspace{0.5mm}
      AzTEC-14-E  & (7) & -- &  -- & -- & --&--&$>3.03$ \\ \vspace{0.5mm}
      AzTEC-14-W  & (7) &  -- &  -- & -- &--&$ 1.30^{+ 0.12}_{- 0.36}$ & -- \\ \vspace{0.5mm}
      AzTEC-15    & (7) &  -- & -- & -- & --&$3.01^{+ 0.29}_{- 0.37}$ & --\\ \vspace{0.5mm}
 %%%%%
     AzTEC-16 & (9) & -- & -- & -- &1.505 & $1.09^{+0.08}_{-0.06}$ & -- \\ \vspace{0.5mm}
%%%
     J1000+0234  & (10)& -- & -- & -- & 4.542 & $4.45^{+0.08}_{-0.08}$ & -- \\
       AzTEC/C1  & (11)&   COSLA-89 & -- & $12.4\pm3.7$ & -- & $5.6\pm1.2$\\ \vspace{0.5mm}
      Cosbo-1$^*$       & (12) & COSLA-1 & -- & $13.8\pm1.5$ & -- & -- & $3.83^{+0.68}_{-0.49}$\\ \vspace{0.5mm}
      Cosbo-3 & (8), (11)  & COSLA-2  & -- & $13.1\pm2.6$ & 2.490 & $1.9^{+0.9}_{-0.5}$ & -- \\ \vspace{0.5mm}
      Cosbo-8 &  (11) &    -- & -- & -- & -- & $3.1\pm0.5$ & -- \\ \vspace{0.5mm}
      Cosbo-14    & (12) & -- & -- & -- & -- & -- & -- \\ 
\hline
\end{tabular}\\
%Fluxes are deboosted fluxes \\
$^+$ Photometric redshifts drawn from the total $\chi^2$ distribution as described in \s{sec:redsft} \ and {\em not} corrected for any systematic offsets\\
$^*$Formally this source is not detected in optical, near- and mid-IR maps/catalogs, therefore we here use the mm-to-radio flux based redshift here, which is consistent with the photometric redshift given by Aravena et al.\ (2010)\\
(1) Younger et al.\ (2007)\\
(2) Younger et al.\ (2009)\\
(3) \smo\ et al.\ (2011)\\
(4) Balokovi\'{c} et al., in prep\\
(5) Capak et al.\ (2010); Riechers et al.\ (2010)\\
(6) Karim et al., in prep\\
(7) Younger et al.\ (2009)\\
(8) Riechers et al., in prep.\\
%(7) Sheth et al., subm.\\
(9) Sheth et al., in prep.\\
(10) Capak et al.\ (2009), Schinnerer et al.\ (2009)\\
(11) \smo\ et al., 2012.\\
(12) Aravena et al.\ (2010)\\
%(12) this work
%(3) Younger et al.\ (2010)\\
\end{table*}

\section{Data}
\label{sec:data}

\subsection{The COSMOS Project}

The Cosmic Evolution Survey (COSMOS) is an imaging and
spectroscopic survey of an equatorial 2\sqdeg\ field (Scoville et al.\
2007). The field has been observed with most major space- and
ground-based telescopes over most of the electromagnetic spectrum.  
The COSMOS Project has obtained very deep broad-band (u*BVgrizJHK) and medium and narrow-band
imaging data in over 30 optical to near-infrared bands.   Additionally, there is GALEX, Spitzer IRAC/MIPS,
Herschel PACS/SPIRE, HST/ACS, XMM-Newton, VLA (1.4~GHz and 320~MHz), GMRT (600 \& 200
MHz) data, as well over 25,000 optical spectra
(Capak et al.\ 2007, Sanders et al.\ 2007, Scoville et al.\ 2007, Leauthaud et al.\ 2007, Koekemoer et al.\ 2009, Frayer et al.\ 2009, 
Hasinger et al.\ 2007, Zamojski et al.\ 2007, Taniguchi et al.\ 2007; Lilly et al.\ 2007, 2009; Le Floc'h et al.\ 2009; McCracken et al.\ 2010; 
Trump et al.\ 2007, Schinnerer et al.\ 2004, 2007, 2010; \smo\ et al., in prep).
The inner square degree of COSMOS has also been observed in X-rays at a 
higher resolution and sensitivity with Chandra (Elvis et al.\ 2009) and
at mm and submm wavelengths with AzTEC, BOLOCAM, MAMBO and LABOCA (Aretxaga et al.\ 2010;
Navarrete et al., in prep.; Bertoldi et al.\ 2007, Aguirre et al., in
prep., Scott et al.\ 2008). 

Particularly relevant for the work presented here are the deep
UltraVista observations of COSMOS which reach $5\sigma$ ($2"$ aperture AB magnitude)
sensitivities of 24.6, 24.7, 23.9, and 23.7 in Y, J, H, and Ks bands respectively (McCracken et al.\ 2012), as
well as the VLA 1.4~GHz observations which reach a rms of 7-12$~\mu$Jy/beam (Schinnerer et al.\ 2004,
2007, 2010). We use the updated UV-MIR COSMOS photometric catalog (Capak et al.\ 2007) including all available UV-MIR photometric observations. % Unpublished Herschel data exist and a joint analysis will be included in a future publication. 
 
\subsection{Submillimeter galaxies in the COSMOS field}

The COSMOS field was mapped at mm or submm wavelengths with MAMBO at the IRAM 30m (0.11~deg$^{2}$; 1.2mm, $11"$ angular resolution; Bertoldi et al.\ 2007), BOLOCAM at the CSO (0.27~deg$^{2}$;
1.1mm; $31"$  angular 
resolution; Aguirre et al., in prep), AzTEC at the JCMT (0.15~deg$^{2}$;
1.1mm; $18"$  angular 
resolution; Scott et al.\ 2008), AzTEC at ASTE (0.72~deg$^2$; 1.1~mm;
$34"$  angular 
resolution; Aretxaga et al.\ 2011), and LABOCA at APEX (0.7~deg$^{2}$;
870~$\mu$m; $27''$  angular 
resolution, Navarette et al., in prep.). To properly determine
the multi-wavelength counterparts of the SMGs identified in these surveys, 
numerous interferometric and spectroscopic follow-up efforts have been made 
(Younger et al.\ 2007, 2008, 2009; Capak et al.\ 2008, 2010;
Schinnerer et al.\ 2008; Riechers et al.\ 2010; Aravena et al.\ 2010a;
\smo\ et al.\ 2011, 2012.; Karim et al.\ 2012, in prep.).  To date a
sample of 24 interferometrically identified COSMOS SMGs has been established prior to our observations 
 (Table~\ref{tab:interf}). For 11 of those spectroscopic redshifts are available, either from a dedicated COSMOS optical spectroscopic follow-up campaign using Keck~II/DEIMOS (Capak et al., in prep., Karim et al., in prep.), or from CO line observations with mm interferometers (Schinnerer et al.\ 2008; Riechers et al.\ 2010, in prep.; Balokovi\'{c} et al., in prep., Karim et al., in prep., Sheth et al., in prep).

%% The PREVIOUS 2 SECTIONS ARE SUPER-REPITITIVE

\section{PdBI follow-up of LABOCA-COSMOS SMGs} 
\label{sec:obs}

\begin{table*}
%\centering
%\begin{center}
%\rotatebox{90}
\caption{LABOCA sources observed with the PdBI}
\label{tab:cosla}
%\vskip 10pt
{\scriptsize
\begin{tabular}{|lll|ccc|ccc|ccc|}%{|c|l|l|c|c|l|c|c|c|c|c|c|}
\hline
    \multicolumn{3}{|c|}{LABOCA}       &   \multicolumn{3}{|c|}{AzTEC}   &  \multicolumn{3}{|c|}{MAMBO} \\ %& \multicolumn{3}{|c|}{BOLOCAM}   \\
    \,\,\, source & \,\,\,\,\,  \,\,\,\,\,  \,\, position & \,\,\,$F_\mathrm{870\mu m}$ & source & separation & $F_\mathrm{1.1mm}$ & source & separation & $F_\mathrm{1.2mm}$ \\
    \,\,\,\, name & \,\,\,\,\,  \,\,\,\,\, \,\,\,\,\,[J2000] &  \,\,\,\,\,[mJy] & name & arc.\ sec. & [mJy]  & name & arc.\ sec. & [mJy] \\ %& name & distance & flux \\
\hline 
COSLA-5  &  $10~00~59.6 \,\,\, +02~17~5.7$   & $12.5\pm2.6$  & -- & -- &    --    &        Cosbo-12  & 9.9 &$4.78\pm1.0$ \\ %&  BOLOCAM-24 &  15.4 & $(6.1-2.4)\pm1.9$ \\ 
COSLA-6    & $10~01~23.5 \,\,\,  +02~26~11.1$ & $16.0\pm3.3$  & -- & -- &  --  &--& -- &  -- \\ %&  -- & -- \\ 
COSLA-8    & $10~00~25.6 \,\,\,  +02~15~1.7$   & $6.9\pm1.6$  & -- &  -- &  -- & -- & -- & -- \\ %& --& -- \\
COSLA-9    & $10~00~14.2   \,\,\,  +01~56~40.5$  & $14.4\pm3.3$   & AzTEC/C8   &   3.8 & $8.7\pm1.1$ &-- & -- & -- \\ %& --& -- \\
COSLA-10   & $10~00~8.6   \,\,\,  +02~13~9.7$    & $6.6\pm1.7$  & -- & -- &     --      &    Cosbo-6   &  7.6  &$5.00\pm0.9$\\ %& -- & -- & -- \\
COSLA-11   & $10~01~14.1   \,\,\,  +01~48~12.4$  & $19.4\pm4.5$  & -- & -- &  --  & -- &  -- &  -- \\ %& -- \\ 
COSLA-12   & $10~00~30.2   \,\,\,  +02~41~37.6$  & $17.6\pm4.2$  & -- &   -- & -- &-- & -- & -- \\ %& --& -- \\
COSLA-13   & $10~00~32.2   \,\,\,  +02~12~38.4$   & $7.7\pm1.9$ & AzTEC/C145  & 9.3      &   $3.3^{+1.1}_{-1.2}$ &    Cosbo-5  &   3.5 &$5.11\pm0.9$\\ %& -- & -- & -- \\
COSLA-14   & $09~59~57.4   \,\,\,  +02~11~31.6$  & $7.9\pm2.1$ & AzTEC/C176  & 0.9    & $3.0\pm1.2$    &     Cosbo-10  & 8.0 &$5.88\pm1.1$\\ %& -- & -- & -- \\
COSLA-16   & $10~00~51.4   \,\,\,  +02~33~35.7$    & $14.0\pm3.6$   & -- & -- &  --  & -- &  -- &  -- \\ 
COSLA-17   & $10~01~36.4   \,\,\,  +02~11~2.9$   & $12.5\pm3.2$  & AzTEC/C12  & 6.2 & $7.5^{+1.0}_{-1.1}$ &-- & -- & -- \\ %& --& -- \\
COSLA-18   & $10~00~43.2   \,\,\,  +02~05~22.0$ & $10.0\pm2.6$   & AzTEC/C98  & 4.4 & $3.8^{+1.1}_{-1.2}$ & -- & -- & -- \\ %& --& -- \\
COSLA-19   & $10~00~7.7    \,\,\,  +02~11~42.7$    & $6.7\pm1.8$    & AzTEC/C34 & 8.9  &$5.3^{+1.1}_{-1.2}$ &        Cosbo-4 &  6.4 &$5.55\pm0.9$\\ %& -- & -- & -- \\
COSLA-23   & $10~00~10.1   \,\,\,  +02~13~33.3$   & $6.4\pm1.6$   &  -- & -- &          --&                 Cosbo-2 &  4.7 &$5.77\pm0.9$\\ %& -- & -- & -- \\
COSLA-25   & $09~58~51.5   \,\,\,  +02~15~53.7$   & $13.4\pm3.8$    & -- & -- &  --  & -- &  -- &  -- \\ %& -- \\ 
%COSLA-27   & $10~00~13.6   \,\,\,  +02~12~8.4$      & $6.6\pm1.7$  & AzTEC/C76   &    6.1  &$4.2\pm1.1$ &-- & -- & -- \\ %& --& -- \\
COSLA-30   & $09~58~47.7   \,\,\,  +02~21~7.4$      & $14.4\pm4.2$    & -- & -- &  --  & -- &  -- &  -- \\ %& -- \\ 
COSLA-33   & $10~00 ~9.2   \,\,\,  +02~19~11.6$      & $5.3\pm1.8$    & -- & -- &  --  & -- &  -- &  -- \\ %& -- \\ 
COSLA-35   & $10~00~23.4   \,\,\,  +02~21~55.5$    & $8.2\pm2.2$ & AzTEC/C38   &   6.4 &$5.1^{+1.2}_{-1.1}$ & -- & -- & -- \\ %& --& -- \\
COSLA-38  & $10~00~12.1   \,\,\,  +02~14~57.2$     & $5.8\pm1.6$        &             --&--&      --&       Cosbo-19    &   13.06 &$2.95\pm0.9$\\ %& -- & -- & -- \\
COSLA-40  & $09~59~26.3   \,\,\,  +02~20 ~6.0$    & $11.1\pm3.4$  & AzTEC/C117   &     13.8 &$3.7^{+1.1}_{-1.2}$ & -- & -- & -- \\ %& --& -- \\
COSLA-47   & $10~00~33.1   \,\,\,  +02~26~6.9$     & $9.0\pm2.8$ & AzTEC/C80    &  13.4 &$4.1\pm1.1$ & -- & -- & -- \\ %& --& -- \\
COSLA-48  & $10~00~24.7   \,\,\,  +02~17~42.3$    & $6.1\pm1.7$ & AzTEC/C160      & 10.0   &$3.1\pm1.2$    &         Cosbo-7  & 12.0 &$5.00\pm0.9$\\ %& -- & -- & -- \\
COSLA-50   & $10~00~19.0   \,\,\,  +02~16~54.0$  & $5.6\pm1.6$    & -- & -- &  --  & -- &  -- &  -- \\ %& -- \\ 
COSLA-51   & $10~00~11.5   \,\,\,  +02~12~7.1$     & $6.2\pm1.7$    & --   &    -- & -- & -- & -- & -- \\ %& --& -- \\
COSLA-54   & $09~58~38.3   \,\,\,  +02~14~2.5$     & $11.6\pm4.1$   & AzTEC/C13    & 8.4  &$8.7^{+1.3}_{-1.4}$& -- & -- & -- \\ %& --& -- \\
COSLA-62  & $10~01~53.2   \,\,\,  +02~20~9.5$      & $12.5\pm3.6$   & --  &     -- & -- & -- & -- & -- \\ %& --& -- \\
COSLA-128  & $10~01~38.3   \,\,\,  +02~23~36.1$   & $11\pm3.5$   & --   & -- & -- & -- & -- & -- \\ %& --& -- \\
COSLA-161  & $10~00~15.6   \,\,\,  +02~12~36.0$    & $5.2\pm1.7$  & AzTEC/C158    & 15.6  &$3.2^{+1.1}_{-1.2}$&       Cosbo-13S  &        5.9 &$1.37\pm0.9$\\ %& -- & -- & -- \\
\hline
\end{tabular}\\
}
%Fluxes are deboosted fluxes 
\vspace{1cm}
\end{table*}

%\begin{sidewaystable}
%\vspace{15cm}
\begin{table*}
\centering
%\begin{center}
%\rotatebox{90}
\caption{PdBI detections}
\label{tab:pdbi}
%\vskip 10pt
{\scriptsize
\begin{tabular}{|lccc|cc|ccc|}
\hline
\,\,\,\,\,source  &  position   &  $F_\mathrm{1.3mm}$& S/N &   \multicolumn{2}{|c|}{LABOCA}  &  \multicolumn{3}{|c|}{redshift}  \\
\,\,\,\,\,\,name  &  [J2000]  & [mJy] & & dist. [$"$] &  S/N$_\mathrm{870\mu m}$ &spectroscopic & photometric$^+$ & mm-to-radio \\
\hline
\vspace{0.5mm}
COSLA-5  &  10 00 59.521~ +02 17 02.57  & 2.04 $\pm$    0.49 & 4.1 &    3.4 &  5.0 &     -- & $0.85^{+0.07}_{-0.06}$ & -- \\ \vspace{0.5mm}
{\bf COSLA-6-N } &  10 01 23.640~ +02 26 08.42 &   2.66 $\pm$    0.49 &  5.4  &    3.4 & 4.7  &  --& -- & $4.01^{+1.51}_{-0.83}$\\ \vspace{0.5mm}
{\bf COSLA-6-S } &  10 01 23.570~ +02 26 03.62  &   3.08 $\pm$    0.65 &  4.8   &    7.6 &  4.7 & -- & $0.48^{+ 0.19}_{-0.22}$ & -- \\ \vspace{0.5mm}
COSLA-8  &  10 00 25.550~ +02 15 08.44  &  2.65 $\pm$    0.62&  4.2  &    6.8 &   4.6  & -- & $1.83^{+ 0.41}_{-1.31}$ & -- \\ \vspace{0.5mm}
COSLA-9-N & 10 00 13.750~ +01 56 41.54 & $1.69\pm0.47$& 3.2 & 7.0 & 4.5  & -- &  $2.62^{+ 0.60}_{-2.02}$ & -- \\
COSLA-9-S & 10 00. 13.829~ +01 56 38.64 & $1.87\pm0.58$ & 3.2 & 5.8 & 4.5 & -- & $1.90^{+ 0.26}_{-0.31}$ & -- \\
COSLA-11-N  &  10 01 14.260~ +01 48 18.86  &  2.15 $\pm$    0.62&  3.5  &    6.9 &   4.4  & -- & $0.75^{+0.23}_{-0.25}$ & -- \\ \vspace{0.5mm}
COSLA-11-S  &  10 01 14.200~ +01 48 10.31 &   1.43 $\pm$    0.48 &  3.0  &    2.6 &  4.4  & -- & $3.00^{+0.14}_{-0.07}$ & -- \\ \vspace{0.5mm}
COSLA-13  &  10 00 31.840~ +02 12 42.81  & 2.38 $\pm$    0.61&  3.8  &    7.0 &  4.3 & 2.175 & $2.11^{+ 0.14}_{-0.12}$ & -- \\ \vspace{0.5mm}
COSLA-16-N  &  10 00 51.585~ +02 33 33.56  & 1.39 $\pm$    0.32&   4.3 & 3.5 & 4.2  & -- & $2.16^{+0.12}_{-0.25}$ & -- \\ \vspace{0.5mm}
COSLA-16-S  &  10 00 51.554~ +02 33 32.09  & 1.19 $\pm$    0.33&   3.6 & 4.3 & 4.2  & -- & -- & $2.40^{+0.62}_{-0.51}$    \\ \vspace{0.5mm}
COSLA-16-E  &  10 00 51.780~ +02 33 33.58  &  2.26 $\pm$    0.58&  3.9   &    6.0 &   4.2  &-- &  $1.25^{+3.03}_{-1.15}$ & -- \\ \vspace{0.5mm}
{\bf COSLA-17-S } &  10 01 36.772~ +02 11 04.87 &   3.02 $\pm$    0.57 &  5.3  &    5.9 &  4.2  & -- & $0.70^{+0.21}_{-0.22}$ & -- \\ \vspace{0.5mm}
{\bf COSLA-17-N } &  10 01 36.811~ +02 11 09.66&  3.55 $\pm$    0.77  &  4.6 &    9.1 &   4.2  & -- & $3.37^{+0.14}_{-0.22}$ & -- \\ \vspace{0.5mm}
COSLA-18  &  10 00 43.190~ +02 05 19.17 &   2.15 $\pm$    0.48 &  4.5  &    2.8 &  4.2  & -- & $2.90^{+0.31}_{-0.43}$ & -- \\ \vspace{0.5mm}
COSLA-19  &  10 00 08.226~ +02 11 50.677 & 3.17 $\pm$ 0.76& 4.1  & 11.2 & 4.1  & -- & -- & $3.98^{+1.62}_{-0.90}$  \\ \vspace{0.5mm}
{\bf COSLA-23-N } &  10 00 10.161~ +02 13 34.95 &    3.42 $\pm$    0.47 &  7.3   &    1.9 & 3.9  & -- & $4.00^{+0.67}_{-0.90}$ & -- \\ \vspace{0.5mm}
{\bf COSLA-23-S } &  10 00 10.070~ +02 13 26.87  &  3.70 $\pm$    0.60&  6.2  &    6.4 &   3.9  & -- & $2.58^{+1.52}_{-2.48}$ & -- \\ \vspace{0.5mm}
COSLA-33 & 10 00 9.580~ +02 19 13.86 & $1.78\pm0.58$& 3.1 & 6 & 3.8  & -- & $3.27^{+ 0.22}_{-0.20}$ & --\\
COSLA-35  &  10 00 23.651~ +02 21 55.22  &  2.15 $\pm$    0.51&  4.2  &    3.7 &   3.8  & -- & $1.91^{+1.75}_{-0.64}$ & -- \\ \vspace{0.5mm}
COSLA-38  &  10 00 12.590~ +02 14 44.31  &    8.19 $\pm$    1.85&  4.4   &   14.8 & 3.7  & -- & $2.44^{+0.12}_{-0.11}$ & -- \\ \vspace{0.5mm}
COSLA-40  &  09 59 25.909~ +02 19 56.40  &    3.41 $\pm$    1.02&  3.4 &   11.3 &  3.7  & -- & $1.30^{+0.09}_{-0.11}$ & -- \\ \vspace{0.5mm}
{\bf COSLA-47}  &  10 00 33.350~ +02 26 01.66  &    3.11 $\pm$    0.59&  5.3  &    6.4 &  3.6  & -- & $2.36^{+0.24}_{-0.24}$ & -- \\ \vspace{0.5mm}
{\bf COSLA-54  } &  09 58 37.989~ +02 14 08.52  &   3.26 $\pm$    0.65&  5.0  &    7.6 &   3.6  & -- & $2.64^{+0.38}_{-0.26}$ & -- \\ \vspace{0.5mm}
{\bf COSLA-128 } &  10 01 37.990~ +02 23 26.50  &   4.50 $\pm$    0.94&  4.8 &   10.7 &   3.1  & -- & $0.10^{+ 0.19}_{-0.00}$ & -- \\ \vspace{0.5mm}
COSLA-161  &  10 00 16.150~ +02 12 38.27 &    2.54 $\pm$    0.74 &  3.4  &    8.5 &  3.1  & 0.187 & $0.19^{+0.05}_{-0.03}$ & --  \\ 
\hline
\end{tabular}
}
\\
%Fluxes are deboosted fluxes \\
$\mathrm{S/N}>4.5$ detections are marked bold-faced\\
$^+$ Photometric redshifts drawn from the total $\chi^2$ distribution as described in \s{sec:redsft} \ and {\em not} corrected for any systematic offsets
\vspace{1cm}
\end{table*}

\subsection{Observations}

The COSMOS-LABOCA observations\footnote{APEX project IDs: 080.A-3056(A), 082.A-0815(A) and 086.A-0749(A)} reach a rms of 1.5~mJy  per
beam (27.6'').  The rms increases towards the edges of the map, and the catalog was extracted from an area of $\sim 0.7$ deg$^2$ (Navarrete et al.\ in prep.). 
Our sample of 28 COSMOS LABOCA sources selected for PdBI follow-up observations (Table~\ref{tab:cosla}) was 
chosen with a requirement that the signal-to-noise in the LABOCA map is S/N$_{870\mu m}\gtrsim 3.8$.  Eight other LABOCA-COSMOS sources had already been observed previously with mm-interferometers\footnote{COSLA-1, COSLA-2, COSLA-3, COSLA-4, COSLA-60, COSLA-73, COSLA-89, and COSLA-158} (see Table~\ref{tab:interf}).

The SMGs in our sample were observed using the PdBI during two nights in Oct./Nov.\ 2007 (COSLA-10, and COSLA-19) and three nights in Oct./Nov.\ 2011 (the remaining 26 SMGs) in C- and D-configurations with 6 working antennas and the updated PdBI system. All observations were done in good/excellent millimeter weather conditions. During our 2007 observations we used the full 2 GHz bandwidth available with the correlator at the PdBI, and the receivers were tuned to 232~GHz and 231.5~GHz for observations of COSLA-19 and COSLA-10, respectively. The total on-source time was 2.3 and 2.2~hrs for COSLA-10, and COSLA-19, respectively. Our 2011 observations were done  using the WideX correlator covering a bandwidth of 3.6~GHz, with receivers tuned to 230~GHz (1.3~mm). These observations were performed in snap-shot mode 
cycling through the 26 SMGs in each track and observing the phase/amplitude calibrator for 2.25 minutes every 19.5 min. The total on-source time reached is $\sim43$~minutes per source. 

Sources J1055+018, J1005+066, J0923+392 were used for phase/amplitude calibration, and MWC349, J0923+392, 3C84 for flux calibration which we consider accurate within 10-20\%. Calibration and editing was done using the GILDAS CLIC package. For each source, the final $uv$ data were collapsed in frequency. The final dirty maps reach an rms noise level of 0.55 mJy beam$^{-1}$ and 0.39 mJy beam$^{-1}$, with FWHM beam sizes of $3.3''\times2.3''$ and $3.0''\times2.1''$ for COSLA-19 and COSLA-10, respectively, and an rms of 0.46 mJy beam$^{-1}$ with FWHM $\sim1.8''\times1.1''$ for the remaining SMGs.

\begin{figure}
\includegraphics[bb = 54 360 486 792, width=\columnwidth]{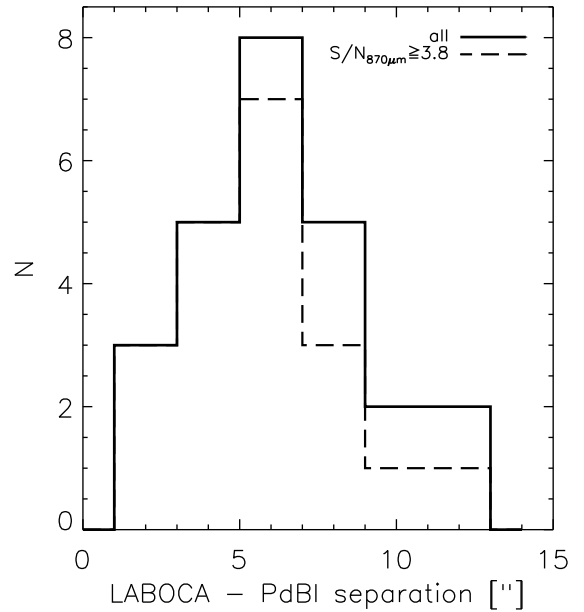}
\caption{ Distribution of separations between the PdBI sources and the corresponding LABOCA-COSMOS sources. \label{fig:separation}}
\end{figure}

\begin{figure}
\includegraphics[bb = 124 360 450 792, scale=0.85]{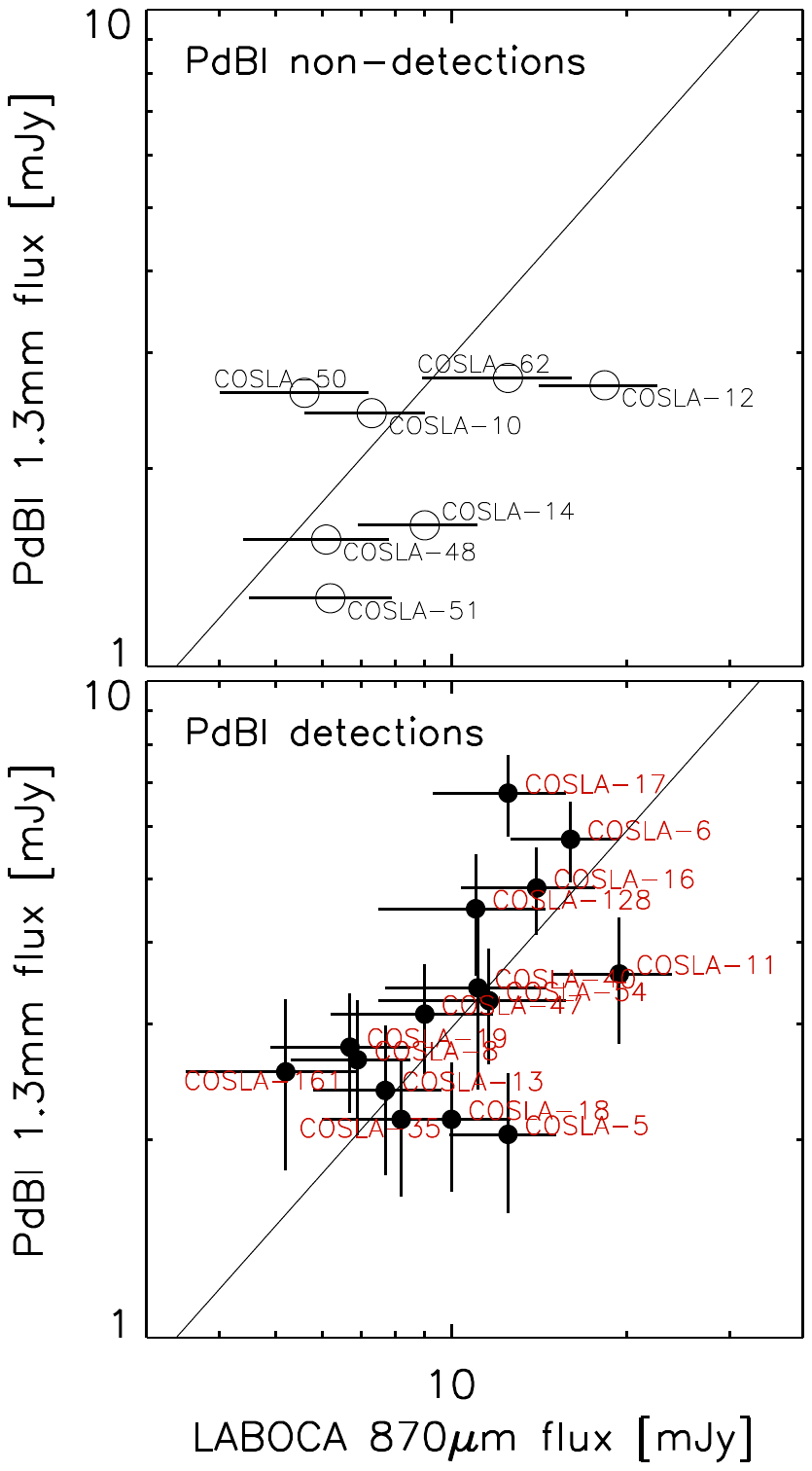}\\
\includegraphics[bb = 80 380 450 812, scale=0.58]{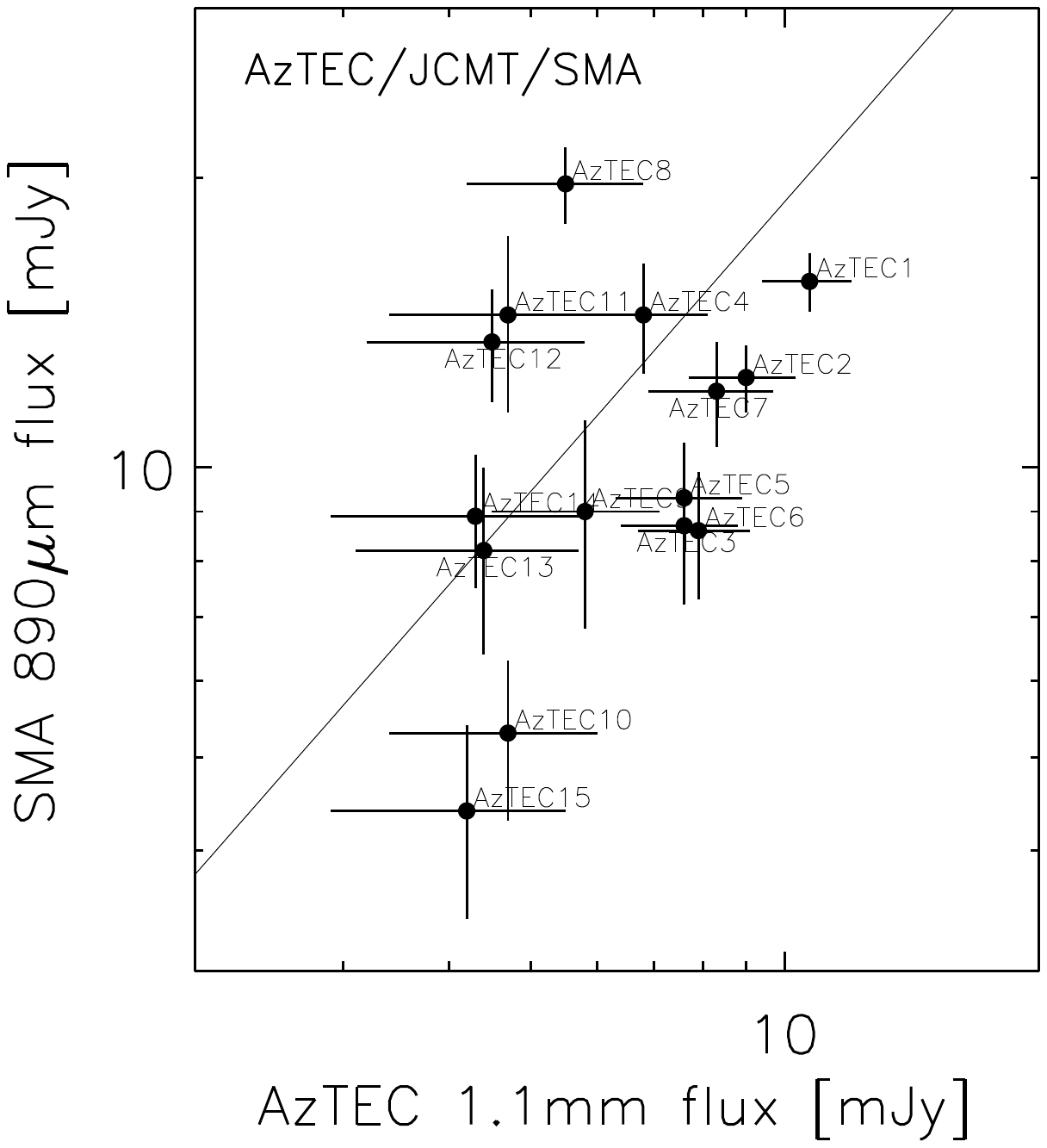}
\caption{ Comparison between LABOCA 870~$\mu$m and PdBI 1.3~mm fluxes for SMGs (indicated in the panel) detected (middle panel) and not detected (top panel) with the PdBI. For LABOCA sources identified as multiple PdBI sources the individual PdBI source fluxes were added, and for the LABOCA sources not detected by PdBI 1.3mm flux estimates for the most likely multi-wavelength counterparts were extracted from the PdBI maps (see \s{sec:nondet} \  for details). The bottom panel shows the comparison between AzTEC 1.1~mm and SMA 890~$\mu$m fluxes (adopted from Younger et al.\ 2007, 2009) for AzTEC/JCMT SMGs in our \mmsample . The solid line in all panels shows the flux ratios for a spectral power law index of 3.\label{fig:flux}}
\end{figure}

\subsection{PdBI mm-sources}

We searched for point sources in the dirty 1.3~mm PdBI maps within a $\sim14"$ radius from the phase center, which about corresponds to the LABOCA map resolution of $27"$. 
Peaks with $\mathrm{S/N} > 4.5$ were considered detections regardless of any multi-wavelength
association. When such peaks were present in sidelobe-contaminated regions, we tested the reality of the sources by cleaning the map by setting a CLEAN box around the brightest peak (see Appendix A for notes on individual sources). For peaks with $3\lesssim\mathrm{S/N}\leq 4.5$ 
we required an associated optical, 
near/mid-IR, or radio source within a radius of $\lesssim1"$.  Assuming a Gaussian 
noise distribution, the $\mathrm{S/N} > 4.5$ requirement implies 
a false detection rate of $\sim0.15\%$  
within a search radius of $14"$. A mm source association with optical, NIR, MIR, or radio sources further decreases the probability that the source is false\footnote{If the source is {\em independently} detected in various bands then the final false match probability is given by the product of the individual-band false match probabilities.}. Given the surface densities of sources present in various catalogs the false match probabilities independently estimated for each band are 12\% (optical), $\sim2\%$ (for each, UltraVista Y, J,  H, Ks, and IRAC 3.6$\mu$m), and 0.017\% (20~cm radio). 

To further constrain the false match probability, we performed a source search in the same way as described above, but on the inverted, i.e. negative maps. We find only one occurrence of a $>4.5\sigma$ (i.e.\ $4.8\sigma$) peak ($\sim10"$ away from the phase center and with no multi-wavelength counterpart) consistent with the above given false match probability expectation. We further find $\sim10\%$ of $3\lesssim\mathrm{S/N}\leq4.5$ peaks matched to multi-wavelength counterparts. This suggests a $\sim10\%$ false match probability for our $3\lesssim\mathrm{S/N}\leq4.5$ sources. Hereafter we consider $\mathrm{S/N}>4.5$ detections as significant and those with $3<\mathrm{S/N}\leq4.5$ as tentative.

The 1.3~mm sources and their properties are summarized in Table~\ref{tab:pdbi}.   Comments on individual sources and their PdBI maps and multi-wavelength stamps are presented in Appendix~A. 
Of the 28 LABOCA sources observed, 9 yielded no detection in the 1.3~mm maps (see next Section).
Six of the 19 detected LABOCA sources break up into multiple sources, so that in total we identify 26 submm  sources.
Nine of these 26 sources have $\mathrm{S/N}>4.5$, 7 have S/N between 4 and 4.5, and 10  between 3 and 4. 
The distribution of separations between the LABOCA source position and the corresponding PdBI source position is shown in \f{fig:separation} . We find a median separation of $6.40"$ for all sources, and $5.95"$ for those with $\mathrm{S/N_{870\mu m}}\geq3.8$. This is consistent with the results based on artificial source tests performed on the LABOCA map. They result in a positional uncertainty for LABOCA sources down to  $\mathrm{S/N_{870\mu m}}=3.8$ of $\sim5.3"$ with an inter-quartile range of $3.1" - 9.8"$  (Navarrete et al., in prep.). 

All detections except COSLA-6-1 and COSLA-6-2 are consistent with point-sources at our resolution.
We extract their fluxes  from the brightest pixel value in the dirty maps. The
flux uncertainty is estimated as the rms noise level in the map. The fluxes for COSLA-6-1 and COSLA-6-2 were obtained by fitting a double Gaussian to the source. All fluxes (tabulated in Table~\ref{tab:pdbi}) \ were corrected for the primary beam response of the PdBI dishes (assuming a Gaussian distribution with HPBW of 21").

Scaling the observed 1.3~mm fluxes to the LABOCA 870~$\mu$m, and where available to the AzTEC 1.1~mm, or MAMBO 1.2~mm fluxes (Table~\ref{tab:cosla}  ), yields consistent values.\footnote{The fluxes were scaled assuming $S_\nu\propto\nu^{2+\beta}$ where $S_\nu$ is the flux density at frequency $\nu$ and $\beta=1$ the dust emissivity index.}  This is shown in \f{fig:flux} \ and described in more detail for each source in Appendix~A. This further strengthens the validity of our detections.

\subsection{Non-detections}
\label{sec:nondet}

Nine LABOCA sources remain undetected within our PdBI observations. The reasons for this could be that i) the LABOCA sources are fainter than  our PdBI  sensitivity limit ($1\sigma\sim0.46$~mJy), ii) the LABOCA sources break up into multiple components at 1.5'' resolution and are all fainter than  our flux limit  or iii) the LABOCA source is spurious. To investigate this further we have made use of the COSMOS multi-wavelength data by assigning statistical counterparts to those LABOCA sources given our radio, 24~$\mu$m, and IRAC data (see \s{sec:pstat} \ for details). For 3/9 sources we find no robust or tentative counterparts while for 6/9 we find either one or several tentative or robust counterparts (see \f{fig:statstampscoslanodet} \ and Appendix~A for details). For the latter sources we have then identified the maximum pixel value within a circular annulus of 1" radius in the 1.3~mm map. If multiple potential counterparts were present, we have summed up the maximum pixel values. Such derived 1.3~mm fluxes, compared to the LABOCA 870~$\mu$m fluxes are shown in \f{fig:flux} . They agree well with the LABOCA fluxes suggesting that the LABOCA sources are not spurious but that at interferometric resolution and sensitivity, the source is breaking up into multiple-components fainter than our 1.3~mm sensitivity limit.  This is also consistent with the results based on artificial source tests performed on the LABOCA map which yield that down to a $\mathrm{S/N_{870\mu m}}=3.8$ $5\pm3$ spurious sources are expected (Navarette et al., in prep.).  

\subsection{Panchromatic properties of PdBI-detected LABOCA-COSMOS SMGs}

Twenty-three of the 26 PdBI-detected LABOCA SMGs can be associated with multi-wavelength counterparts drawn from the deep COSMOS photometric catalog. In addition to the UV to MIR photometry from the COSMOS multi-wavelength catalog we have added deep YJHK imaging from the recent UltraVista Data Release 1.  Their photometry is presented in Table~\ref{tab:phot}.
The COSMOS spectroscopic database (Lilly et al.\ 2007; 2009; Trump et al.\ 2007) provided spectroscopic redshifts for the COSLA-13 and COSLA-161 counterparts. From the 26 SMGs identified interferometrically only COSLA-161 was found to be  associated with X-ray emission in the Chandra-COSMOS data (Elvis et al.\ 2009). 

For each PdBI source we extracted the 1.4~GHz flux from the VLA-COSMOS Deep map (Schinnerer et al.\ 2010) using the AIPS task MAXFIT (see Table~\ref{tab:phot}).  Thirteen of the 26 sources ($\sim50\%$ with a Poisson error of $\pm14\%$)  are associated with a $>3\sigma$ radio peak, where the average rms noise level is $rms_\mathrm{1.4GHz}=9~\mu$Jy/beam. Nine sources are detected with $\mathrm{S/N_{1.4GHz}}>4$. This radio detection fraction does not depend on the significance of the PdBI-source: from those with $\mathrm{S/N}>4.5$ we find five of nine have a radio counterpart whereas from those with $\mathrm{S/N}\geq5$,  three of six show a radio counterpart.  %KS: This previous sentence makes no sense to me.

\section{Statistical samples of SMGs in the COSMOS field identified at intermediate ($\lesssim2"$) resolution }
\label{sec:statsamples}

Our PdBI observations yielded 26 (9 significant $\mathrm{S/N}>4.5$ and 17 tentative, $3<\mathrm{S/N}\leq4.5$) source detections at 1.3~mm. Combined with previous mm-interferometric detections of SMGs in the COSMOS field this adds to 50 SMGs detected with mm-interferometers. To date this is the largest interferometric SMG sample. It can be utilized, e.g., for a critical assessment of statistical counterpart identification methods, and to measure the redshift distribution of SMGs with unambiguously determined multi-wavelength counterparts.

In the following we examine two statistically significant samples of COSMOS SMGs detected at mm-wavelengths at $\lesssim2"$ resolution:

\begin{enumerate}
\item[] {\bf \mmsample :} 15 SMGs drawn from the 1.1~mm AzTEC/JCMT-COSMOS survey at $18"$ angular resolution (AzTEC-1 to AzTEC-15; see Table~\ref{tab:interf}) that  form a ($\mathrm{S/N_{1.1mm}}>4.5$) flux-limited ($F_\mathrm{1.1mm}\gtrsim4.2$~mJy), 1.1~mm sample. All 15 SMGs were followed-up and detected with the SMA at 890~$\mu$m,  yielding 17 interferometric sources (as two were found to be multiples; Younger et al.\ 2007, 2009). More details about the multi-wavelength photometry of the counterparts are provided in Appendix B.
\vspace{2mm}
\item[] {\bf \submmsample }: LABOCA-COSMOS sources that were identified at $27"$ angular resolution and confirmed through (sub)mm-interferometry  at $\lesssim2"$ resolution (Younger et al.\ 2007; 2009; Aravena et al.\ 2010a; \smo\ et al.\ 2012., this work). 
Thirty six LABOCA sources were followed-up in total with the SMA, CARMA, and PdBI, and 
 9 resulted in no detection within the PdBI observations down to a depth of $\sim0.46$~mJy/beam. The remaining 27 yielded 16 significant ($\mathrm{S/N}>4.5$) and 18 tentative ($3<\mathrm{S/N}\leq4.5$) interferometric (sub)mm-detections. For the less significant detections we required an association with a source seen at other wavelengths (see Table~\ref{tab:pdbi} \ and Table~\ref{tab:interf}). The 16 {\em significant} detections form the least biased sample, and we hereafter refer to this subsample as the least-biased \submmsample .
\end{enumerate}

The sources in the 1.1mm- and 870$\mu$m-selected samples are summarized in Table~\ref{tab:interf} and Table~\ref{tab:statsamples}. Five SMGs belong to both samples\footnote{AzTEC-1/COSLA-60, AzTEC-2/COSLA-4, AzTEC-8/COSLA-73, AzTEC-9/COSLA-3, AzTEC-13/COSLA-158} (see Table~\ref{tab:interf}). 
Hereafter we will use these two samples to investigate blending, counterpart properties, and the redshift distribution of SMGs.  For clarity a master table of all interferometrically observed SMGs in the COSMOS field is given in Table~\ref{tab:master}.

\begin{table}
%\centering
%\begin{center}
%\rotatebox{90}
\caption{Statistical samples of  SMGs with $\lesssim2"$ angular resolution mm-detections in the COSMOS field}
\label{tab:statsamples}
%\vskip 10pt
\begin{tabular}{|ll|ll|}
\hline
1.1mm-selected & \,\,\,best & least-biased- & \,\,\,best \\
sample & redshift & 870$\mu$m-selected & redshift \\
& & sample & \\
\hline
AzTEC-1 & 4.64$^+$ & COSLA-1& $3.83^{+0.68}_{-0.49}$$^*$\\ \vspace{0.5mm}

AzTEC-2 & 1.125$^+$ & COSLA-2 & 2.490$^+$\\ \vspace{0.5mm}

AzTEC-3 & 5.299$^+$ & COSLA-3 & 1.357$^+$ \\ \vspace{0.5mm}

AzTEC-4 & $4.93^{+ 0.43}_{- 1.11}$$^\#$ &  COSLA-4 & 1.125$^+$\\ \vspace{0.5mm}

AzTEC-5 & 3.971$^+$ &  COSLA-6-N & $4.01^{+1.51}_{-0.83}$$^*$ \\ \vspace{0.5mm}

AzTEC-6 & 0.802$^+$ &  COSLA-6-S & $0.48^{+ 0.19}_{-0.22}$\\ \vspace{0.5mm}

AzTEC-7 & $2.30\pm0.10$ &  COSLA-17-S & $0.70^{+0.21}_{-0.22}$\\ \vspace{0.5mm}

AzTEC-8 & $3.179^+$ &  COSLA-17-N & $3.54^{+0.14}_{-0.22}$$^\#$\\ \vspace{0.5mm}

AzTEC-9 & 1.357$^+$ &  COSLA-23-N & $4.20^{+0.67}_{-0.90}$$^\#$\\ \vspace{0.5mm}

AzTEC-10 & $2.79^{+1.86}_{-1.29}$ &  COSLA-23-S & $2.58^{+1.52}_{-2.48}$\\ \vspace{0.5mm}

AzTEC-11$^{**}$ & 1.599$^+$&  COSLA-47 & $2.36^{+0.24}_{-0.24}$\\ \vspace{0.5mm}
~~AzTEC-11N$^{**}$ & $1.51^{+0.41}_{-0.92}$ &  COSLA-54 & $2.64^{+0.38}_{-0.26}$\\ \vspace{0.5mm} 
~~AzTEC-11S$^{**}$ & $>2.58$ &  COSLA-60 & 4.64$^+$\\ \vspace{0.5mm}

AzTEC-12 & $2.54^{+ 0.13}_{- 0.33}$&  COSLA-73 & $3.179^+$ \\ \vspace{0.5mm}

AzTEC-13 & $>3.59^*$&  COSLA-128 & $0.10^{+ 0.19}_{-0.00}$\\ \vspace{0.5mm}

AzTEC-14-E & $>3.03^*$&  COSLA-158 & $>3.59^*$\\ 

AzTEC-14-W & $ 1.30^{+ 0.12}_{- 0.36}$&  & \\

AzTEC-15 & $3.17^{+ 0.29}_{- 0.37}$$^\#$&  & \\
\hline
\end{tabular}\\
Five SMGs belong to both samples; AzTEC-1/COSLA-60, AzTEC-2/COSLA-4, AzTEC-8/COSLA-73, AzTEC-9/COSLA-3, AzTEC-13/COSLA-158 \\
$^{**}$ Here we keep the nomenclature given by Younger et al.\ (2009). Note however that AzTEC-11-S is the northern component of the AzTEC-11 SMG, and AzTEC-11-N is its southern component (see Tab.~1 in Younger et al.\ 2009)\\
$^+$ spectroscopic redshift (see Table~\ref{tab:interf} for references)\\
$^*$ mm-to-radio flux ratio based redshift \\
$^\#$ photometric redshift corrected for the systematic offset of $0.04(1+z)$, see \f{fig:photz} , with errors drawn from the total $\chi^2$ distribution
\end{table}

%%%%%%%%%%%%%%%%%%%%%%%%%%%%%%%%%%%%%%%%%
%%% MASTER TABLE %%%%%%%%%%%%%%%%%%%%%%%%%%%%%
%%%%%%%%%%%%%%%%%%%%%%%%%%%%%%%%%%%%%%%%%
\begin{table*}
\caption{Master table of interferometrically observed SMGs in the COSMOS field}
\label{tab:master}
{\scriptsize
\begin{tabular}{lcccc}
\hline
    Source       &  other names &
     \multicolumn{2}{c}{~~~~~~~(sub)mm-interferometry} & statistical  \\
     & & observed & detected & interferometric  sample \\
    \hline
    \vspace{0.5mm}
      AzTEC-1$^{(1,5)}$ & COSLA-60$^{(4)}$, AzTEC/C5$^{(2)}$  & SMA, CARMA, PdBI & $\surd$ & 1.1mm, 870$\mu$m, least-biased-870$\mu$m\\
      AzTEC-2$^{(1,5)}$ & COSLA-4$^{(4)}$, AzTEC/C3$^{(2)}$ & SMA, CARMA & $\surd$ &  1.1mm, 870$\mu$m, least-biased-870$\mu$m\\
      AzTEC-3$^{(1,5)}$     &AzTEC/C138$^{(2)}$ & SMA, CARMA, PdBI & $\surd$ & 1.1mm \\
      AzTEC-4$^{(1,5)}$     & AzTEC/C4$^{(2)}$& SMA & $\surd$ & 1.1mm \\
      AzTEC-5$^{(1,5)}$     &AzTEC/C42$^{(2)}$ & SMA & $\surd$ & 1.1mm \\
      AzTEC-6$^{(1,5)}$     & AzTEC/C106$^{(2)}$  & SMA & $\surd$ & 1.1mm \\
      AzTEC-7$^{(1,5)}$     & & SMA & $\surd$ & 1.1mm \\
      AzTEC-8$^{(1,6)}$ & COSLA-73$^{(4)}$, AzTEC/C2$^{(2)}$ & SMA & $\surd$ &  1.1mm, 870$\mu$m, least-biased-870$\mu$m\\
     AzTEC-9$^{(1,6)}$ & COSLA-3$^{(4)}$, AzTEC/C14$^{(2)}$ & SMA & $\surd$ &  1.1mm, 870$\mu$m, least-biased-870$\mu$m\\
      AzTEC-10$^{(1,6)}$    & & SMA & $\surd$ & 1.1mm \\
      AzTEC-11$^{(1,6)}$ & AzTEC-11-N$^{(6)}$, AzTEC-11-S$^{(6)}$, AzTEC/C22$^{(2)}$ & SMA & $\surd$ & 1.1mm \\
      AzTEC-12$^{(1,6)}$    &  AzTEC/C18$^{(2)}$ & SMA & $\surd$ & 1.1mm \\
      AzTEC-13$^{(1,6)}$ & COSLA-158$^{(4)}$ & SMA & $\surd$ & 1.1mm, 870$\mu$m, least-biased-870$\mu$m \\
     AzTEC-14$^{(1,6)}$ & AzTEC-14-E$^{(6)}$, AzTEC-14-W$^{(6)}$ & SMA & $\surd$ & 1.1mm \\
      AzTEC-15$^{(1,6)}$    & AzTEC/C10$^{(2)}$  & SMA & $\surd$ & 1.1mm \\
 %%%%%
     AzTEC-16$^{(1,13)}$ &  -- & CARMA  & $\surd$ & --\\
%%%
     J1000+0234$^{(1,14)}$  &  AzTEC/C17$^{(2)}$ & VLA  & $\surd$ & --\\
      AzTEC/C1$^{(2,7)}$  &    COSLA-89$^{(4)}$ & CARMA  & $\surd$ & 870$\mu$m\\
      Cosbo-1$^{(3,16)}$       &  COSLA-1$^{(4)}$, AzTEC/C7$^{(2)}$ & SMA  & $\surd$ & 870$\mu$m, least-biased-870$\mu$m\\
      Cosbo-3$^{(3,7)}$ & COSLA-2$^{(4)}$, AzTEC/C6$^{(2)}$  & CARMA  & $\surd$ & 870$\mu$m, least-biased-870$\mu$m\\
      Cosbo-8$^{(3,7)}$ &  AzTEC/C118$^{(2)}$   & CARMA   & $\surd$ & -- \\
      Cosbo-14$^{(3,16)}$    & & SMA  & $\surd$ & --\\
 %%%
COSLA-5$^{(4,17)}$  &   Cosbo-12$^{(3)}$  & PdBI   & $\surd$ & 870$\mu$m \\
COSLA-6$^{(4,17)}$    & COSLA-6-N$^{(17)}$, COSLA-6-S$^{(17)}$ & PdBI   & $\surd$ & 870$\mu$m, least-biased-870$\mu$m\\
COSLA-8$^{(4,17)}$    & & PdBI   & $\surd$ & 870$\mu$m\\
COSLA-9$^{(4,17)}$    & AzTEC/C8$^{(2)}$, COSLA-9-N$^{(17)}$, COSLA-9-S$^{(17)}$ & PdBI   & $\surd$ & 870$\mu$m\\
COSLA-10$^{(4,17)}$   &   Cosbo-6$^{(3)}$  & PdBI &  -- & --\\
COSLA-11$^{(4,17)}$   & COSLA-11-N$^{(17)}$, COSLA-11-S$^{(17)}$& PdBI   & $\surd$ & 870$\mu$m\\
COSLA-12$^{(4,17)}$   & & PdBI & -- & --\\
COSLA-13$^{(4,17)}$   & AzTEC/C145$^{(2)}$,  Cosbo-5$^{(3)}$  & PdBI   & $\surd$ & 870$\mu$m\\
COSLA-14$^{(4,17)}$   & AzTEC/C176$^{(2)}$, Cosbo-10$^{(3)}$ & PdBI & -- & --\\
COSLA-16$^{(4,17)}$   & COSLA-16-N$^{(17)}$, COSLA-16-S$^{(17)}$& PdBI   & $\surd$ & 870$\mu$m\\
COSLA-17$^{(4,17)}$   & AzTEC/C12$^{(2)}$, COSLA-17-N$^{(17)}$, COSLA-17-S$^{(17)}$ & PdBI   & $\surd$ & 870$\mu$m, least-biased-870$\mu$m\\
COSLA-18$^{(4,17)}$   & AzTEC/C98$^{(2)}$ & PdBI   & $\surd$ & 870$\mu$m\\
COSLA-19$^{(4,17)}$   & AzTEC/C34$^{(2)}$,   Cosbo-4$^{(3)}$ & PdBI   & $\surd$ & 870$\mu$m\\
COSLA-23$^{(4,17)}$   &     Cosbo-2$^{(3)}$, COSLA-23-N$^{(17)}$, COSLA-23-S$^{(17)}$ & PdBI   & $\surd$ & 870$\mu$m, least-biased-870$\mu$m\\
COSLA-25$^{(4,17)}$   & & PdBI & -- & -- \\
COSLA-30$^{(4,17)}$   & & PdBI & -- & --\\
COSLA-33$^{(4,17)}$   & & PdBI   & $\surd$ & 870$\mu$m\\
COSLA-35$^{(4,17)}$   & AzTEC/C38$^{(2)}$ & PdBI   & $\surd$ & 870$\mu$m\\
COSLA-38$^{(4,17)}$  &    Cosbo-19$^{(3)}$ & PdBI   & $\surd$ & 870$\mu$m\\
COSLA-40$^{(4,17)}$  & AzTEC/C117$^{(2)}$ & PdBI   & $\surd$ & 870$\mu$m\\
COSLA-47$^{(4,17)}$   & AzTEC/C80$^{(2)}$ & PdBI   & $\surd$ & 870$\mu$m, least-biased-870$\mu$m\\
COSLA-48$^{(4,17)}$  & AzTEC/C160, Cosbo-7$^{(3)}$& PdBI & -- & -- \\
COSLA-50$^{(4,17)}$   & & PdBI & -- & --\\
COSLA-51$^{(4,17)}$   & & PdBI & --& --\\
COSLA-54$^{(4,17)}$   & AzTEC/C13$^{(2)}$  & PdBI   & $\surd$ & 870$\mu$m, least-biased-870$\mu$m\\
COSLA-62$^{(4,17)}$  & & PdBI & -- & --\\
COSLA-128$^{(4,17)}$  & & PdBI   & $\surd$ & 870$\mu$m, least-biased-870$\mu$m\\
COSLA-161$^{(4,17)}$  & AzTEC/C158$^{(2)}$, Cosbo-13S$^{(3)}$ & PdBI   & $\surd$ & 870$\mu$m, least-biased-870$\mu$m\\
\hline
\end{tabular}\\
(1) Scott et al.\ (2008); 
(2) Aretxaga et al.\ (2011); 
(3) Bertoldi et al.\ (2007); 
(4) Navarette et al., in prep.; 
(5) Younger et al.\ (2007); 
(6) Younger et al.\ (2009); 
(7) \smo\ et al.\ (2011); 
(8) Balokovi\'{c} et al., in prep.; 
(9) Capak et al.\ (2010); Riechers et al.\ (2010); 
(10) Karim et al., in prep.; 
(11) Younger et al.\ (2009); 
(12) Riechers et al., in prep.; 
(13) Sheth et al., in prep.; 
(14) Capak et al.\ (2009); Schinnerer et al.\ (2009); 
(15) \smo\ et al., 2012; 
(16) Aravena et al.\ (2010); 
(17) this work
}
\end{table*}

\section{Properties of  single-dish detected SMGs when mapped at intermediate angular resolution}
\label{sec:counterparts}

In this section we investigate the multiplicity of SMGs detected at intermediate ($\lesssim2"$) angular resolution, and the statistical multi-wavelength counterpart association that is commonly applied to single-dish detected SMGs. 

\subsection{Blending: Single-dish SMGs breaking-up into multiple sources}

In the \mmsample  \ of the 15 AzTEC sources mapped with the SMA, AzTEC-14 clearly breaks up into two sources within the AzTEC beam when observed at $\sim2"$ angular resolution (AzTEC-14-E and AzTEC-14-W), while AzTEC-11 shows extended structure and is best fit by a double Gaussian (see Appendix~B and Younger et al.\ 2009 for details). Thus, in the \mmsample \ only two of 15 (13\% with a Poisson uncertainty of 9\%) single-dish sources are blended, i.e., they break up into multiple components when observed at intermediate angular resolution. The comparison between the single-dish 1.1~mm AzTEC and the interferometric SMA 890~$\mu$m fluxes for these 15 sources, shown in the bottom panel of \f{fig:flux} , suggests that although the agreement is reasonable, it is possible that some faint companions were missed thus potentially increasing the fraction of multiples in this sample.

For the \submmsample \ of 36 LABOCA-COSMOS SMGs followed-up and 27 out of these detected with interferometers, 6 SMGs\footnote{COSLA-6, COSLA-9, COSLA-11, COSLA-16, COSLA-17, COSLA-23} ($22\%\pm9\%$)  break up into multiple sources when observed with interferometers. This is within the statistical uncertainties of the results for the \mmsample .  Three more LABOCA SMGs detected by PdBI\footnote{COSLA-3, COSLA-5, and COSLA-47} may also consist of multiple components (see Appendix~A, Aravena et al.\ 2010b and \smo\ et al.\ 2012 for details), and the P-statistics (see next section) suggests that at least four of the LABOCA sources not detected by our PdBI observations\footnote{COSLA-10, COSLA-12, COSLA-48, and COSLA-50} are potential blends. This suggests a fraction of $\gtrsim6/36\approx17\%$, potentially rising up to $\sim40\%$ of  LABOCA sources blended within the single-dish beam. This is consistent with the fraction obtained if only the least-biased-\submmsample \ is considered (see Table~\ref{tab:statsamples}).

\subsection{Counterpart assignment methods to single-dish detected SMGs}

Here we perform a statistical counterpart assignment for
the SMGs detected at low angular resolution in our 1.1mm-  and 870$\mu$m-selected samples, and compare them with the exact positions obtained from the interferometers.
 
\subsubsection{$P$-statistic}
\label{sec:pstat}

The most common way to associate single-dish identified SMGs 
with counterparts in higher resolution maps is through the
$P$-statistic (Downes et al.\ 1986), i.e., 
the corrected Poisson probability that, e.g., a radio source is
identified by chance in a background of randomly distributed radio/IR
sources (Downes et al. 1986; Ivison et al. 2002, 2005). For a
potential radio counterpart of flux density $S$ at distance $r$
from the SMG position, 
$P_c=1-\exp(-P_S [1+\ln(P_S/P_{3\sigma})])$, where  
$P_S=1-\exp(-\pi  r^2 n_S)$ 
is the raw probability to find a source brighter than $S$ within a distance $r$ from the (sub-)mm source, 
$n_S$ is the local density of sources brighter than the candidate,
and $P_{3\sigma} = \pi r^2 n_{3\sigma}$ is the critical Poisson level, with n$_{3\sigma}$
being the source surface density above the $3\sigma$ detection level.
Robust counterparts are considered those with $P_c \leq 0.05$, 
while tentative counterparts have $0.05<P_c<0.2$.

The commonly used samples search for SMG counterparts are from 
radio, 24$\mu$m, and/or IRAC flux or color-selected data (e.g.,
Pope et al.\ 2005, Biggs et al.\ 2011, Yun et al.\ 2012).   The maximum search radius
is adjusted to the positional uncertainty of the SMG.

\begin{figure*}
\hspace{-0.6mm}\includegraphics[scale=0.65]{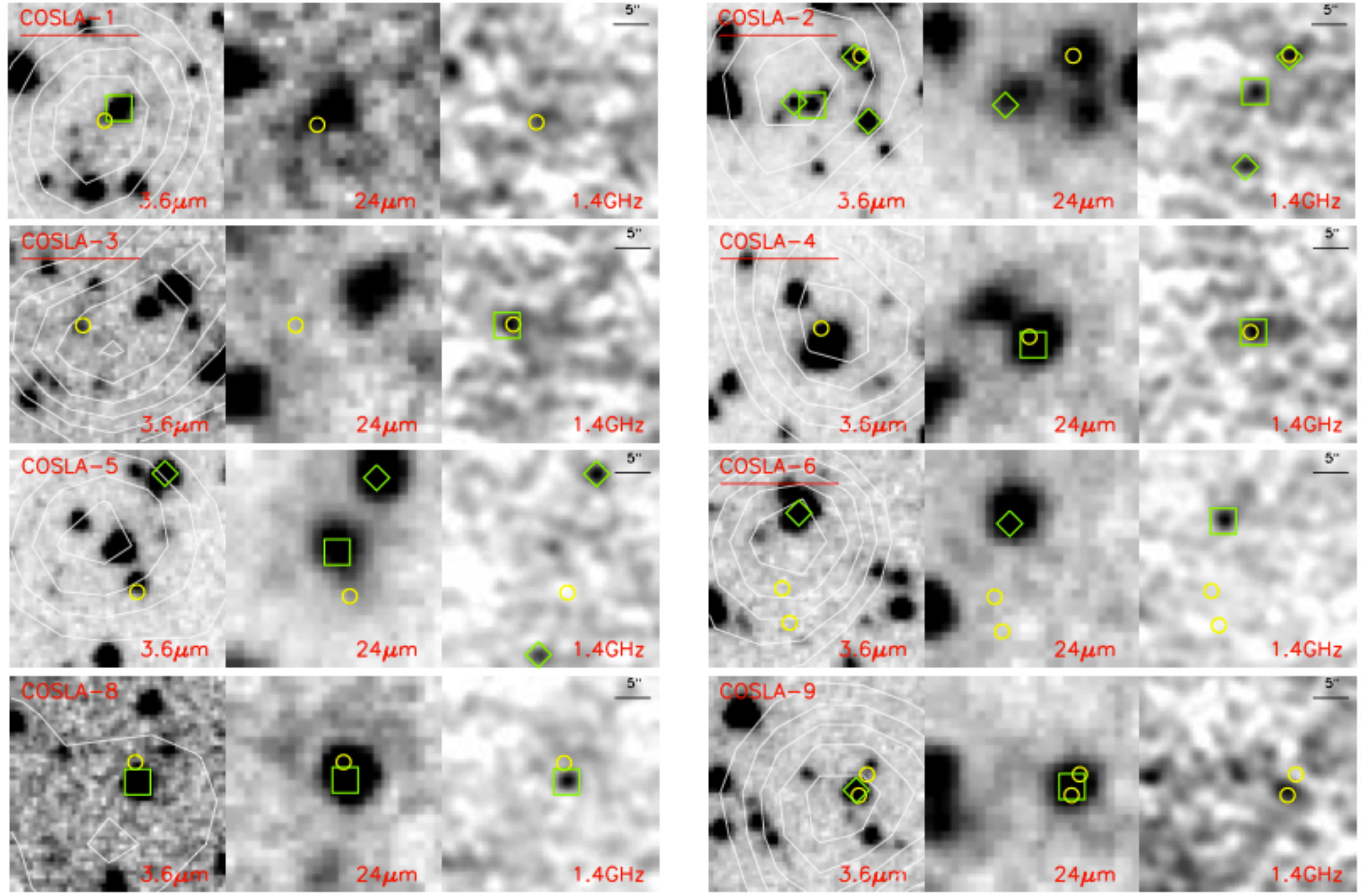}
\includegraphics[scale=0.65]{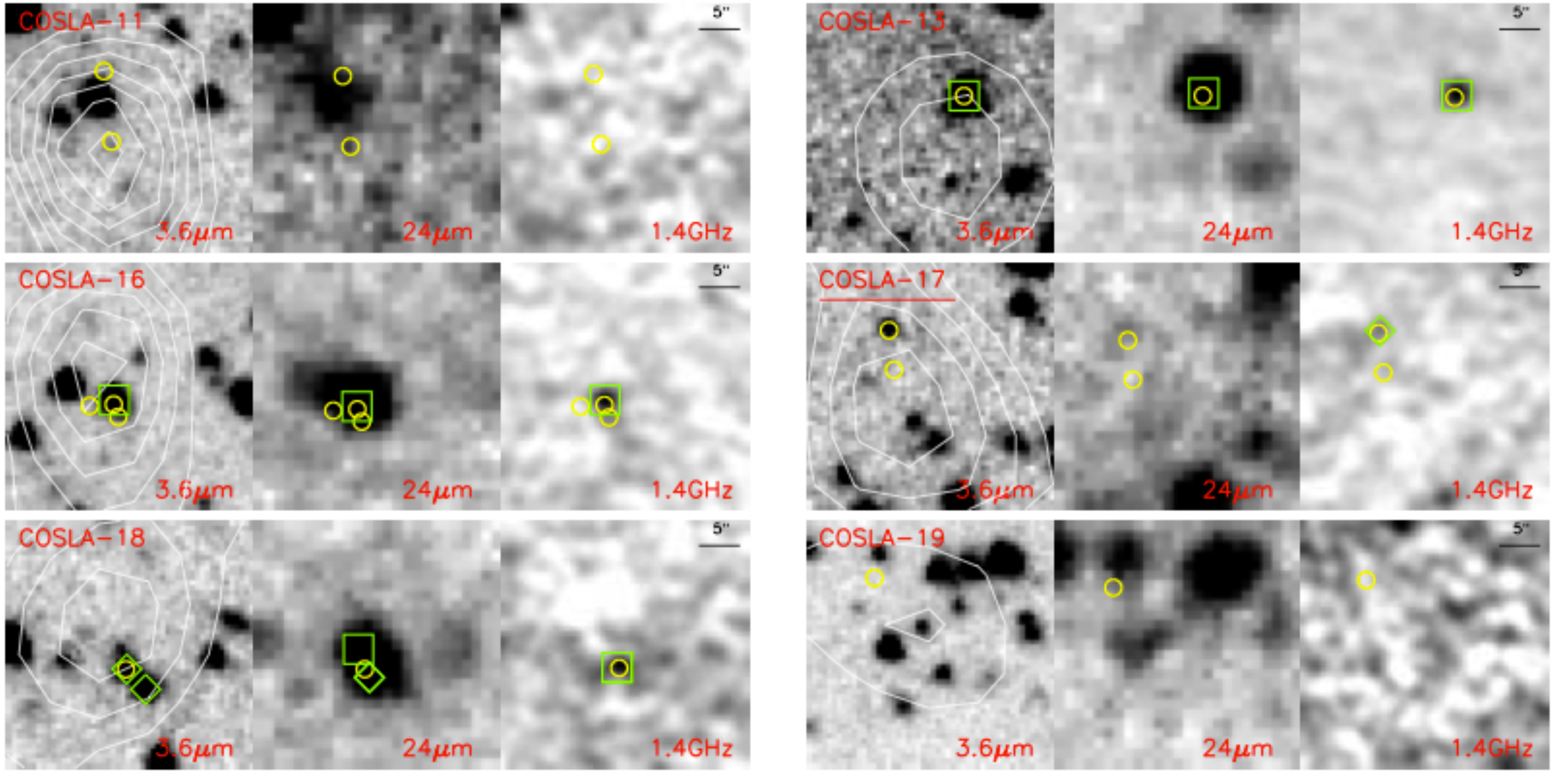}\\
\caption{ $3.6~\mu$m, $24~\mu$m, and $20$~cm stamps ($30"\times 30"$ area) for LABOCA COSMOS sources detected by mm-interferometers at $\lesssim2"$ resolution (see Table~\ref{tab:pdbi} ). The bands and sources are indicated in the panels and the names of sources detected with interferometers at $\mathrm{S/N}>4.5$ are underlined. The thick yellow circle, $2"$ in diameter, indicates the mm-interferometer position. Robust (square) and tentative (diamond) counterparts determined via P-statistic in each particular ($3.6~\mu$m, $24~\mu$m, and $20$~cm) band are also shown (see text for details; see also Table~\ref{tab:statcountcosla} ).  For each source LABOCA contours in $1\sigma$ steps starting at $2\sigma$ (with locally determined rms)  are overlaid onto the 3.6~$\mu$m stamp. \label{fig:statstampscosla1}}
\end{figure*}

\begin{figure*}
\hspace{-0.6mm}\includegraphics[scale=0.65]{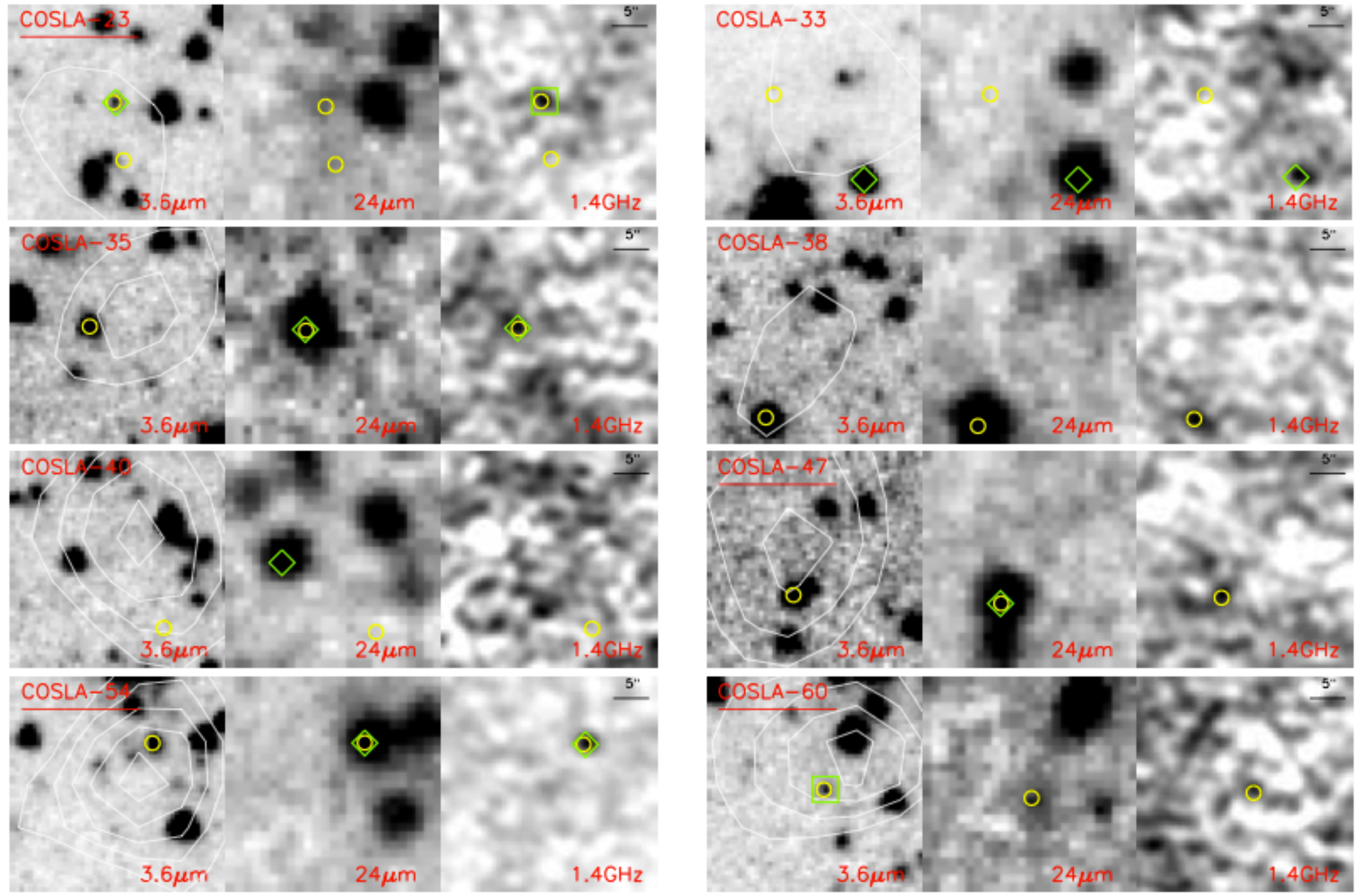}
\vspace{-0.1mm}\includegraphics[scale=0.65]{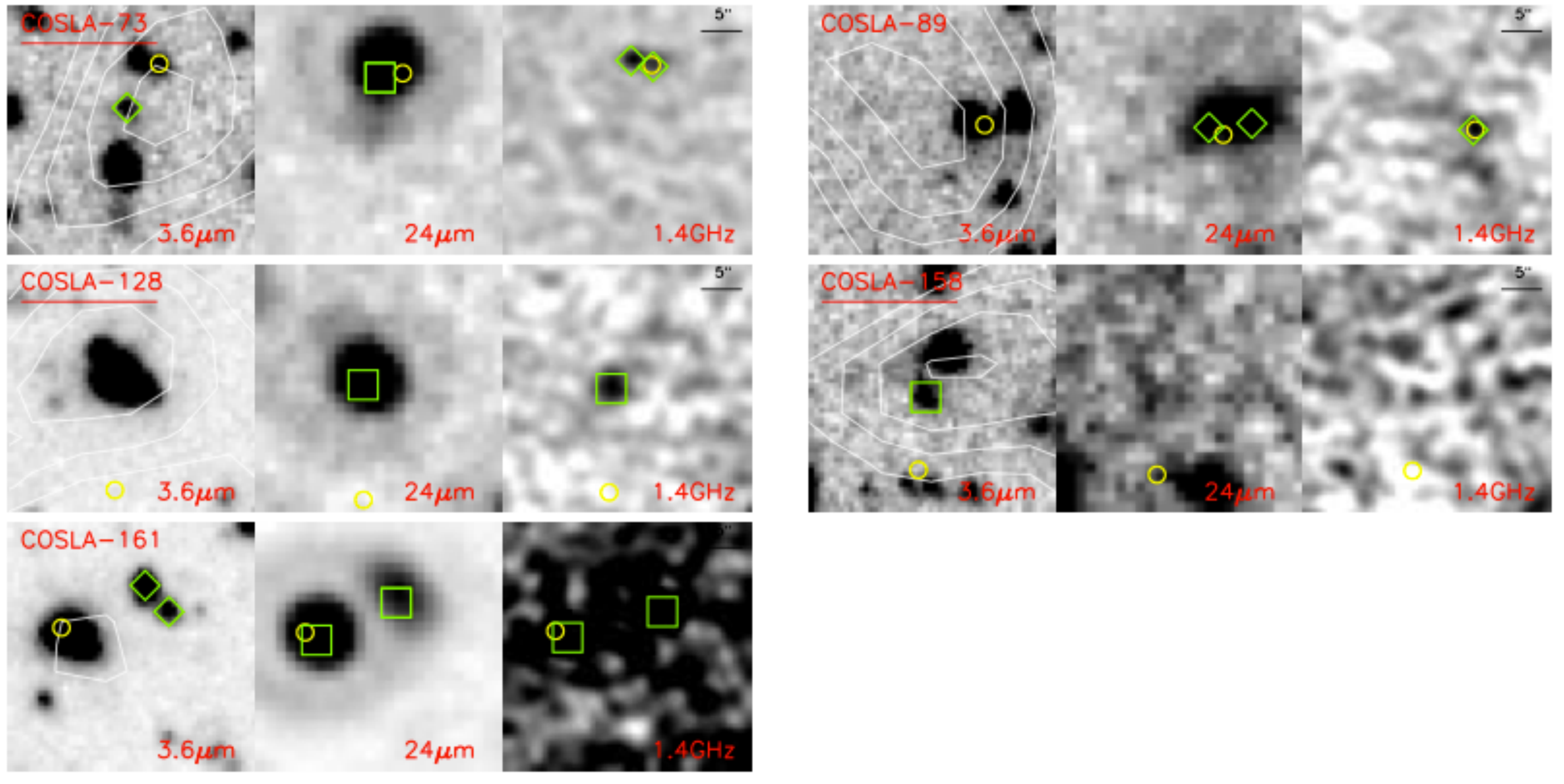}
\caption{ \f{fig:statstampscosla1} \ continued. \label{fig:statstampscosla2} }
\end{figure*}

With search radii of $9"$, and $13.5"$ for the AzTEC and LABOCA SMGs, respectively, we independently computed the P-statistics for the potential radio, 24~$\mu$m, and IRAC color selected ($m_\mathrm{3.6\mu m}-m_\mathrm{4.5\mu m}\geq0 $) counterparts and display those in 
Tables \ref{tab:statcountcosla} and \ref{tab:statcountaztec}, and Figs.\ \ref{fig:statstampscosla1}, \ref{fig:statstampscosla2}, and \ref{fig:statstampsaztec}.


\begin{figure*}
\includegraphics[scale=0.65]{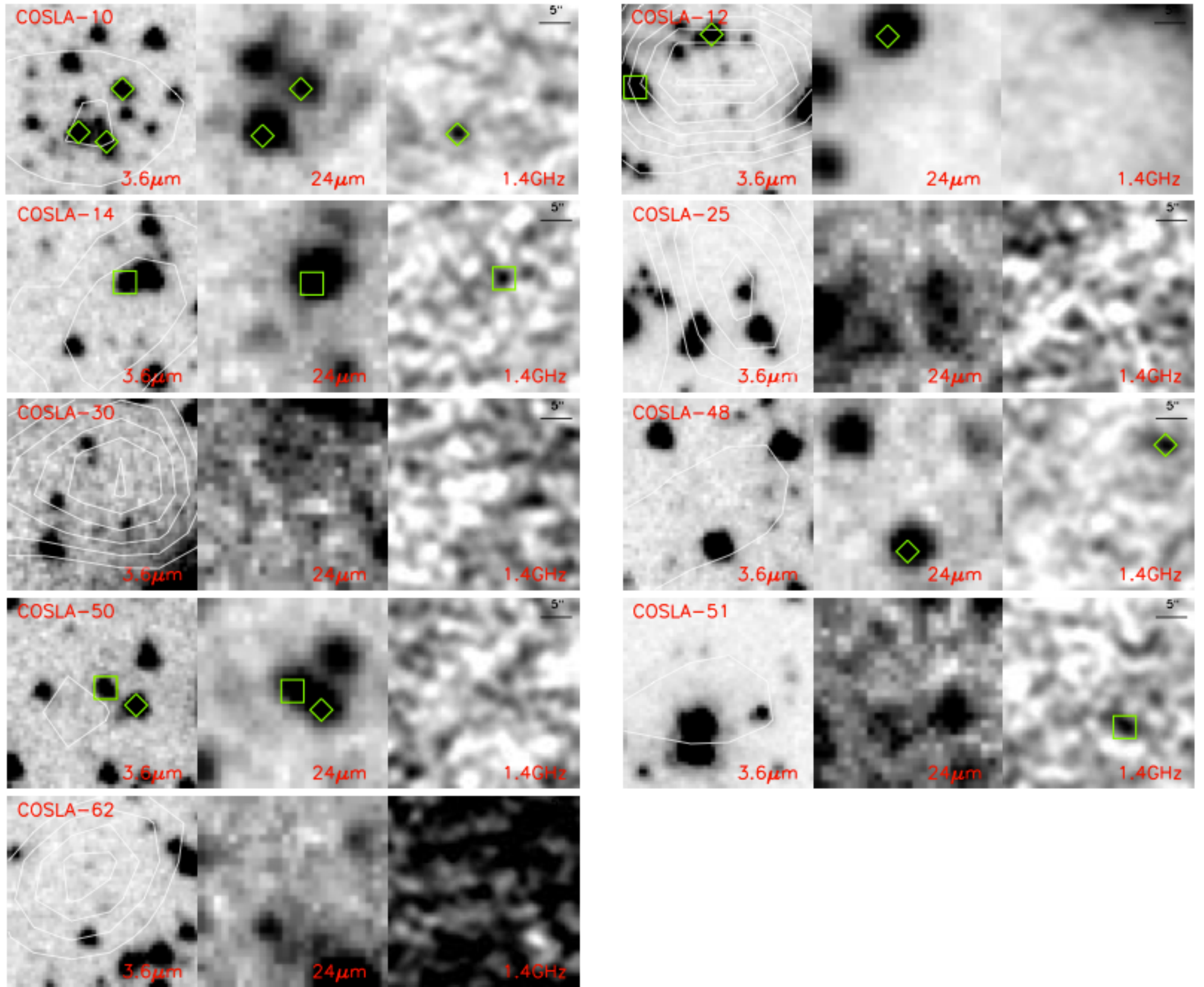}
\caption{ Same as \f{fig:statstampscosla1} \ but for LABOCA sources not detected within our PdBI observations. \label{fig:statstampscoslanodet}}
\end{figure*}

%%%%% AzTEC/JCMT stamps

\begin{figure*}
\hspace{-0.6mm}\includegraphics[scale=0.65]{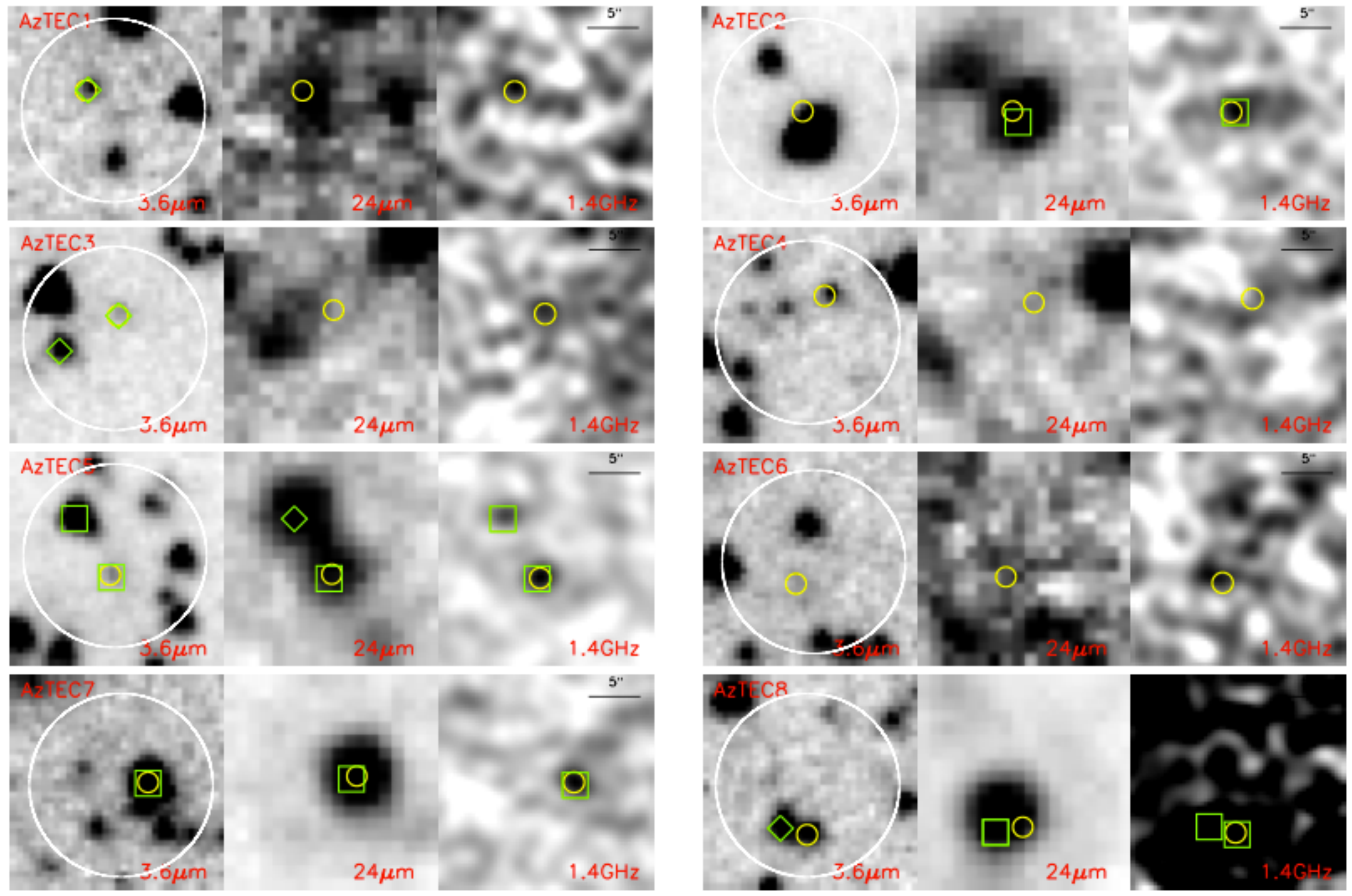}
\vspace{-0.1mm}\includegraphics[scale=0.65]{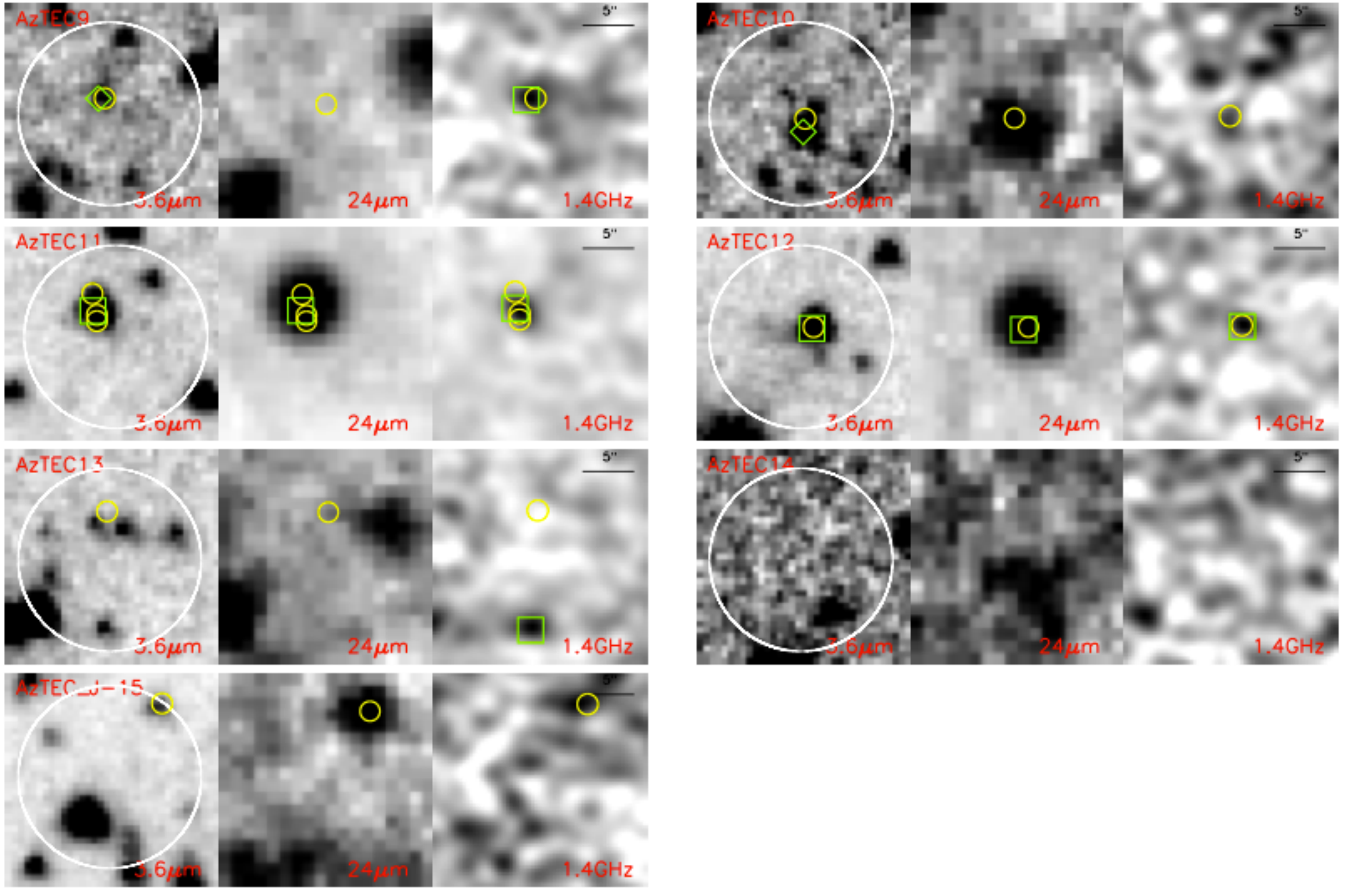}
\caption{Same as \f{fig:statstampscosla1} , but for AzTEC/JCMT/SMA COSMOS sources in our \mmsample \ (see Table~\ref{tab:interf} and Table~\ref{tab:statcountaztec}). The AzTEC/JCMT beam is indicated by the circle in the 3.6~$\mu$m stamp. 
 \label{fig:statstampsaztec} }
\end{figure*}

\begin{table*}
\centering
%\begin{center}
%\rotatebox{90}
\caption{LABOCA sources observed with mm-interferometers at $\lesssim2"$ resolution, and with  counterparts identified via P-statistic. Robust statistical P-counterparts ($p \leq 0.05$) are marked italic, tentative P-counterparts ($0.05 \leq p \leq 0.2$) are shown in regular font, while counterparts identified via  mm-interferometry are marked bold-faced.}
\label{tab:statcountcosla}
%\vskip 100pt
\begin{tabular}{|cccccccccc|}
\hline
          name &  VLA ID$^*$ & d$_\mathrm{VLA}$ & MIPS ID$^*$ & d$_\mathrm{MIPS}$ & IRAC ID$^*$ & d$_\mathrm{IRAC}$ & P$_\mathrm{VLA}$ & P$_\mathrm{MIPS}$  & P$_\mathrm{IRAC}$  \\
           & COSMOSVLA* & [$"$] &  & [$"$] && [$"$] & & &  \\[2pt]
\hline & & & & & & & & &\\[-2ex]
         \bf{\em  COSLA-1 }   &  -- & -- &       --&    --&  \bf{\em   181505 }  &  \bf{\em   0.54  } & --&    --&     \bf{\em   0.005  } \\[0.5ex]
\hline  & & & & & & & & &\\[-2ex] 
{\bf  COSLA-2} &      {\bf $\_$J100056.94+022017.5} &      {\bf 9.84} &  {\bf 15949} &        {\bf 9.89} & {\bf 198261} &      {\bf 9.47} &         {\bf 0.083} &      {\bf 0.273} &      {\bf 0.121} \\ 
                &     -- &  --&      {\em 13173} &  {\em 3.33} &   {\em 197942} &   {\em 1.13} &  -- &   {\em 0.071 }&   {\em 0.005}  \\ 
            &    --&   -- &  15948 &       7.20 &       197682 &       7.45 &  -- &      0.262 &      0.065 \\
           &    {\em     $\_$J100057.27+022012.6} &  {\em      3.12 }&   -- &  --&   -- &  -- &    {\em  0.027} &   -- &   --\\
           &  DP$\_$J100057.35+022002.0   &       7.82 &       --&  --&       197036 &       7.78 &      0.075 &  --&      0.538 \\[0.5ex]
\hline & & & & & & & & &\\[-2ex]   
           \bf{\em COSLA-3 } &      \bf{\em  $\_$J095957.30+022730.4} &       \bf{\em 5.94} &   -- &   --&       \bf{\em 225725} &     \bf{\em   5.70} &    \bf{\em   0.046} &  --&     \bf{\em  0.347} \\ [0.5ex]
\hline & & & & & & & & &\\[-2ex]   
           \bf{\em COSLA- 4}&   \bf{\em   DP$\_$J100008.02+022612.1} &    \bf{ \em1.00} &       \bf{\em    9851} &      \bf{\em   1.05} &  -- & -- &   \bf{ \em 0.002} &  -- &       \bf{\em 0.006} \\ [0.5ex]
\hline & & & & & & & & &\\[-2ex]   
           COSLA-5 &     $\_$J100059.24+021719.1  &      13.79 &        17275 &      13.53 &       187234 &      13.60 &      0.105 &      0.091 &      0.084\\
             &      -- &   --&        {\em  17272} &      {\em   1.80 }&     --&    -- &   --&   -- &     {\em   0.006 } \\ 
            &         $\_$J100059.78+021653.9 &      13.25 &      --&  --&       185375 &      13.49 &      0.167 &   --&      0.277\\ [0.5ex]
\hline & & & & & & & & &\\[-2ex]   
            {\em COSLA-6 }&     {\em   $\_$J100123.52+022618.1 } &     {\em    6.46} &     {\em     16498 }&    {\em     6.68 }&     {\em    221331 } &    {\em     6.85 } &     {\em   0.045} &    {\em    0.110 }&       {\em 0.073 } \\ [0.5ex]
\hline & & & & & & & & &\\[-2ex]   
           {\em  COSLA-8 } &      {\em  $\_$J100025.52+021505.8 }&     {\em    2.54 }&       {\em   11883} &     {\em    2.47} &       {\em  178641} &      {\em   2.51} &   {\em     0.012} &    {\em    0.010}&       {\em 0.024 }\\ [0.5ex]
\hline & & & & & & & & &\\[-2ex]   
   \bf{\em COSLA-9 } & --  &  --&      \bf{\em   15193} &    \bf{\em    5.71} &     \bf{\em   109636} &     \bf{\em   5.33} &   -- &    \bf{\em   0.045} &    \bf{\em   0.087} \\ [0.5ex]
\hline & & & & & & & & &\\[-2ex]   
   \bf{\em  COSLA-13} &      \bf{\em  $\_$J100031.82+021243.1} &      \bf{\em  5.84} &  \bf{\em  11821} &   \bf{\em  5.77 }& \bf{\em  169468} &     \bf{\em   5.49} &       \bf{\em    0.023} &  \bf{\em      0.030} & \bf{\em      0.046 }\\ [0.5ex]
\hline & & & & & & & & &\\[-2ex]   
        \bf{\em     COSLA-16 }&   \bf{\em   $\_$J100051.58+023334.3 }&   \bf{\em      2.63} &    \bf{\em       6490} &    \bf{\em     2.86} &    \bf{\em     248076 }&   \bf{\em      2.67} &   \bf{\em     0.010} &   \bf{\em     0.018 }&   \bf{\em     0.029 }\\ [0.5ex]
\hline & & & & & & & & &\\[-2ex]   
          COSLA-17 &      {\bf $\_$J100136.80+021109.9} &     {\bf 8.64} &     --&    --&      {\bf 163233} &     {\bf  8.80} &    {\bf  0.127} &   -- &    {\bf  0.532 } \\ [0.5ex]
\hline & & & & & & & & &\\[-2ex] 
        \bf{\em   COSLA-18} &    \bf{\em   $\_$J100043.20+020519.2} &   \bf{\em       2.84} &      \bf{\em     11637 }&    \bf{\em      1.31 }&     \bf{\em     142009 }&   \bf{\em       2.81} &      \bf{\em  0.024} &      \bf{\em   0.013} &     \bf{\em    0.055 } \\ 
          &      -- &   --&   11636 &       3.82 &       141453 &       5.57 &  --&      0.060 &      0.185 \\ [0.5ex]
\hline & & & & & & & & &\\[-2ex] 
        \bf{\em    COSLA-23} &     \bf{\em  $\_$J100010.12+021334.9 }&    \bf{\em     1.65} &      -- &     -- &       \bf{\em  172879} &     \bf{\em    1.63} &   \bf{\em     0.016} &   -- &       \bf{\em 0.104} \\ [0.5ex]
\hline & & & & & & & & &\\[-2ex] 
          COSLA-33 & $\_$J100008.73+021902.4  &      11.45 &         9597 &      11.55 &       193342 &      11.73 &      0.108 &      0.155 &      0.058  \\ [0.5ex]
\hline & & & & & & & & &\\[-2ex] 
         {\bf COSLA-35 }&  {\bf    $\_$J100023.65+022155.3 }&   {\bf    4.16 }&      {\bf   1749 }&   {\bf    4.15 }&  {\bf     204426 }&    {\bf   4.06} &   {\bf   0.070 }&   {\bf   0.062 }&  {\bf    0.211} \\ [0.5ex]
\hline & & & & & & & & &\\[-2ex] 
          COSLA-40 &      -- &  --&   11997 &       6.65 &       197365 &       6.12 &  --&      0.127 &      0.242 \\[0.5ex]
\hline & & & & & & & & &\\[-2ex] 
        {\bf  COSLA-47} &     -- &    --&  {\bf 9849} &    {\bf   6.20} &    {\bf   219900} &    {\bf   6.13} &    -- &   {\bf   0.094} &  {\bf    0.249 }\\ [0.5ex]
\hline & & & & & & & & &\\[-2ex] 
           \bf{\em  COSLA-54} &       \bf{\em  $\_$J095837.96+021408.5} &      \bf{\em    7.91 }&      \bf{\em      9392} &      \bf{\em    7.72} &     \bf{\em     175095} &      \bf{\em    7.88} &     \bf{\em    0.052} &      \bf{\em   0.160} &    \bf{\em     0.250} \\ [0.5ex]
\hline & & & & & & & & &\\[-2ex] 
        \bf{\em   COSLA-60} &  -- &    -- &    --&    -- &  \bf{\em 233568} &     \bf{\em   1.65} &    --&  --&     \bf{\em  0.043 }\\ [0.5ex]
\hline & & & & & & & & &\\[-2ex] 
        \bf{\em   COSLA-73} &     \bf{\em  $\_$J095959.33+023440.8 }&     \bf{\em   8.62} &      \bf{\em   17463} &    \bf{\em    7.55 }&    \bf{\em    252264} &    \bf{\em    8.78} &    \bf{\em   0.078 }&      \bf{\em 0.031} &   \bf{\em    0.429} \\
          &      {\em $\_$J095959.50+023441.5} &   {\em    8.64} &     {\em   17463 }&     {\em  7.55} &     {\em  252508} &   {\em    8.71 }&    {\em  0.057} &    {\em  0.031} &    {\em  0.364 }\\  
          &     --&  -- & -- &    -- &  251986 &       2.73 &    -- &  --&      0.128 \\[0.5ex]
\hline & & & & & & & & &\\[-2ex] 
   {\bf COSLA-89} &   {\bf $\_$J100141.77+022713.0 }&   {\bf 5.96 }&   {\bf  16255} &    {\bf 4.65 }&      --&  -- &   {\bf  0.113 } &  -- &     {\bf  0.068 } \\ 
          &   --&   --&        16256 &       9.84 &    -- &  -- &  --&      0.189 &  -- \\ [0.5ex]
\hline & & & & & & & & &\\[-2ex] 
    {\em   COSLA-128} & \em DP$\_$J100137.96+022339.1 &     {\em   1.68} &      {\em   16495} &     {\em   2.52} &      --&    --&      {\em 0.005} &  --&    {\em   0.008}  \\ [0.5ex]
\hline & & & & & & & & &\\[-2ex] 
        {\em  COSLA-158 }&    -- &   -- & -- &   --& {\em  247857} &       {\em 1.31} &  -- &    --&   {\em    0.036} \\[0.5ex]
 \hline & & & & & & & & &\\[-2ex] 
      {\em COSLA-161} & $\_$J100015.28+021240.6 &   {\em    6.53} &    {\em    17233} &   {\em    6.43} &    {\em   169172} &   {\em    6.27 }&    {\em  0.017 }&  {\em    0.029} &   {\em   0.190} \\  
         & \bf{ \em $\_$J100016.05+021237.4 } &     \bf{ \em  7.08} &    \bf{ \em    17235 }&    \bf{ \em   7.16} &       --&   --&      \bf{ \em0.017} &   -- &   \bf{ \em   0.010}  \\ [0.5ex]
\hline
          \end{tabular}\\
          $^*$ The radio, MIPS/24$\mu$m and IRAC catalogs are available at {\tt http://irsa.ipac.caltech.edu/data/COSMOS/tables/} %(scosmos_mips_24_GO3_200810.tbl, scosmos_irac_200706.tbl)
\vspace{1cm}
\end{table*}

\begin{table*}
\centering
\caption{AzTEC/JCMT/SMA SMGs with identified robust/tentative counterparts based on the P-statistics. Robust statistical P-counterparts ($p \leq 0.05$) are marked italic, tentative P-counterparts ($0.05 \leq p \leq 0.2$) are shown in regular font, while counterparts identified via  mm-interferometry are marked bold-faced.}
\label{tab:statcountaztec}
\begin{tabular}{|cccccccccc|}
\hline
          name &  VLA ID$^*$ & d$_\mathrm{VLA}$ & MIPS ID$^*$ & d$_\mathrm{MIPS}$ & IRAC ID$^*$ & d$_\mathrm{IRAC}$ & P$_\mathrm{VLA}$ & P$_\mathrm{MIPS}$  & P$_\mathrm{IRAC}$  \\
           & COSMOSVLA* & [$"$] &  & [$"$] && [$"$] & & &  \\[0.5ex]
\hline & & & & & & & & &\\[-1.7ex] 
        {\bf AzTEC-1} &    -- &  -- & --&   -- & {\bf 233568 }&  {\bf  3.39 }&     -- &  --&  {\bf  0.095} \\  [0.5ex]
\hline & & & & & & & & &\\[-1.7ex] 
         \bf{\em  AzTEC-2 }&    \bf{\em  DP$\_$J100008.02+022612.1} &    \bf{\em  0.12} & \bf{\em   9851 }&    \bf{\em      1.06} &  \bf{\em   -- }&   \bf{\em  -- }&     \bf{\em     0.000}&  \bf\bf{\em   -- }&  \bf{\em     0.006} \\ [0.5ex]
\hline & & & & & & & & &\\[-1.7ex] 
\bf{AzTEC-3}  & -- &    -- &         --&    -- &       \bf{254678} &     \bf{  2.30} &    --&    -- &     \bf{ 0.119 }  \\ 
  & --  &   -- &        --&    -- &       254530 &       5.71 &   --&   -- &      0.093 \\ [0.5ex]
\hline & & & & & & & & &\\[-1.7ex] 
        \bf{\em    AzTEC- 5} &   \bf{\em  $\_$J100019.77+023204.3 }&   \bf{\em    1.75} &  \bf{\em 10042}&    \bf{\em      2.12 } & \bf{\em 242438} &   \bf{\em    1.86 }&     \bf{\em     0.004} &  \bf {\em    0.036 }&   \bf{\em    0.036 } \\ 
            &   {\em   $\_$J100019.99+023210.1} &    {\em   5.64} & {\em  10043 } &   {\em    5.70} & {\em 242872} &   {\em    5.44} &     {\em     0.039} &   {\em    0.070 }&   {\em    0.022 }\\ [0.5ex]
\hline & & & & & & & & &\\[-1.7ex] 
  \bf{\em  AzTEC-7 } &  \bf{\em  $\_$J100018.05+024830.2 } &  \bf{\em   3.02} &    \bf{\em  15453} &  \bf{\em   2.36} &   \bf{\em  304354} &        \bf{\em  2.90} &    \bf{\em     0.006} &      \bf{\em   0.006 }&   \bf{\em      0.011}\\[0.5ex]
\hline & & & & & & & & &\\[-1.7ex] 
 \bf{\em AzTEC-8  }&  \bf{\em $\_$J095959.33+023440.8}  &    \bf{\em     5.00} &     \bf{\em     17463} &    \bf{\em     6.07 }&    \bf{\em     252264} &      \bf{\em   4.82} &     \bf{\em   0.027} &     \bf{\em   0.019} &    \bf{\em    0.210}  \\ 
                & {\em  $\_$J095959.50+023441.5 } &    {\em     5.03 }&    {\em      17463} &      {\em   6.07} &     {\em    252508 }&  {\em       5.01 }&      {\em  0.021} &    {\em    0.019} &    {\em    0.178 } \\ [0.5ex]
\hline & & & & & & & & &\\[-1.7ex] 
\bf{\em  AzTEC-9 } & \bf{\em $\_$J095957.30+022730.4}  &     \bf{\em    1.73 }&     -- &    -- &     \bf{\em    225725} &     \bf{\em    1.63} &   \bf{\em     0.006} &   -- &   \bf{\em     0.068}  \\ [0.5ex]
\hline & & & & & & & & &\\[-1.7ex] 
          \bf{AzTEC- 10} &     --&  -- &    -- &     -- & \bf{ 274390 }&   \bf{   1.59} &    -- &   --&  \bf {   0.064} \\ [0.5ex]
\hline & & & & & & & & &\\[-1.7ex] 
      \bf{\em     AzTEC- 11} &    \bf{\em  $\_$J100008.93+024010.7} &   \bf{\em    3.42 }&  \bf{\em  6883 }&     \bf{\em     3.49} & \bf{\em 272725 }&   \bf{\em    3.31 }&  \bf{\em        0.021 }&   \bf{\em    0.019 }&  \bf{\em     0.034 }\\ [0.5ex]
\hline & & & & & & & & &\\[-1.7ex] 
      \bf{\em     AzTEC- 12} &    \bf{\em  $\_$J100035.29+024353.2} &    \bf{\em   1.55} &  \bf{\em  2586 } &      \bf{\em    1.07} & \bf{\em 286894} &   \bf{\em    1.38 }&     \bf{\em     0.005 }&  \bf{\em     0.004} &   \bf{\em    0.008 }\\ [0.5ex]
\hline & & & & & & & & &\\[-1.7ex] 
          {\em  AzTEC- 13} &     {\em  $\_$J095937.10+023308.4} &    {\em    7.03 }&     -- &   -- &       --&    --&   {\em    0.035 }&  -- &   -- \\ [0.5ex]
\hline
           \end{tabular}\\
          $^*$ The radio, MIPS/24$\mu$m and IRAC catalogs are available at {\tt http://irsa.ipac.caltech.edu/data/COSMOS/tables/}
\vspace{1cm}
\end{table*}

%%% Summary of P-statistics
\begin{table*}
\centering
%\begin{center}
%\rotatebox{90}
\caption{Summary of P-statistic results compared to intermediate resolution mm-mapping}
\label{tab:psummary}
{\scriptsize
\begin{tabular}{lllllllll}
\hline & & & & & & & &\\[-1.7ex] 
sample & \multicolumn{2}{c}{radio fraction}  & \multicolumn{2}{c}{24~$\mu$m} & \multicolumn{2}{c}{$m_\mathrm{3.6\mu m}-m_\mathrm{4.5\mu m}\geq0 $} &  \multicolumn{2}{c}{combined}\\[0.5ex]
& P-statistic$^*$  & correct ID  & P-statistic$^*$  & correct ID  & P-statistic$^*$  & correct ID & P-statistic$^*$  & correct ID   \\
\hline & & & & & & & & \\[-1.7ex] 
1.1mm-selected$^a$ & 8/15 (53.3\%) & 7/10 (70\%)  & 5/15 (33.3\%) & 5/6 (83.3\%) &  5/15 (33.3\%) & 5/6 (83.3\%) & 8/15 (53.3\%) & 7/10 (70\%) \\ \vspace{0.5mm}
870$\mu$m-selected$^b$ & 11/26 (42.3\%) & 7/12 (58.3\%) & 6/26 (23.1\%) & 4/6 (66.7\%) & 10/26 (38.5\%) & 7/11 (63.6\%) & 17/26 (65.4\%) & 12/18 (66.7\%) \\
\hline
\end{tabular}\\
}
$^*$Only robust ($\mathrm{Pc}\leq 0.05$) counterparts are considered here\\
$^a$Out of 15 bolometer SMGs, 2 robust statistical counterparts are found for each of sources AzTEC-5 and AzTEC-8 (see Table~\ref{tab:statcountaztec}) \\
$^b$Out of 26 bolometer SMGs, 2 robust statistical counterparts are found for each of sources COSLA-73 and COSLA-161 (see Table~\ref{tab:statcountcosla}) \\
\end{table*}

\subsubsection{Radio counterparts}

In our \mmsample \ 9/15 (60\%) bolometer SMGs have radio sources 
(drawn from the Joint Deep and Large radio catalogs with $rms\sim7-12~\mu$Jy/beam; Schinnerer et al.\ 2007, 2010)
within the AzTEC beam. This fraction is consistent with that found in (sub)mm-surveys (e.g.\ Chapman et al.\ 2005). 
Only one mm/SMA source (AzTEC-13) in this sample is not 
associated with a radio source present within the single-dish beam (Younger et al.\ 2009). Furthermore, AzTEC-5 and AzTEC-8 each have two P-robust radio sources within the AzTEC/JCMT $18"$ beam. In both cases only one radio source is associated with the SMA mm-detection.

Correlating with the Joint VLA-COSMOS Large and Deep catalogs, out of the 36 LABOCA SMGs followed-up with interferometers, 23 ($\sim64\%$) have radio sources ($\mathrm{rms_{1.4GHz}}\gtrsim 7-12~\mu$Jy/beam) within the beam. Of these 36, 26 were detected at mm-wavelengths with interferometers (\submmsample ), and out of these 26, 17 (65\%) show radio sources within the LABOCA beam. 

Assigning counterparts to each of these LABOCA sources via P-statistic we find that  (see Table~\ref{tab:statcountcosla},  \f{fig:statstampscosla1} and  \f{fig:statstampscosla2} )
 in 4 cases (2 with $\mathrm{S/N_{1.3mm}}>4.5$)\footnote{COSLA-5, COSLA-6, COSLA-8, COSLA-128} the robust/tentative P-counterpart is not coincident with the interferometric-source. 
Within Poisson uncertainties this  is consistent 
with the results from Younger et al.\  (2009) for the \mmsample.

In our \submmsample \ COSLA-161 has a mm-interferometric detection and two
P-robust radio counterparts. Multiple P-tentative radio counterparts are found also for COSLA-2, COSLA-5, COSLA-17, and COSLA-73 (three out of these 5 are significant interferometric detections). 
In all cases, except for COSLA-5, one of the radio sources is associated with the inetrferometric-source.

Combining the above results for our \submmsample \ we thus find 4 cases where the robust/tentative P-counterpart is not associated with the interferometric source, and 4 more ambiguous cases where from the multiple robust/tentative P-counterparts found for the SMG only one is confirmed by the interferometric source. Taking the 26 single-dish SMGs in the \submmsample \ this amounts to a fraction of $15\pm8\%$ for the first and latter, separately. For the \mmsample \ we find one misidentified and two ambiguous SMG counterparts assigned via P-statistic. Taking the 15 single-dish SMGs in the \mmsample \ this amounts to $7\pm7\%$ and $13\pm7\%$, respectively.

%%%%%%%%%%%%%%%%%%%%%%%%%%%%%%%%%%%%%%%%%%%%%%

\subsubsection{Radio, 24~$\mu$m and IRAC counterparts}

In this section we investigate the agreement between robust counterparts determined via P-statistic using radio, 24~$\mu$m, and IRAC wavelength regimes, and counterparts identified via intermediate $\lesssim2"$ resolution mm-mapping. 

Where both radio and mid-IR data are available, potential counterparts to single-dish 
detected SMGs are commonly selected by searching 
for P-statistics robust radio and 24~$\mu$m counterparts. 
Where no such source can be identified, counterparts are searched for among color selected IRAC sources ($m_\mathrm{3.6\mu m}-m_\mathrm{4.5\mu m}\geq0 $). In Table~\ref{tab:statcountcosla} and Table~\ref{tab:statcountaztec} we list the resulting P-robust counterparts to the LABOCA and AzTEC samples. 
A summary of the identifications is given in Table~\ref{tab:psummary}.

In the \submmsample  \ we find P-robust counterparts for 17 out of the 26 PdBI-identified SMGs - irrespective whether these identification are correct or not. In total we find 18 P-robust counterparts as COSLA-73 and COSLA-161 both have two P-robust counterparts associated. 
From the 18 statistically identified sources, 12 ($66\%$) are correct identifications 
based on our PdBI detections. This fraction remains similar if we consider only the single-dish detected SMGs with mm-interferometric detections at $\mathrm{S/N}>4.5$, i.e.\ identified without any prior assumptions (i.e.\ multi-wavelength association): 11/13 (85\%) single-dish detected SMGs have P-robust counterparts, in total there are 12 P-robust counterparts (as COSLA-73 is in this sub-sample) and 7 out of these 12 (58\%) match our interferometric detections. This amounts to $\sim50\%$ correct identifications via P-statistic within the samples analyzed  (i.e.\ 7/13 for the least-biased- and 12/26 for the \submmsample ).

In the  \mmsample \ we find P-robust counterparts for 8 of 15 ($53\%$)  SMGs with SMA detections
(Table~\ref{tab:statcountaztec}, \f{fig:statstampsaztec} ). 
Since AzTEC-5 and AzTEC-8 each have two P-robust counterparts, we find 10 P-robust associations in total.
Seven of the 10 ($70\%$) are coincident with the mm-interferometric detections. 
The fraction remains the same if robust {\em and} tentative statistical counterparts are considered.  Within the Poisson uncertainties this is consistent with the results for the \submmsample , i.e.\  only $\sim50\%$ of the single-dish detected SMGs have correct counterparts assigned via P-statistic.

\subsection{The biases of assigning counterparts to
  single-dish detected SMGs} 

Intensive work has been invested into optimizing techniques to
determine counterparts to single-dish detected SMGs identified at low
($\sim10-35"$) angular resolution (e.g.\  Ivison et al. 2002, 2005; Pope et al.\ 2006; Hainline et al.\ 2009; Yun et al.\ 2008, 2012). Deep intermediate-resolution
radio observations, which are less time consuming than similar mm-wave
observations, but are expected to trace the same physical processes (given
the IR-radio correlation; e.g.\ Carilli and Yun 1999; Sargent et al.\ 2010) have proven efficient. However, it was
realized that radio-counterpart assignment biases samples to
low-redshift (e.g.\ Chapman et al.\ 2005; Bertoldi et al.\ 2007).  To
overcome this, 24~$\mu$m- and IRAC color- selected samples have been utilized
(e.g.\ Pope et al.\ 2006; Hainline et al.\ 2009; Yun et al.\ 2008). Generally such methods identify counterparts to $\sim60\%$ of
the parent single-dish SMG sample (e.g.\ Chapman et al.\ 2005; Biggs
et al.\ 2011; Yun et al.\ 2012) and yet the fraction
of misidentifications in these samples remains unclear.  A further
source of bias in such samples is the blending of SMGs within the large
single-dish beams. This may potentially be a severe problem as SMGs
have been shown to cluster strongly (Blain et al.\ 2004; Daddi et al.\ 2009a,b; Capak et al.\ 2010; Aravena et al.\ 2010b; Hickox et
al.\ 2011) and reside in close-pairs (as also 
suggested by simulations; Hayward et al.\ 2011). Here we provide 
detailed insight into these issues based on statistical samples.

We generate two unique (870$\mu$m- and 1.1mm-selected) SMG
samples with counterparts to LABOCA/APEX and AzTEC/JCMT COSMOS SMGs
identified via intermediate ($\lesssim2"$) resolution mm-mapping (see
\s{sec:statsamples} \ and Table~\ref{tab:statsamples}). Consistent with
results from the literature we have found {\em statistical}
counterparts for $\sim50-70\%$ of the sources in these samples (see
\s{sec:counterparts} , Table~\ref{tab:statcountcosla} and
Table~\ref{tab:statcountaztec}). Comparing these with the
intermediate ($\lesssim2"$) resolution mm-detections, we find a $\sim70\%$ match. If
there were no caveats with the intermediate-resolution mm-detections
this would imply that statistical counterpart assignment methods
utilizing deep radio, 24~$\mu$m and IRAC data (such as the one applied
here) identify correctly counterparts to $\sim50\%$ of the parent
single-dish samples.  Furthermore, it would imply that $\gtrsim15\%$, and possibly up to $\sim40\%$ of single-dish detected
SMGs separate into multiple components, with a median separation of $\sim5"$, when observed at $\lesssim2"$
angular resolution.  The misclassification of statistical assignment is likely intrinsic to the methods applied and also due to the break-up of single-dish SMGs into multiple components. If indeed a large fraction of SMGs are blended within the single dish beams (on scales $<10"$), this could affect the slope of the (sub)mm counts inferred from single-dish surveys as the bright end would be overestimated, while the faint end would be underestimated (see Kova\'{c}s et al.\ 2010 for a more detailed discussion).

We find that radio assignment, relative to near/mid-IR wavelength regimes, is the most efficient tracer of single-dish detected SMG counterparts (see Table~\ref{tab:psummary}). Thus, as already demonstrated by Lindner et al.\ (2011), who find that a 20~cm rms of $\sim2.7-5~\mu$Jy identifies radio counterparts for $\sim90\%$ of SMGs,
 future deep radio maps with EVLA, ASKAP, MeerKAT and SKA will provide efficient tracers of SMG counterparts.

Our samples of LABOCA/APEX
and AzTEC/JCMT SMGs identified via intermediate ($\lesssim2"$) resolution mm-mapping
are not complete, but constitute half of the parent SMG
samples (see Scott et al.\ 2008, Navarette et al., in prep.).  They 
are also subject to their own incompletenesses and false detection rates
within heterogenous data sets (assembled from SMA, PdBI, and CARMA
observations).  Thus, although our analysis suggests that roughly  half of single-dish detected SMGs are correctly identified via
statistical methods, a more robust insight into these issues will have
to await further follow-up observations of complete samples of
single-dish detected SMGs with higher sensitivities than the
ones presented here and with a uniform rms over the full single-dish
beam area.  One would also preferably want to obtain these data at (at least) 
two separate frequencies.

\begin{figure*}
\includegraphics[bb = 130 430 486 602,scale=0.47]{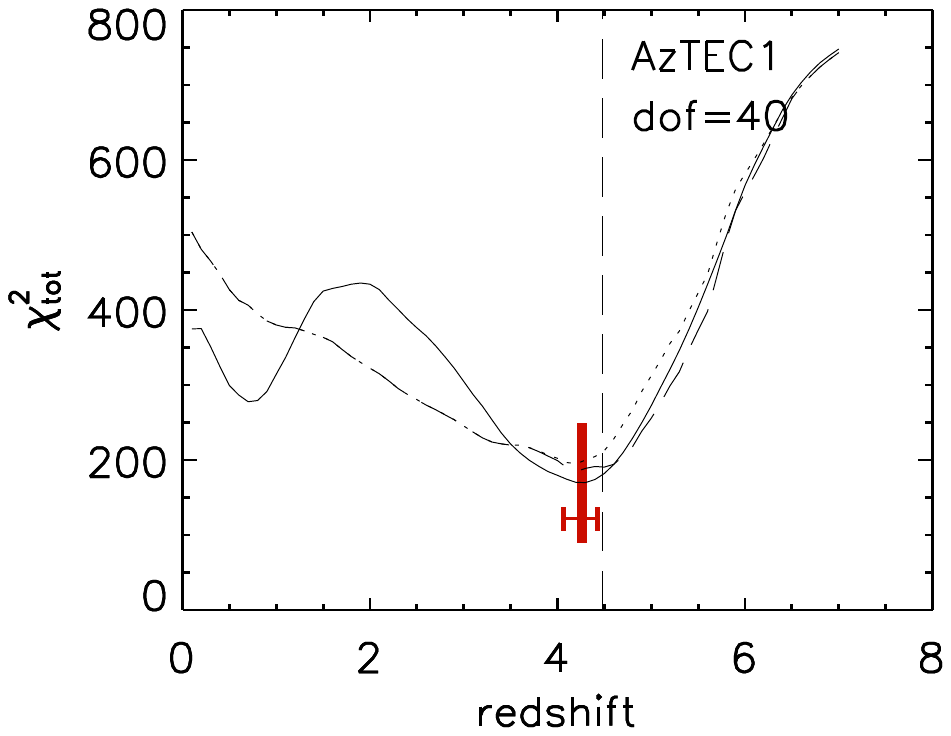}
\includegraphics[bb = 210 430 486 602,scale=0.47]{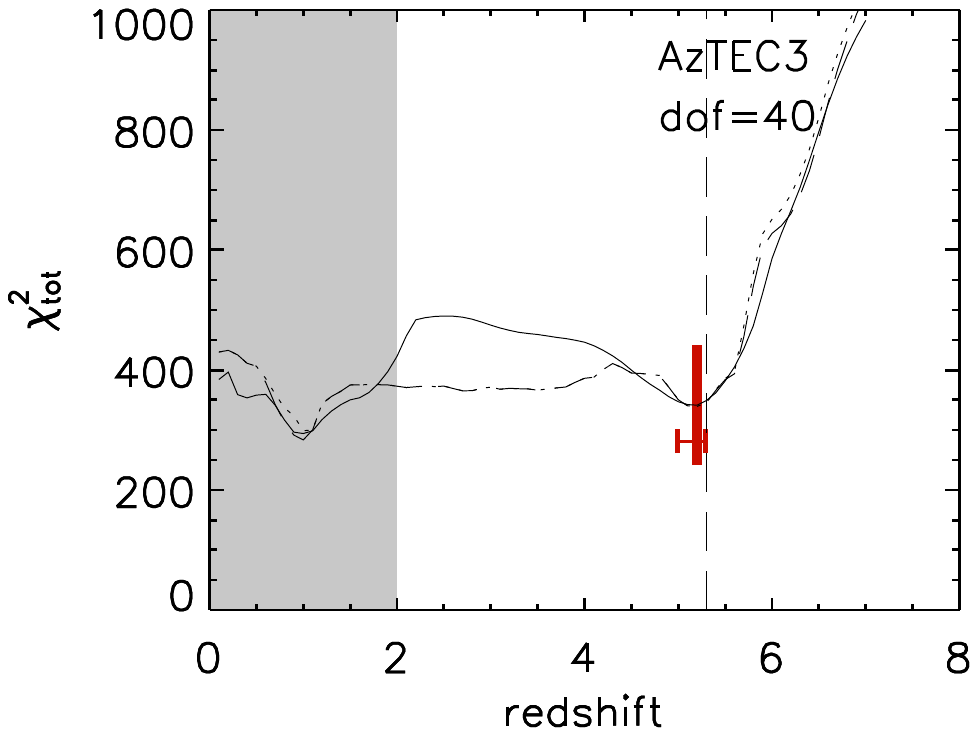}
\includegraphics[bb = 210 430 486 602,scale=0.47]{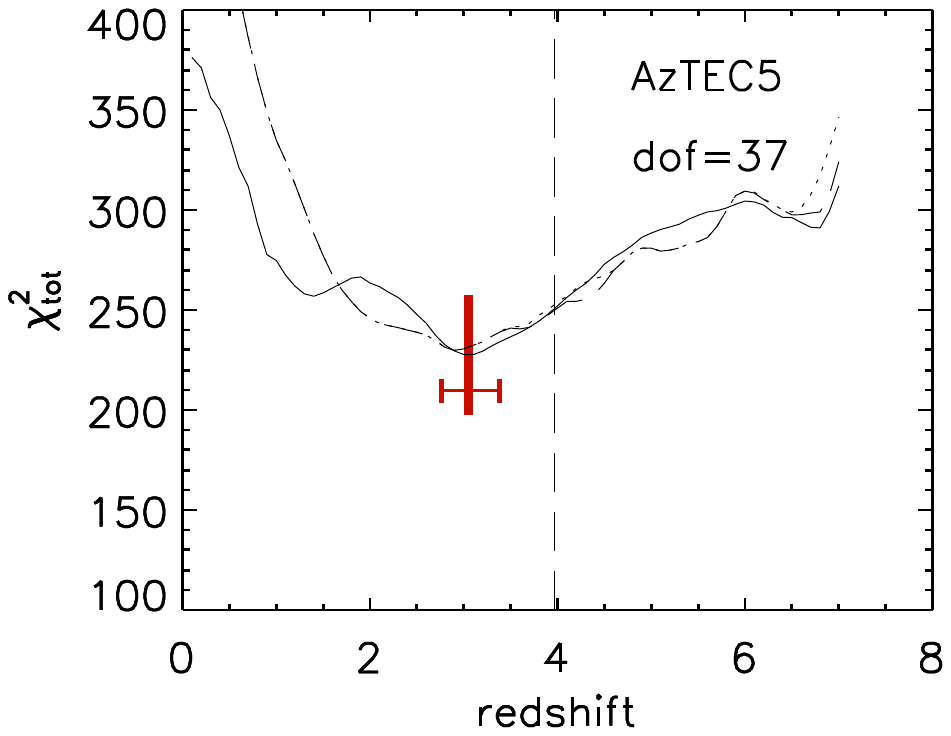}
\includegraphics[bb = 210 430 336 602,scale=0.47]{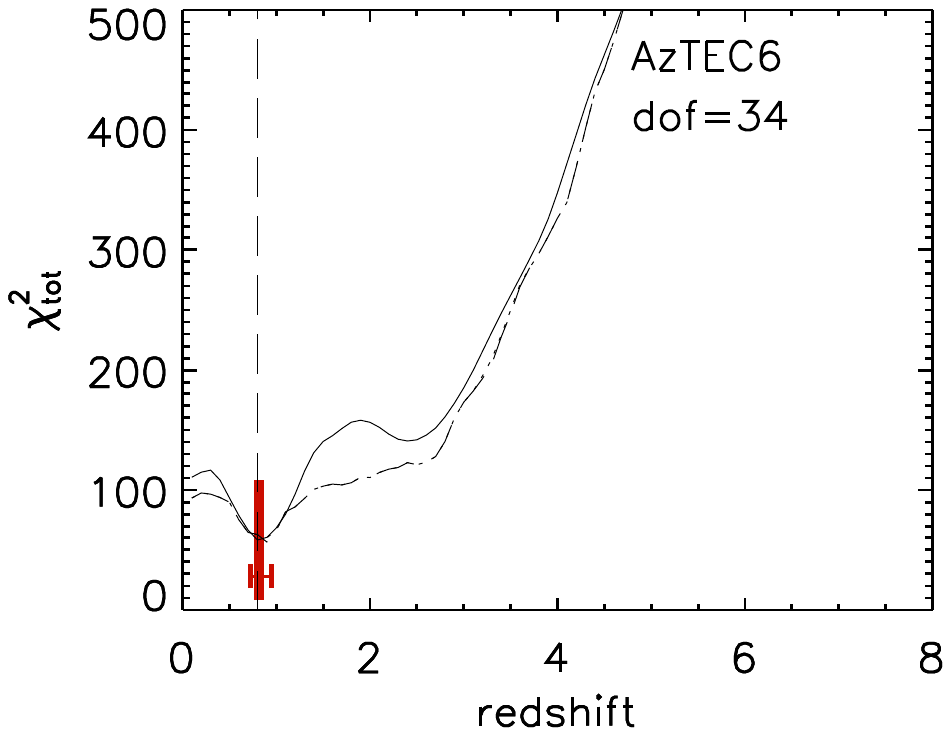}\\
\includegraphics[bb = 130 430 486 622,scale=0.47]{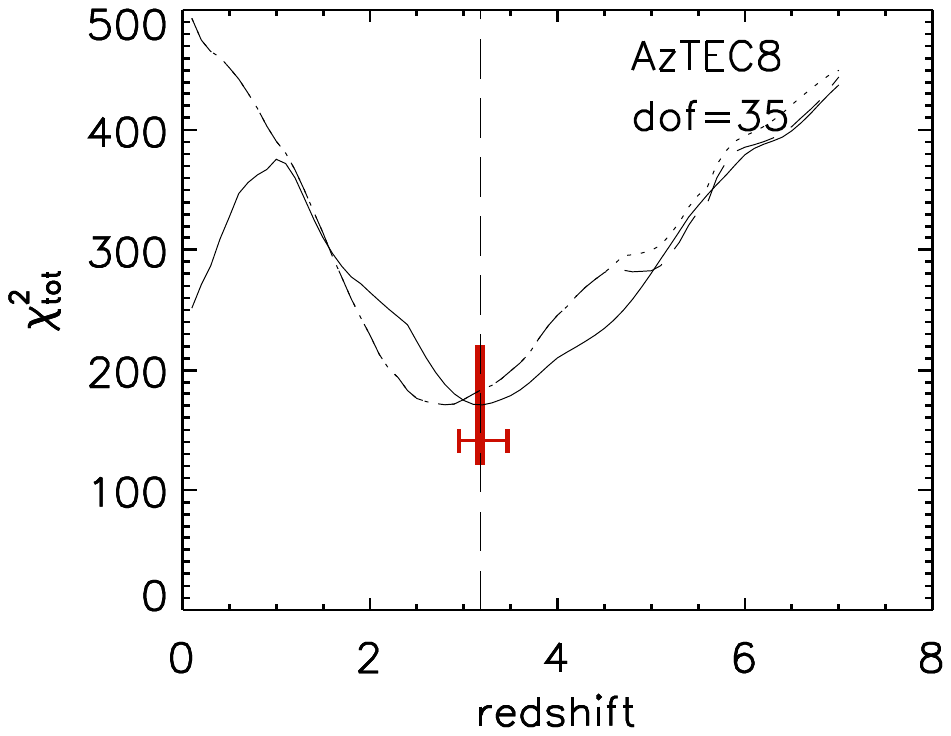}
\includegraphics[bb = 210 430 486 652,scale=0.47]{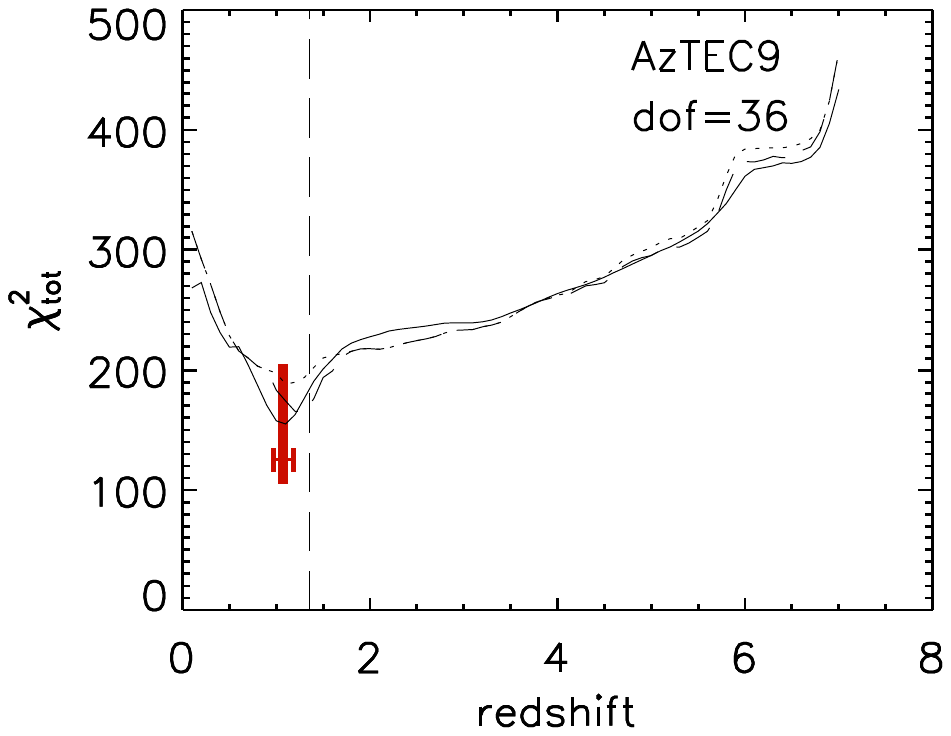}
\includegraphics[bb = 210 430 486 652,scale=0.47]{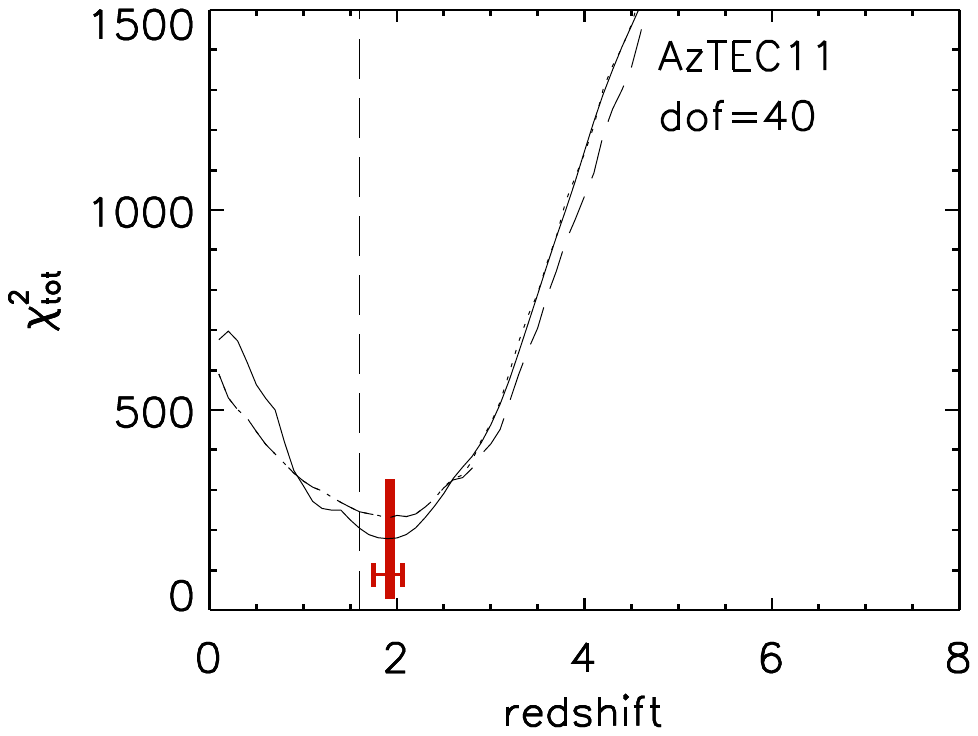}
\includegraphics[bb = 210 430 336 652,scale=0.47]{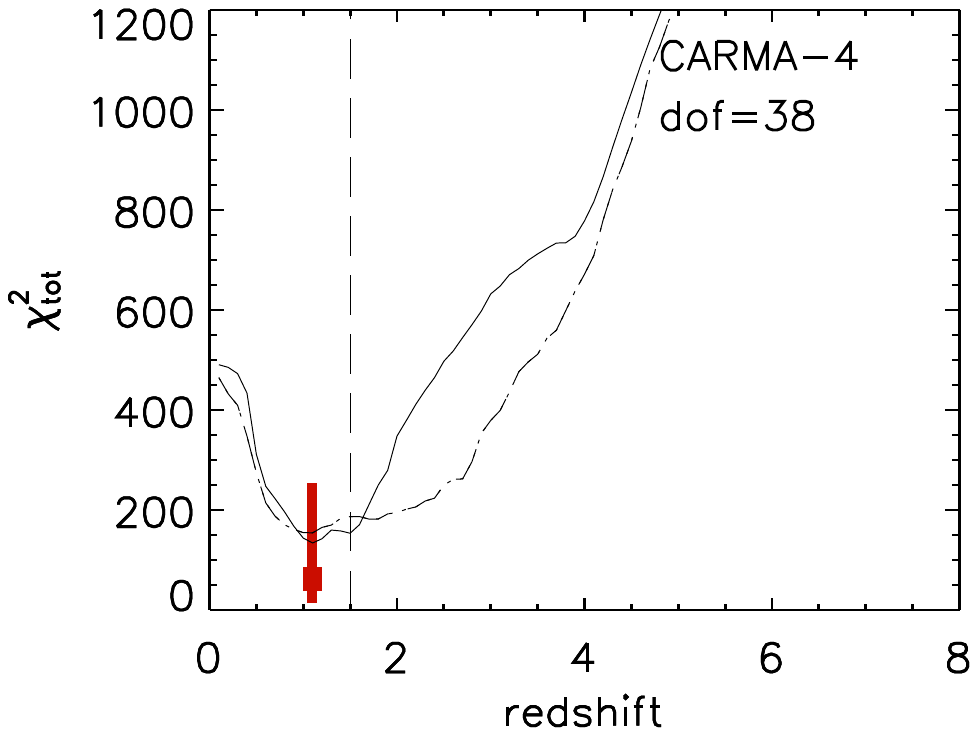}\\
\includegraphics[bb = 130 430 486 652,scale=0.47]{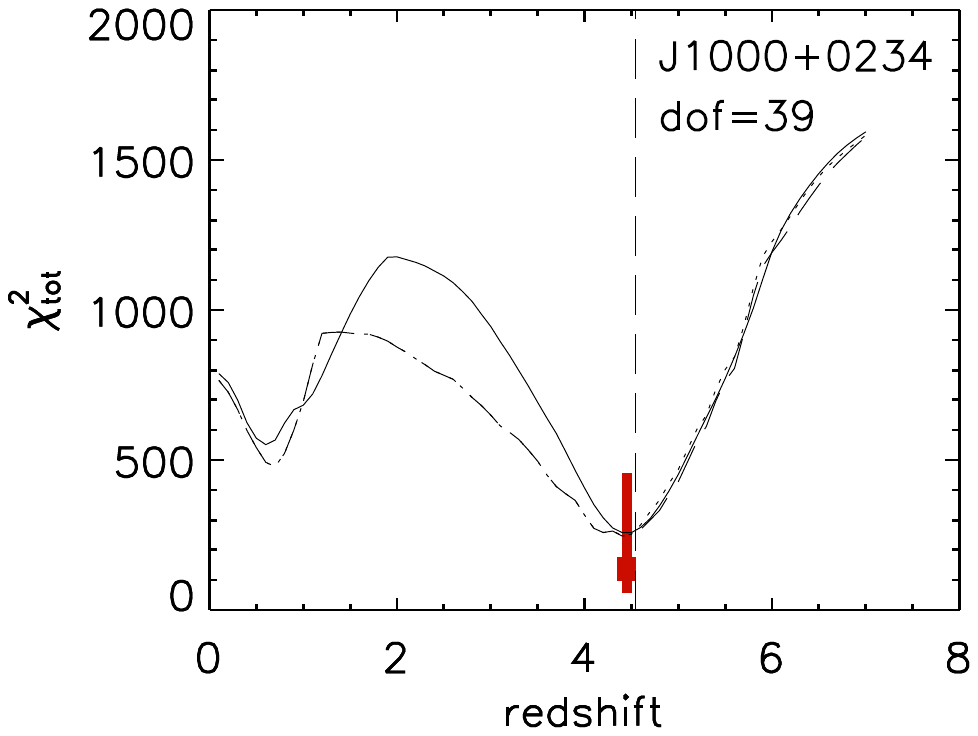}
\includegraphics[bb = 210 430 486 652,scale=0.47]{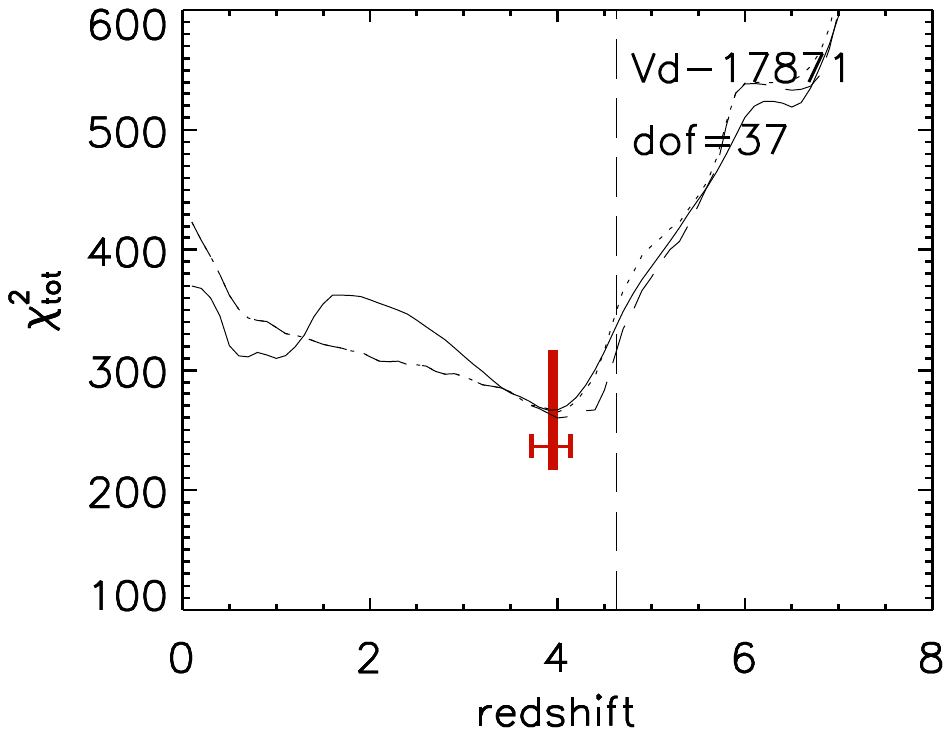}
\includegraphics[bb = 210 430 486 652,scale=0.47]{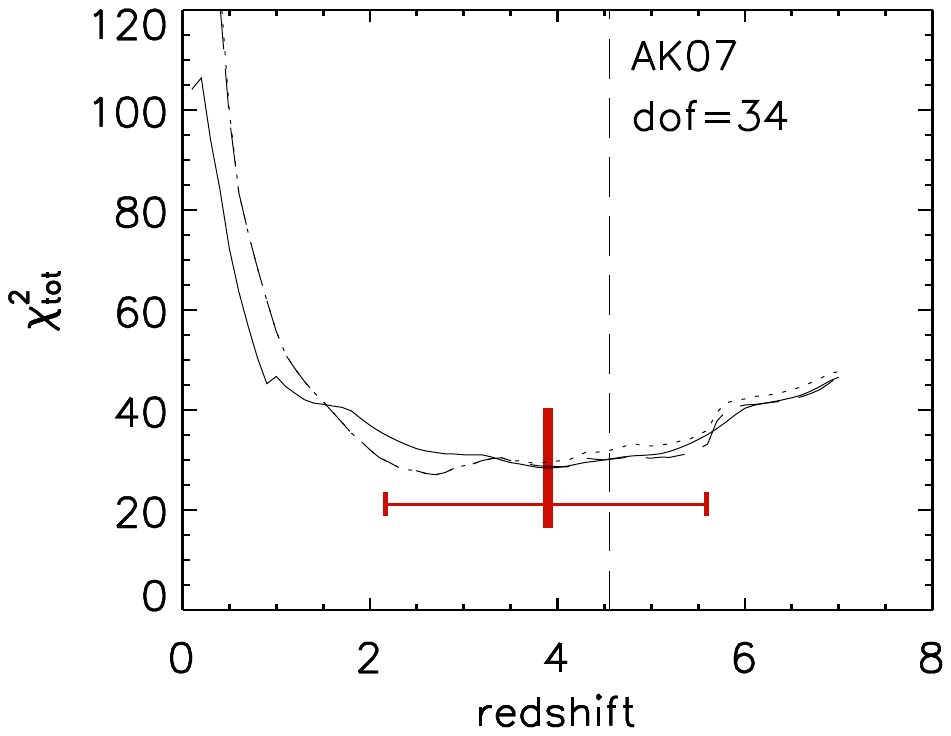}
\includegraphics[bb = 210 430 336 652,scale=0.47]{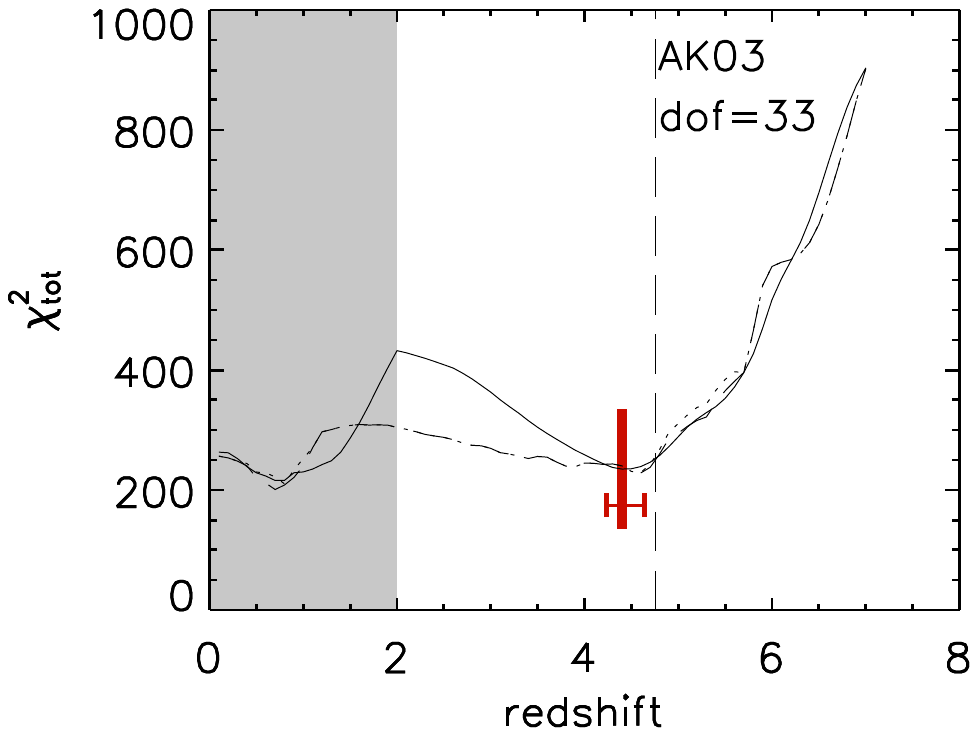}\\
\includegraphics[bb = 130 410 486 652,scale=0.47]{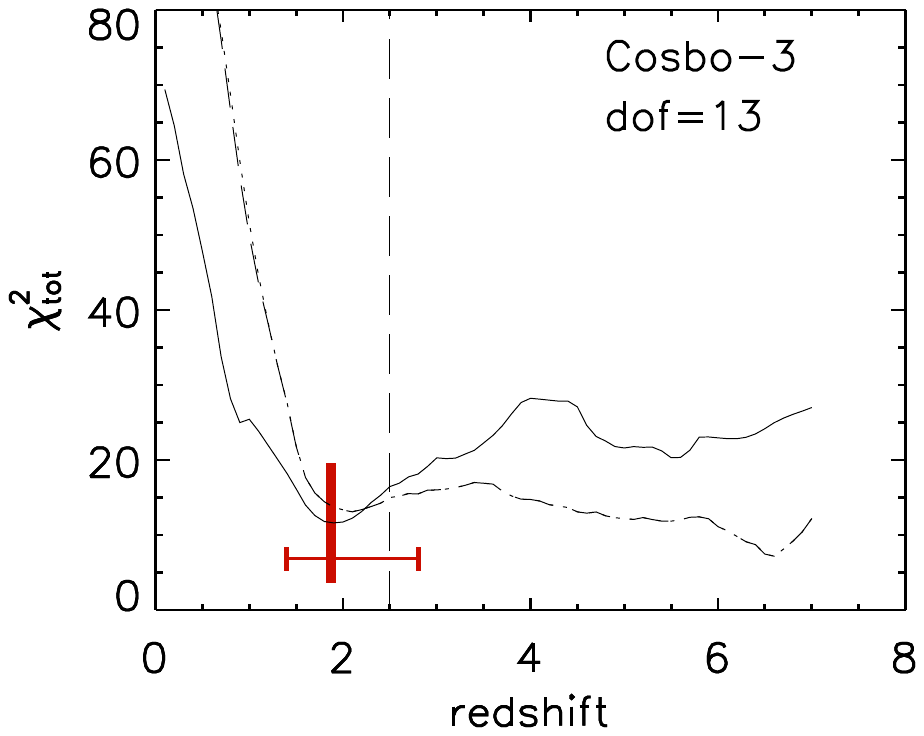}
\includegraphics[bb = 210 410 486 652,scale=0.47]{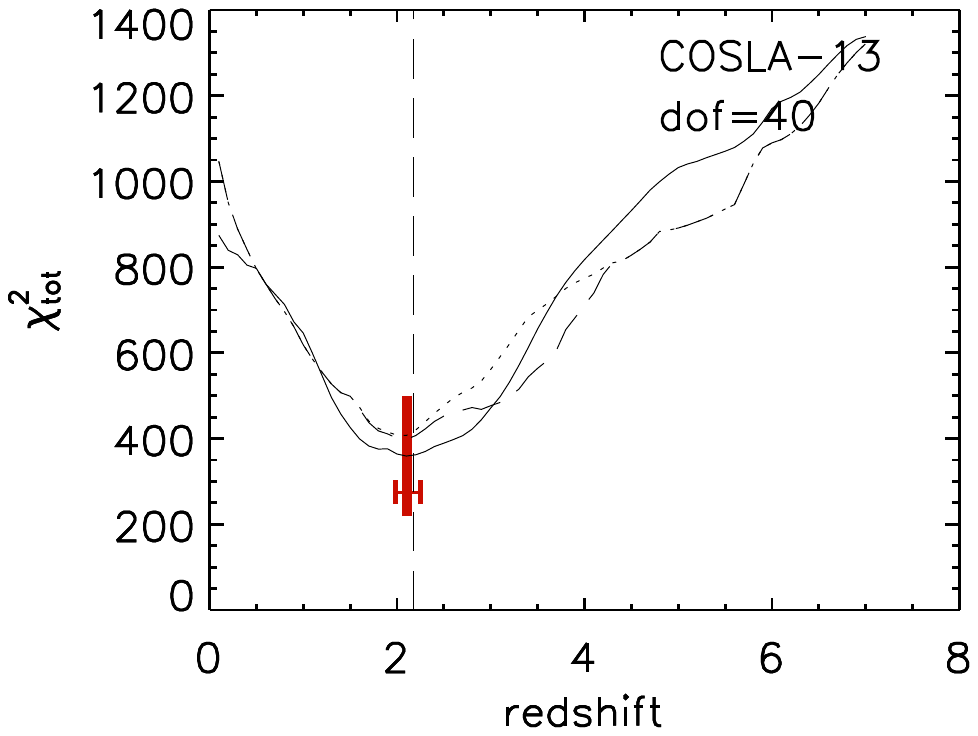}
\includegraphics[bb = 210 410 486 652,scale=0.47]{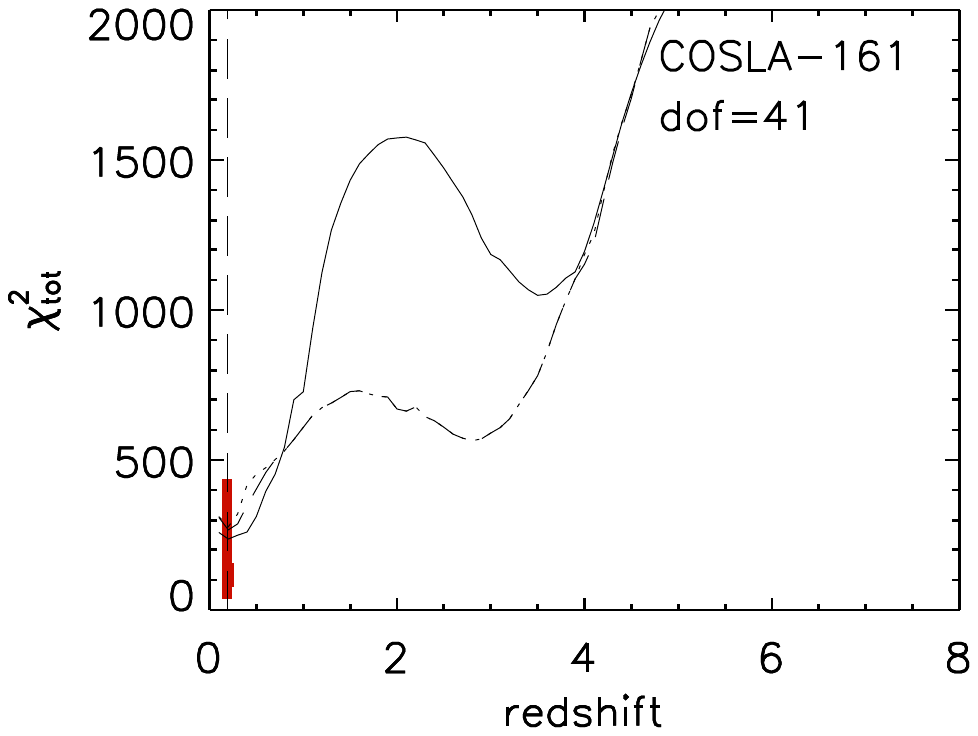}
\includegraphics[bb = 210 410 336 652,scale=0.47]{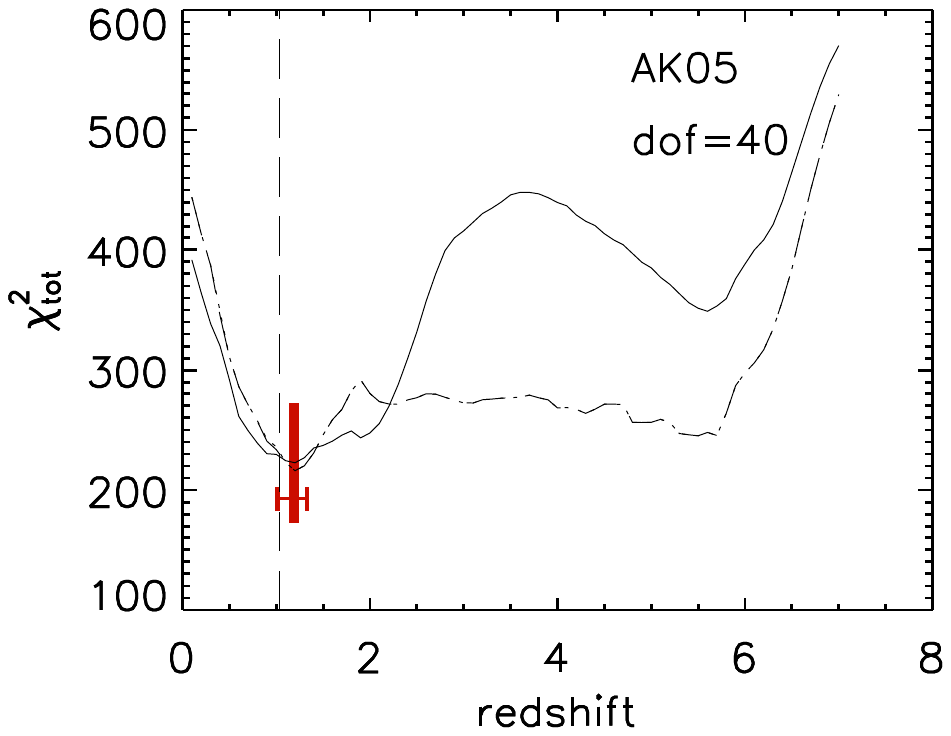}
\caption{Photometric redshift total $\chi^2$ distributions for our SMGs with spectroscopic redshifts. We show results based on  various sets of spectral models (see text for details): 2T
  (dotted lines), 6T (dashed-lines), M (full lines). The spectroscopic
  redshifts are indicated by vertical dashed lines. The source names and the number of degrees of freedom (dof) in the photometric redshift  $\chi^2$ minimization are indicated in each panel. The gray-shaded areas in some of the panels indicate the redshift range ignored for the determination of the best-fit photometric redshift.  The photometric redshift provided by the best model (M) and its uncertainty was taken as the minimum $\chi^2$ value and the 99\% confidence interval, respectively, both indicated in each panel by the thick and thin red lines.}
      \label{fig:photz}
\end{figure*}

\begin{figure}
\includegraphics[bb =  94 410 456 792,width=\columnwidth]{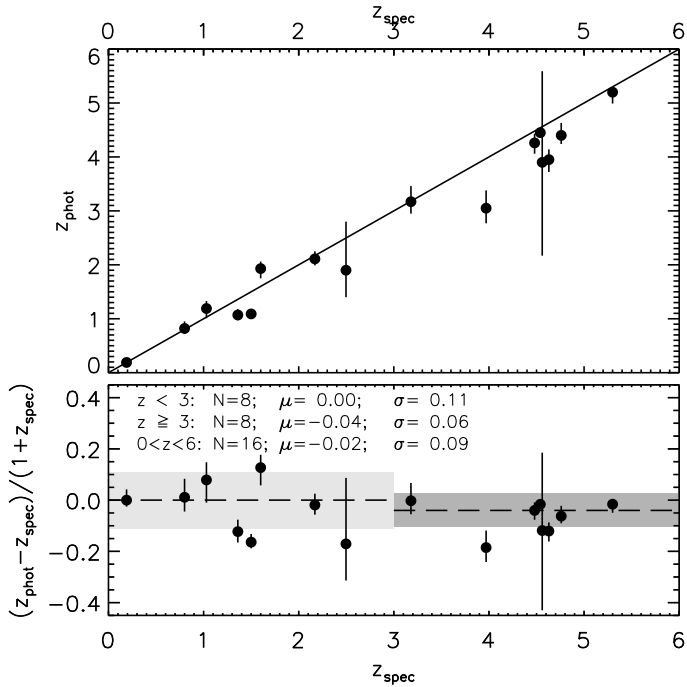}
\caption{Comparison of spectroscopic and photometric redshifts for 16
  starbursts in our COSMOS sample.  The photometric redshifts were determined using the Michalowski spectral templates, and the shown errors are
  $\pm3\sigma$ errors drawn from the $\chi^2$ distributions of the
  photometric-redshift fits (see \f{fig:photz} \ and text for details). The median offset and standard deviation of
  the $\Delta z / (1+z_\mathrm{spec})$ distribution are indicated in
  the bottom panel. Note that for $z\geq3$ we find a slight, $0.04\cdot(1+z)$, systematic underestimate of the photometric redshifts. }
      \label{fig:spec-phot-z}
\end{figure}

\section{Distances to submillimeter galaxies}
\label{sec:redsft}

In this Section we calibrate photometric redshifts for
SMGs based on a sample of 12 SMGs detected via mm-interferometry (and 4 additional high-redshift starburst galaxies) in the COSMOS field with
spectroscopic redshifts spanning a broad redshift range of $z\sim0.1-5.3$
(see  Table~\ref{tab:interf} and Table~\ref{tab:cosla}). We optimize the photometric redshift computation,
and apply it thereafter to the remainder of our SMG sample. 

\subsection{Calibration and computation of photometric redshifts for SMGs}

Photometric redshifts are computed by fitting optimized spectral template libraries to
the spectral energy distribution of a given galaxy, leaving
redshift as a free parameter. The redshift is then determined via a
$\chi^2$ minimization procedure. The quality of the photometric
redshifts will depend on the choice of the spectral library.  To
obtain optimal results for the population of SMGs using Hyper-z, \smo\ et al.\ (2012) tested three sets of 
spectral
model libraries on a sample of eight SMGs in the COSMOS field with counterparts determined via mm-interferometry and with available spectroscopic redshifts:
\begin{itemize}
\item[{\bf 2T}:] Only two -- burst and constant star formation
  history -- templates drawn from the Bruzual \& Charlot (2003)
  library (and provided with Hyper-z).
\item[{\bf 6T}:] Six templates provided by the Hyper-z code: burst, four
  exponentially declining star formation histories (star formation
  rate $\propto e^{-t/\tau}$ where $t$ is time, and $\tau=0.31,\, 1,\,
  3$~and~5~Gyr) and a constant star formation history. This selection of SFH/templates is similar to the approach used by Ilbert et al. (2009) to compute stellar masses with LePhare.
\item[{\bf M}:] Spectral templates developed in GRASIL (Silva et al.\
  1998; Iglesias-P\'{a}ramo et al.\ 2007) and optimized for SMGs by Michalowski et al.\ (2010).
\end{itemize}

They find that all three template libraries yield similar results, while the M templates result in the tightest $\chi^2$ distributions. Here we repeat their analysis using a larger sample containing 12 SMGs in the COSMOS field with counterparts determined via mm-interferometry and available spectroscopic redshifts. We additionally add to this sample 4 sources (Vd-17871, AK03, AK05, AK07), selected in the same way as AzTEC-1, AzTEC-5, and J1000+0234, i.e.\  via criteria identifying high-redshift extreme starbursts (Lyman Break Galaxies  with weak radio emission; Karim et al., in prep.). The photometric redshifts are computed using the entire available COSMOS photometry ($>30$ bands) and the Hyper-z code with a Calzetti et al.\ (2000) extinction law, reddening in the range of $A_V=0-5$, and allowing redshift to vary from 0 to 7.

The results are shown in \f{fig:photz} , where we present the
photometric redshift total $\chi^2$ distributions for the 16 sources in our training-set.
The
overall match between the most probable photometric redshift (corresponding to the minimum $\chi^2$ value) and the
spectroscopic redshift is good. 
We emphasize that the sample used for this analysis is rather
heterogeneous in respect of redshift range, detections in optical
bands, blending, and AGN contribution. For example, Cosbo-3 is a blended source
not detected in images at wavelengths shorter than $1~\mu$m (see \smo\
et al.\ 2012., for details). Constraining its photometric redshift well
(as shown in \f{fig:photz} ) affirms that our deblending techniques
(described in detail in \smo \ et al.\ 2012), as well as photometric
redshift computations work well. Vd-17871 is a weak SMG (with a CO-line detection, and a continuum brightness at 1.2~mm of $\sim2.5$~mJy; Karim et al., in prep.) with substantial AGN contribution identified in the IR (Karim et al., in prep). Even in this case our photometric redshift agrees well with the spectroscopic redshift. 
Note also that within our sample
with spectroscopic redshifts there are no catastrophic redshift outliers.\footnote{The photometric redshift of AzTEC-5 shows the largest deviation from its spectroscopic redshift, but it is still within $2\sigma$ of the $(z_\mathrm{phot}-z_\mathrm{spec})/(1+z_\mathrm{spec})$ distribution (see \f{fig:spec-phot-z} ). } 
For two sources
(AzTEC-3 and AK03) there are two equally probable redshift peaks (i.e.\ $\chi^2_\mathrm{tot}$ minima). In both cases, however, one of those is consistent with the spectroscopic redshift. In particular, in
the case of AzTEC-3 the low redshift peak can be disregarded given
that the galaxy is not detected at 1.4~GHz  given the depth
of the VLA-COSMOS survey. 

In conclusion, comparing the redshift probability distributions given the
2T, 6T, and M models, we find that the Michalowski (M) models
yield the most optimal results (i.e.\ the tightest redshift probability
distributions). Hence, hereafter we will adopt the Michalowski et al.\
(2010) spectral templates for the photometric redshift estimate for our
SMGs. From the redshift probability distribution for a given source we
take the most probable redshift (corresponding to that with minimum $\chi^2$) as the photometric redshift of the
SMG, and derive the 99\% confidence interval from its total $\chi^2$
distribution. The comparison between photometric and spectroscopic
redshifts is quantified in \f{fig:spec-phot-z} \ using the M template
library. As already visible from \f{fig:photz} \ the photometric
and spectroscopic redshifts are in very good agreement. We find a
median of -0.02, and a standard deviation of 0.09 in the overall
$(z_\mathrm{phot}-z_\mathrm{spec})/(1+z_\mathrm{spec})$
distribution. However, from \f{fig:spec-phot-z} \ it is discernible that the systematic offset is higher for higher redshifts. Fitting $z<3$ and $z\geq3$ ranges separately we find a median offset of $0.00$, and $-0.04$, respectively, and a standard deviation of $0.11$, and $0.06$, respectively. 
For comparison, a similar median systematic offset (-0.023) has been found by
Wardlow et al.\ (2011) for their full sample of LESS SMGs with statistically
assigned counterparts. Yun et al. (2012) find a zero offset for
GOODS-South SMGs with statistically assigned counterparts, however
their results suggest a slight systematic underestimate of $z>3$
photometric redshifts (see their Fig.~2), consistent with the
results presented here. This suggests that spectral models used for photometric-redshift estimates could be better optimized for the high-redshift end. This is however beyond the scope of this paper, and here we will correct the ($z\geq3$) photometric redshifts computed for our SMGs for this systematic offset.

\begin{figure*}
\includegraphics[bb = 130 440 486 610,scale=0.465]{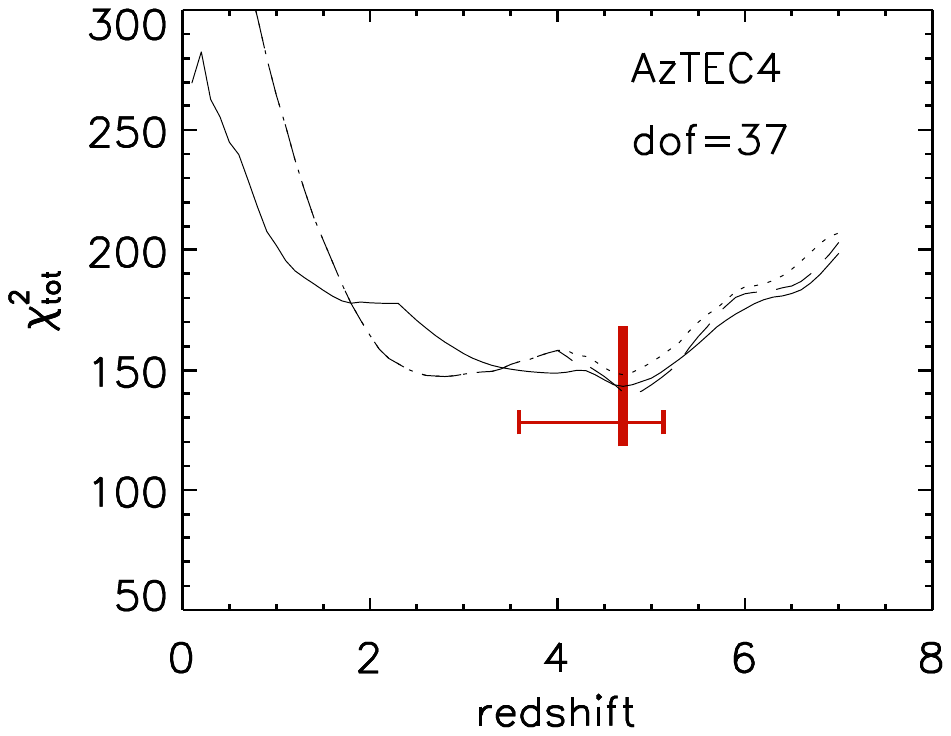}
\includegraphics[bb = 210 440 486 610,scale=0.465]{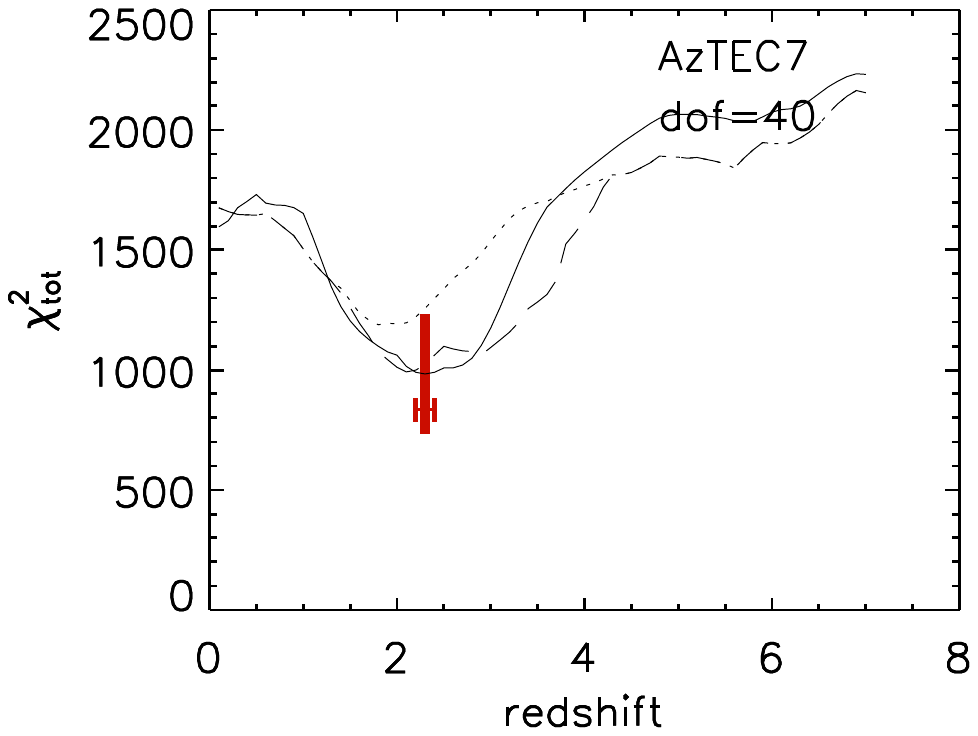}
\includegraphics[bb = 210 440 486 610,scale=0.465]{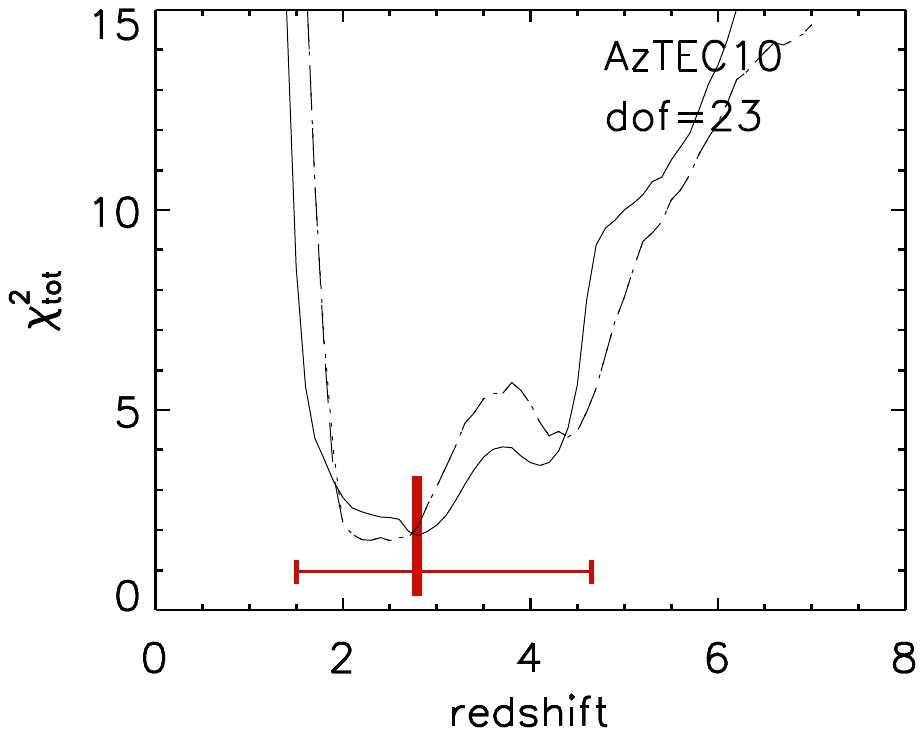}
\includegraphics[bb = 210 440 336 610,scale=0.465]{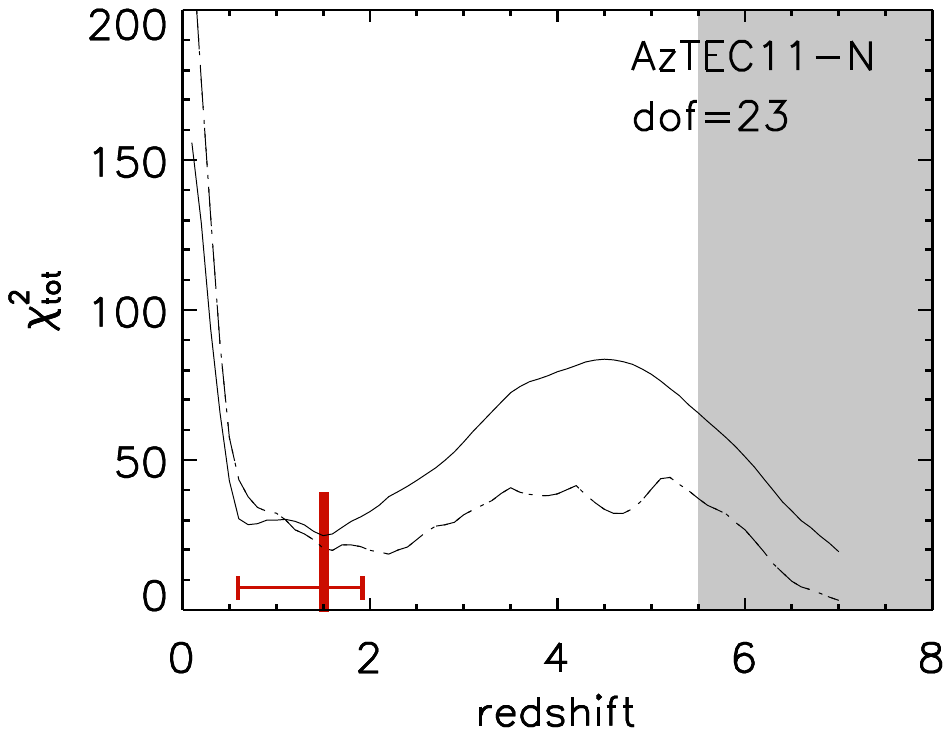}\\
\includegraphics[bb = 130 420 486 652,scale=0.465]{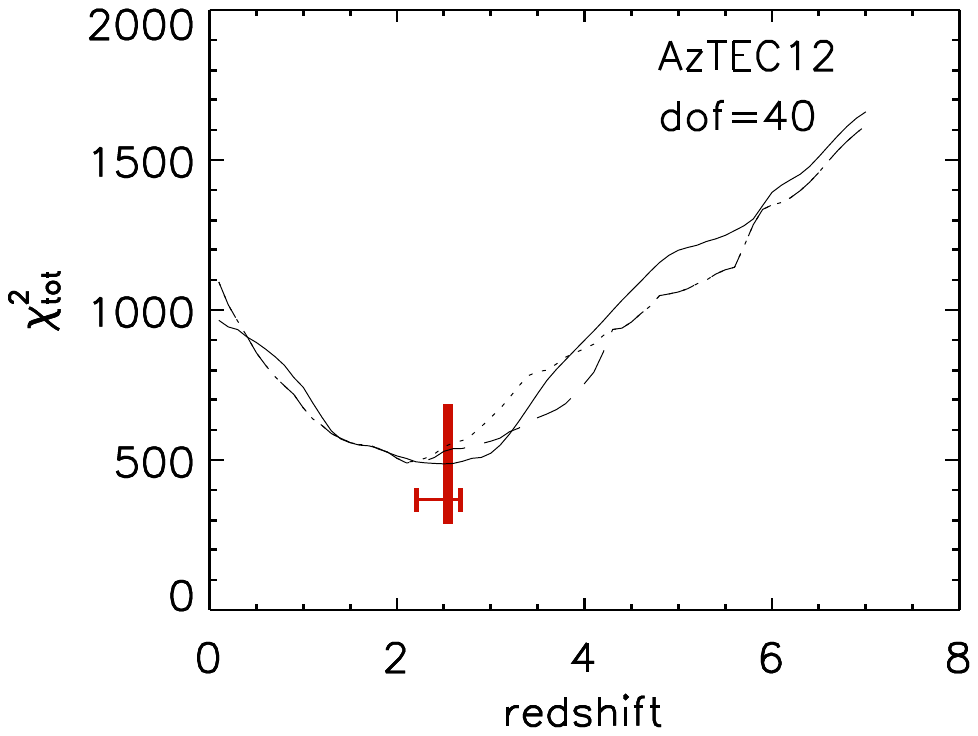}
\includegraphics[bb = 210 420 486 652,scale=0.465]{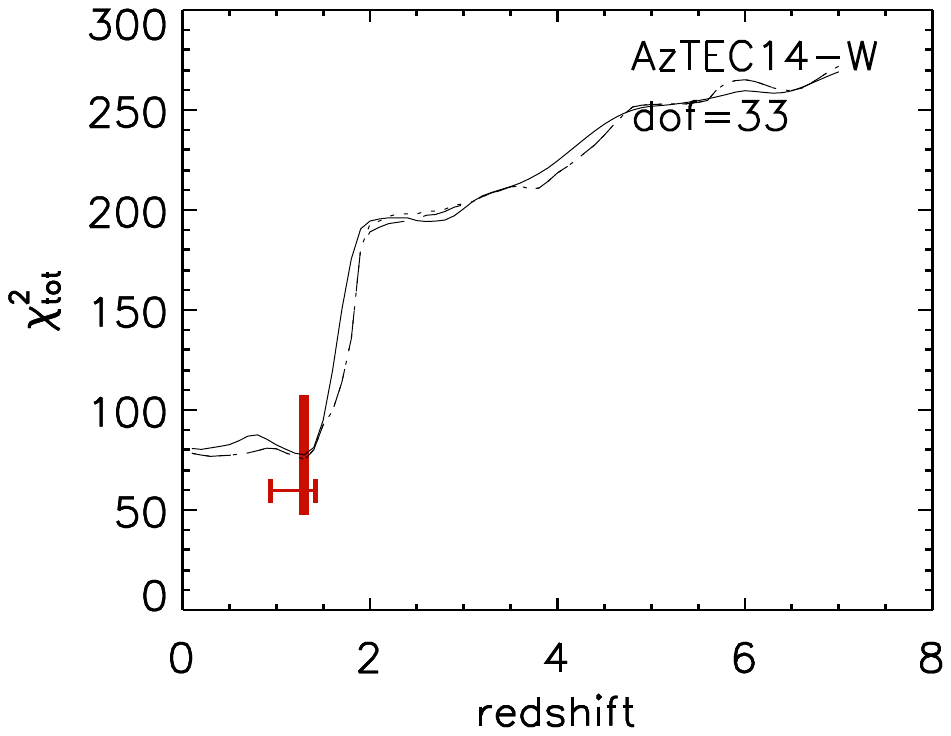}
\includegraphics[bb = 210 420 336 652,scale=0.465]{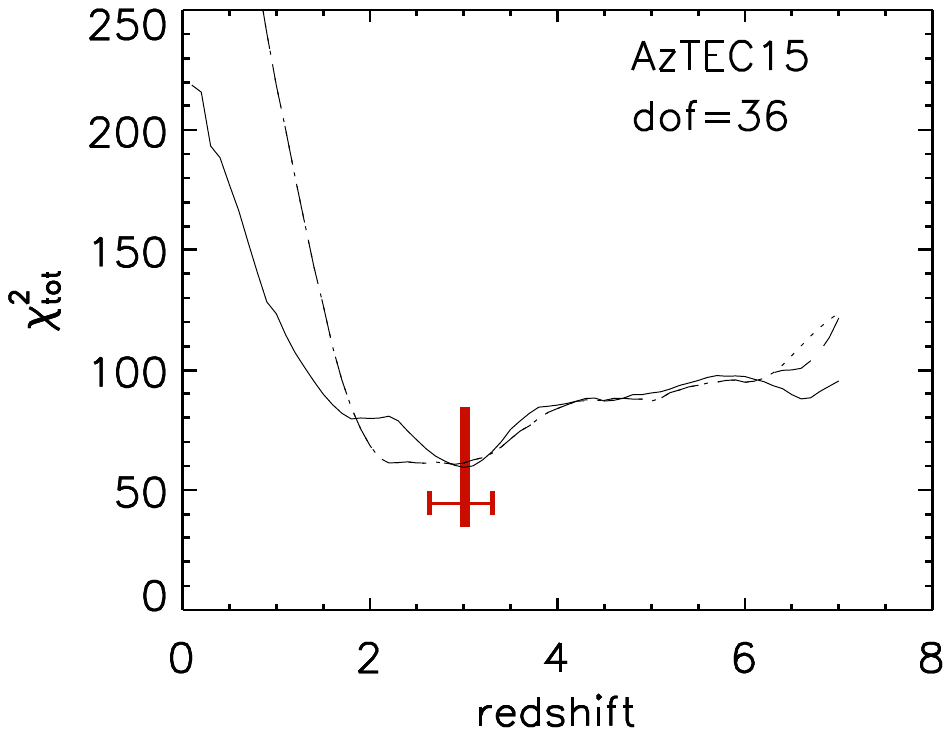}\\
\caption{Same as \f{fig:photz} , but for our AzTEC/JCMT/SMA COSMOS SMGs without spectroscopic redshifts. }
      \label{fig:photz1}
\end{figure*}

\begin{figure*}
\includegraphics[bb = 130 440 486 662,scale=0.465]{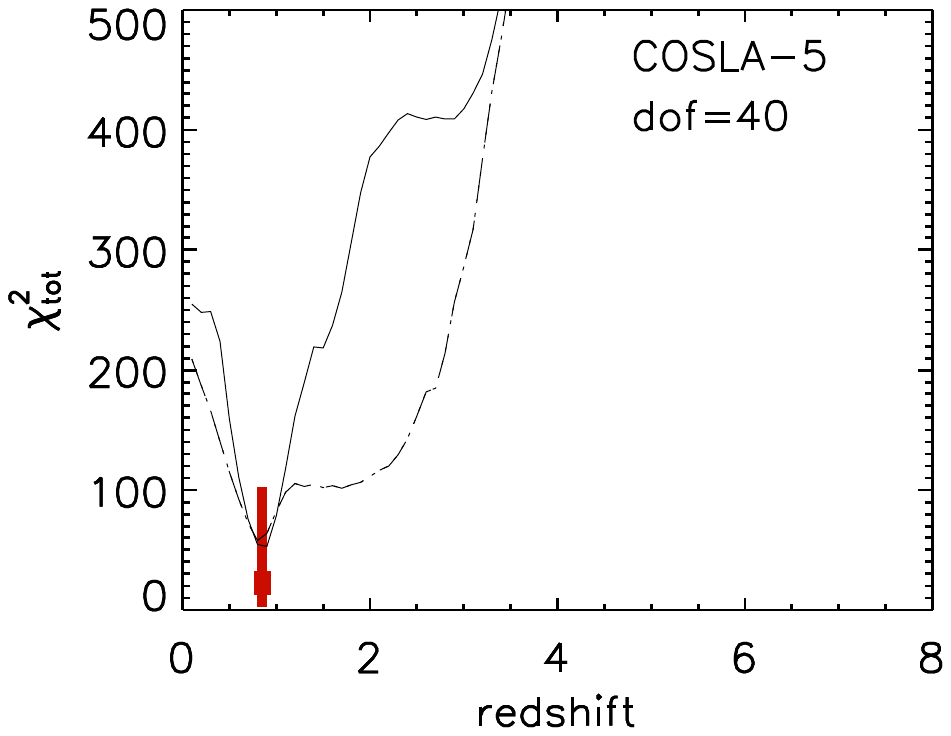}
\includegraphics[bb = 210 440 486 662,scale=0.465]{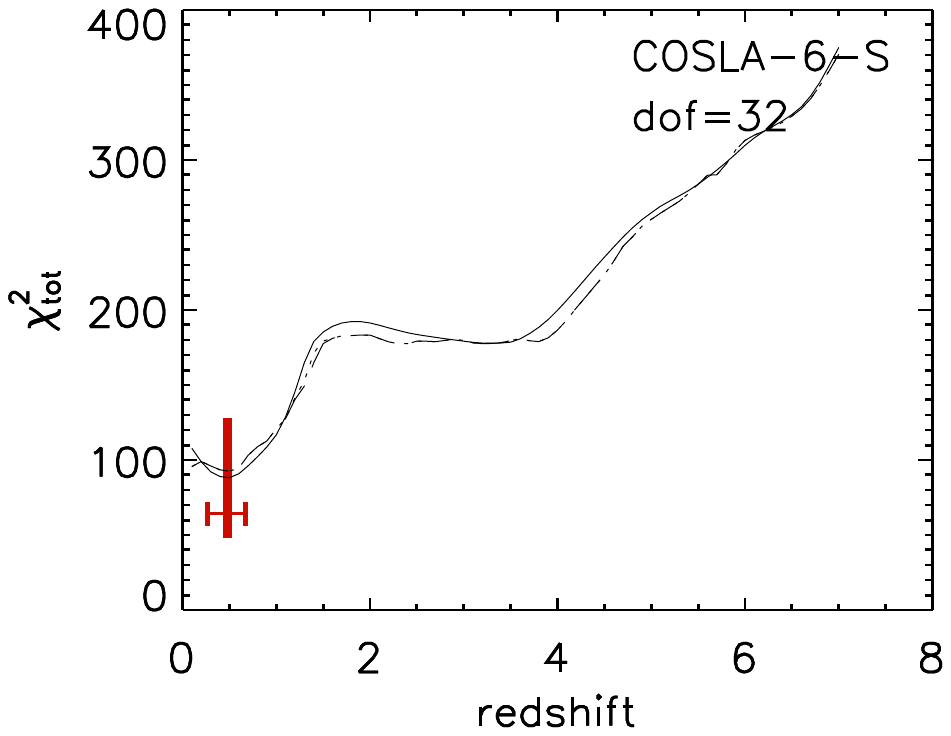}
\includegraphics[bb = 210 440 486 662,scale=0.465]{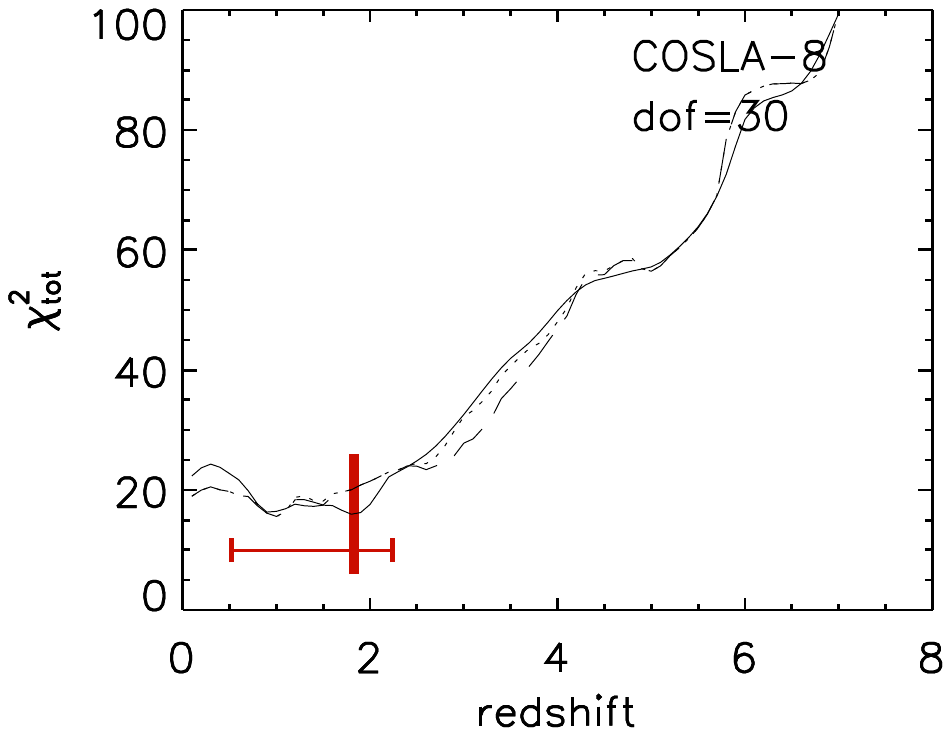}
\includegraphics[bb = 210 440 336 652,scale=0.465]{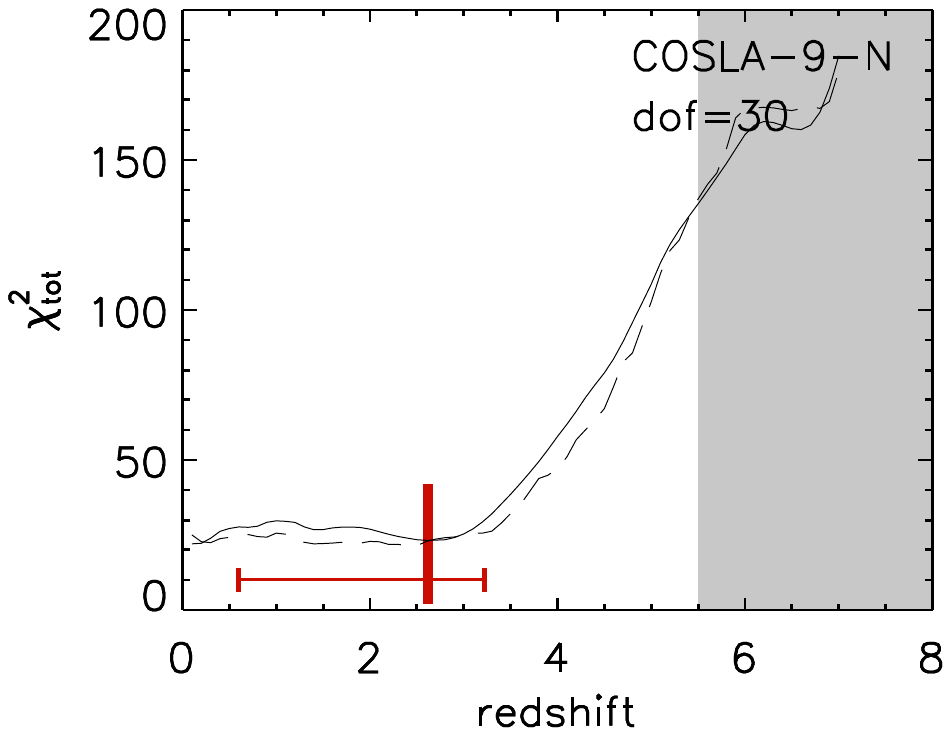}\\
\includegraphics[bb = 130 440 486 652,scale=0.465]{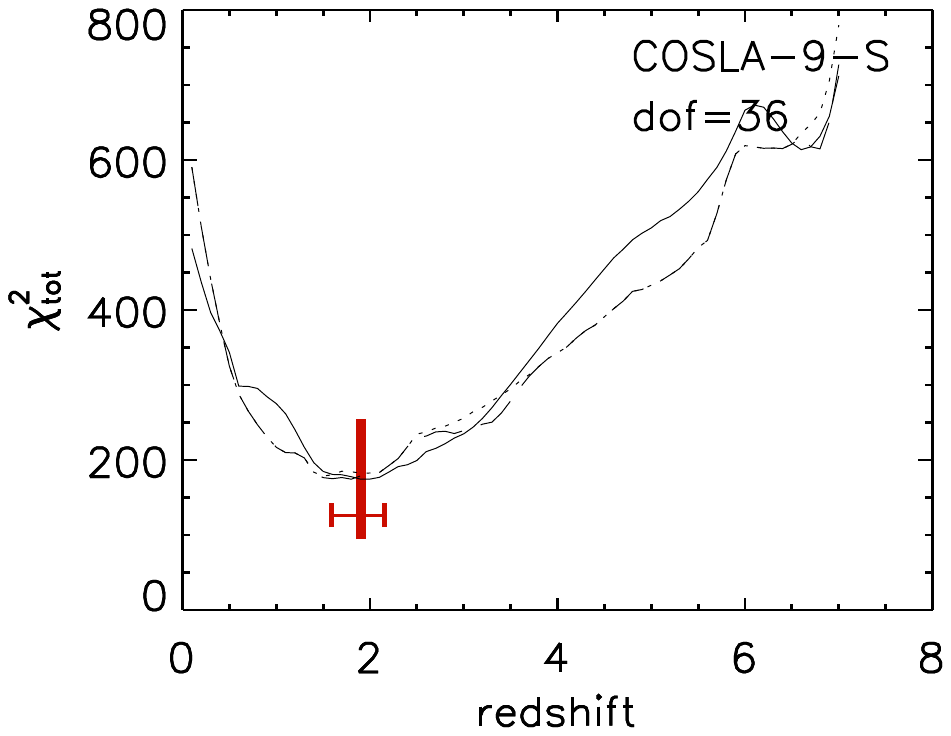}
\includegraphics[bb = 210 440 486 652,scale=0.465]{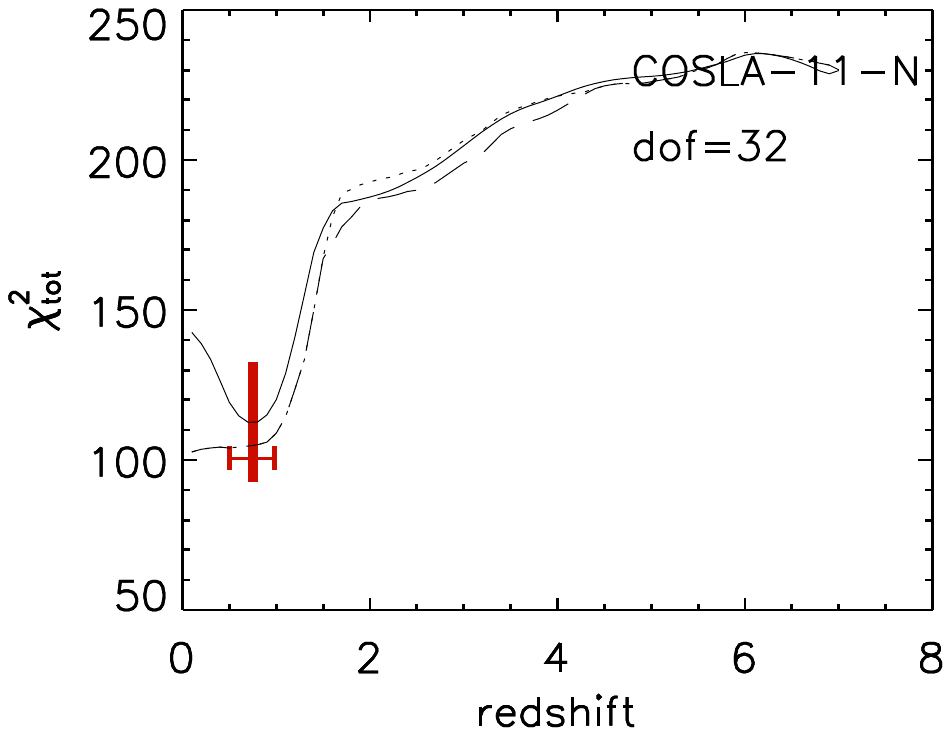}
\includegraphics[bb = 210 440 486 652,scale=0.465]{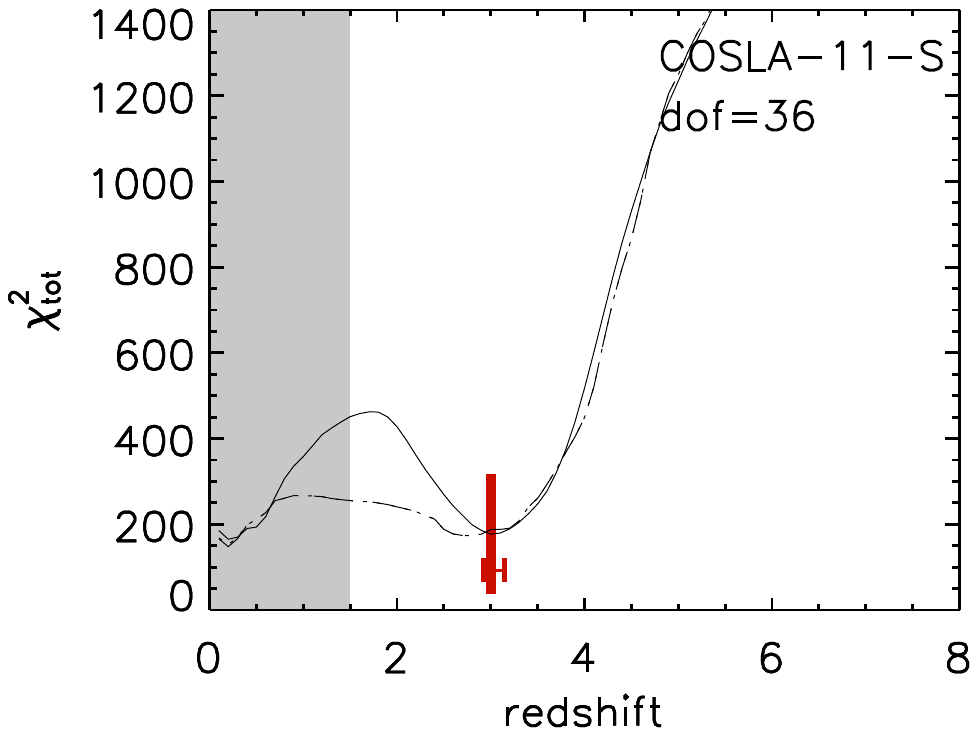}
\includegraphics[bb = 210 440 336 652,scale=0.465]{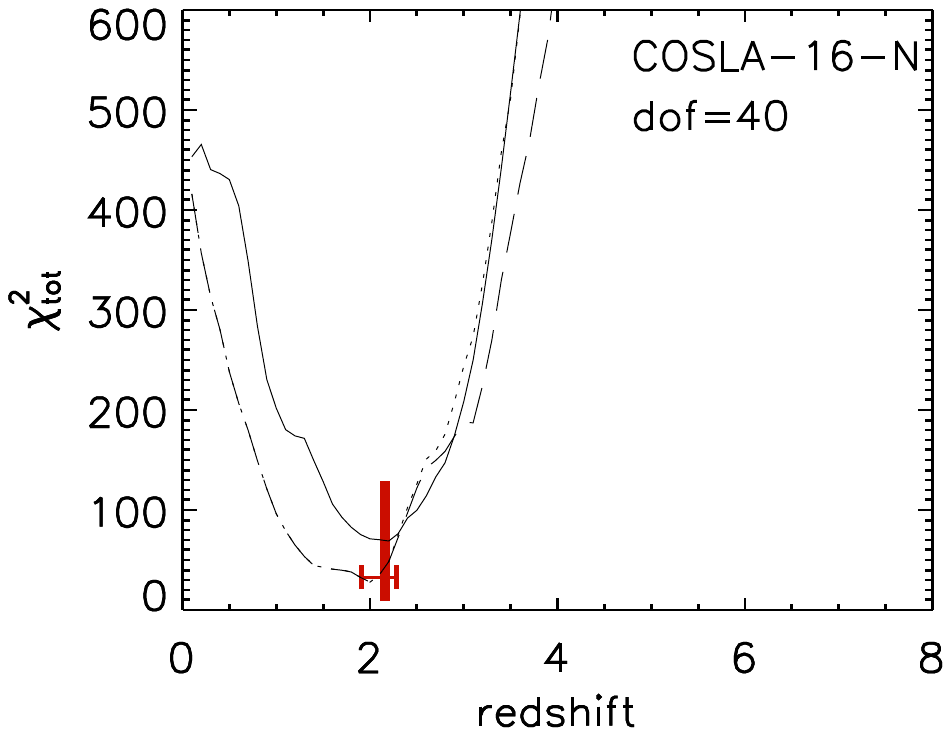}\\
\includegraphics[bb = 130 440 486 652,scale=0.465]{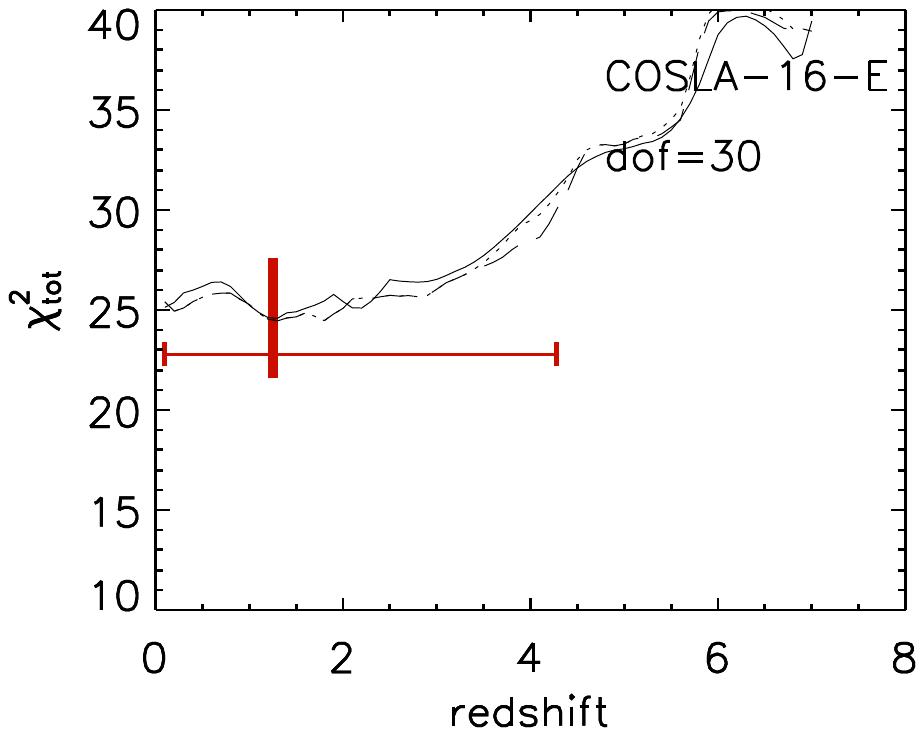}
\includegraphics[bb = 210 440 486 652,scale=0.465]{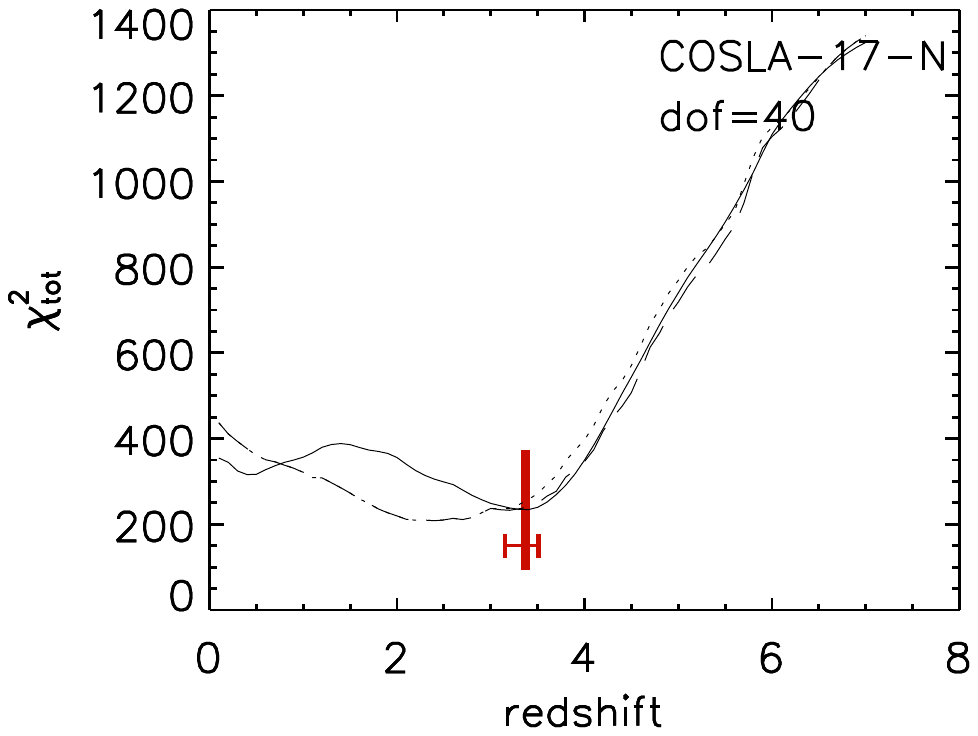}
\includegraphics[bb = 210 440 486 652,scale=0.465]{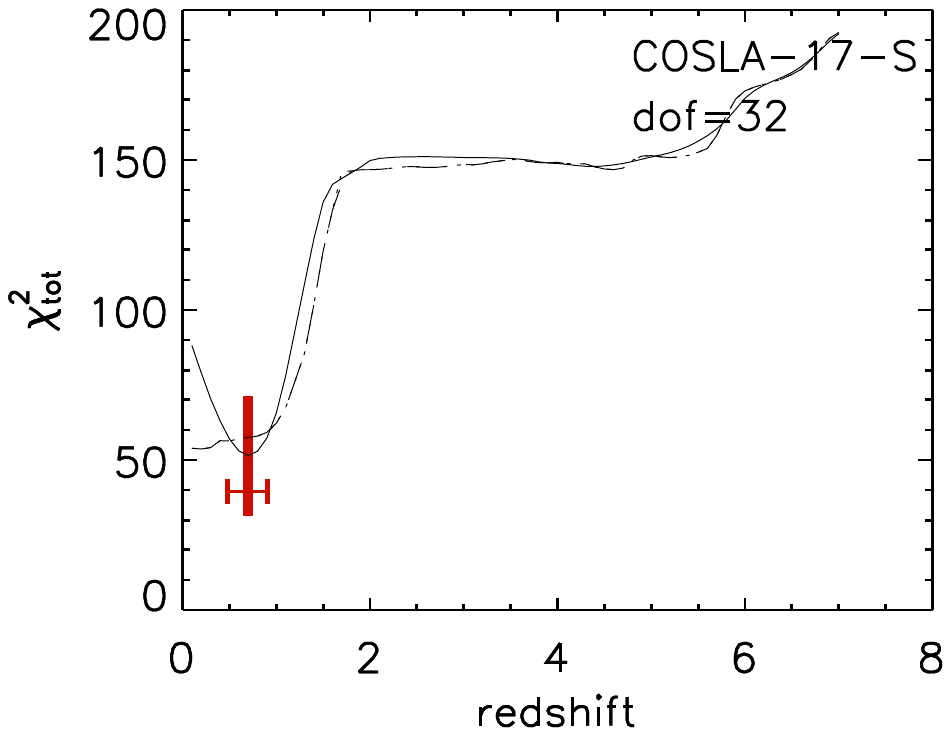}
\includegraphics[bb = 210 440 336 652,scale=0.465]{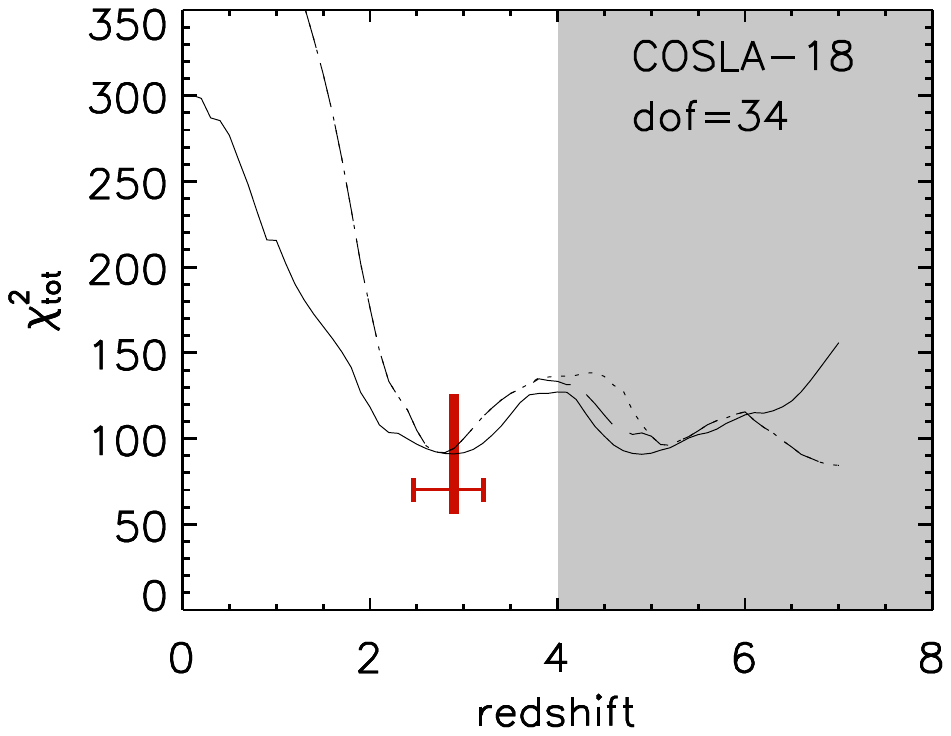}\\
\includegraphics[bb = 130 440 486 652,scale=0.465]{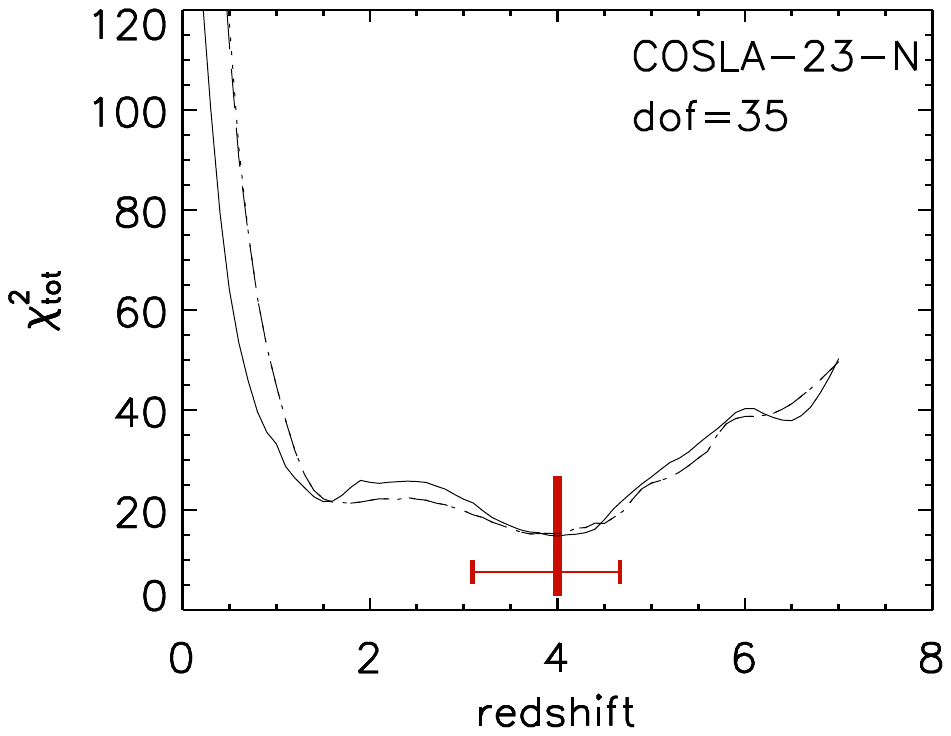}
\includegraphics[bb = 210 440 486 652,scale=0.465]{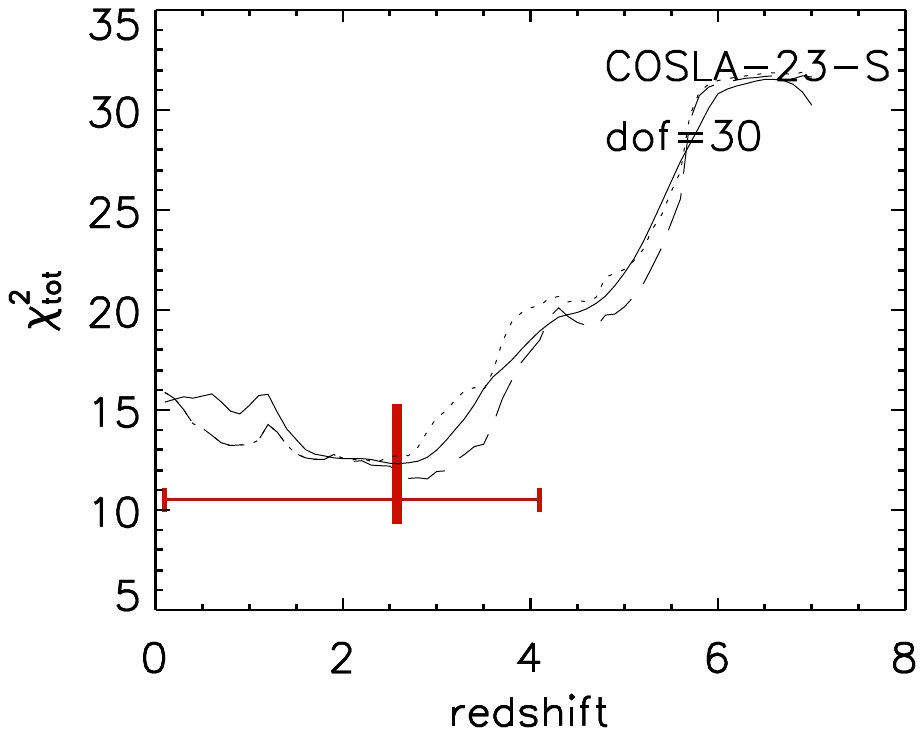}
\includegraphics[bb = 210 440 486 652,scale=0.465]{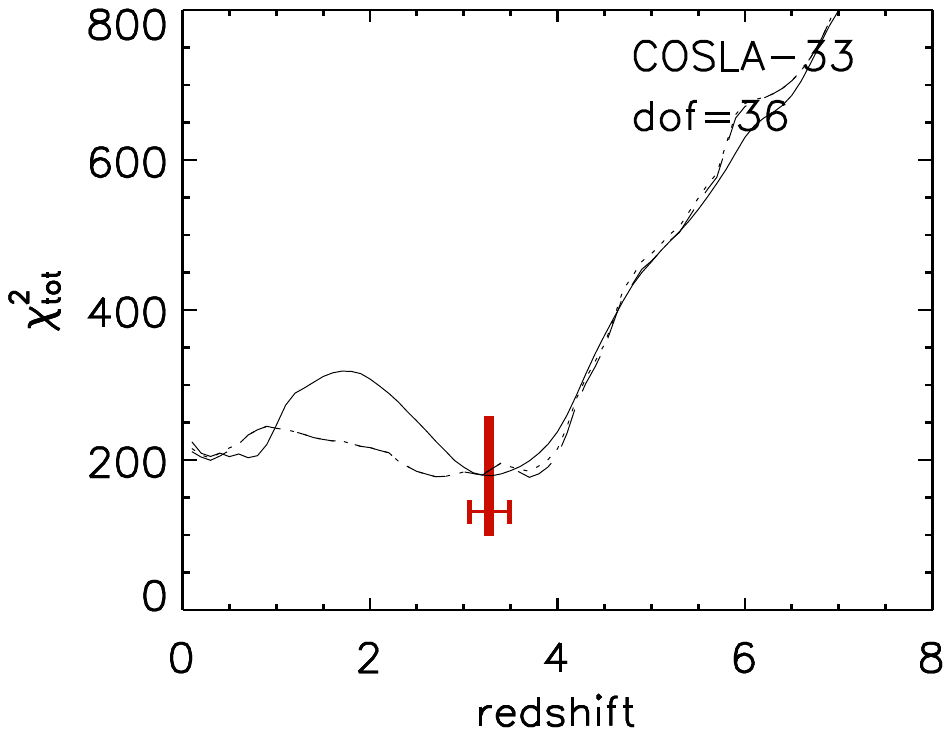}
\includegraphics[bb = 210 440 336 652,scale=0.465]{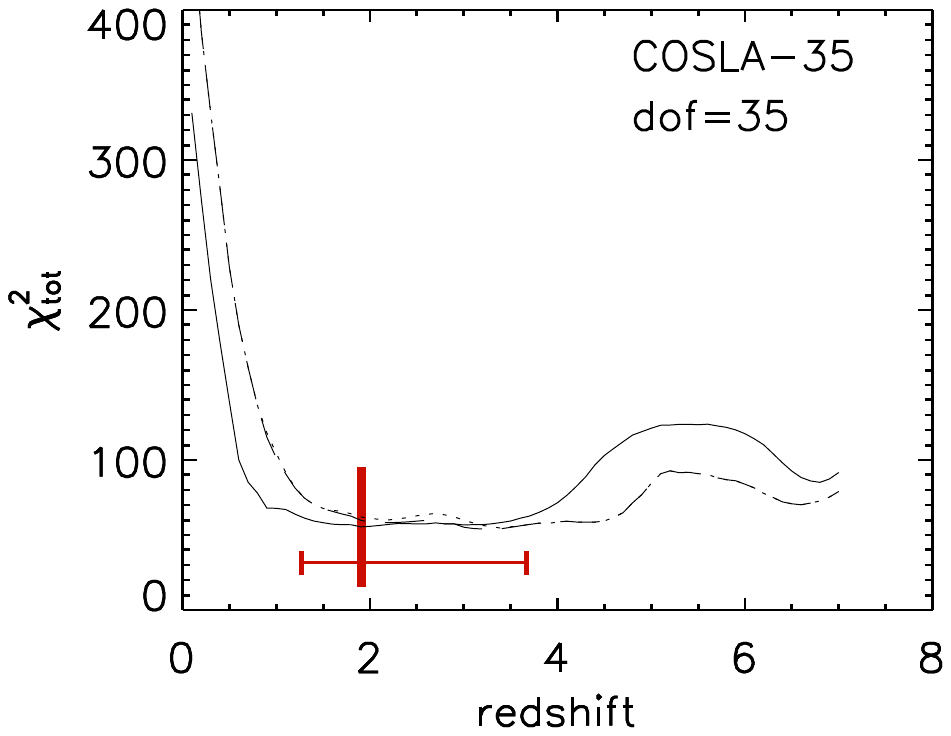}\\
\includegraphics[bb = 130 440 486 652,scale=0.465]{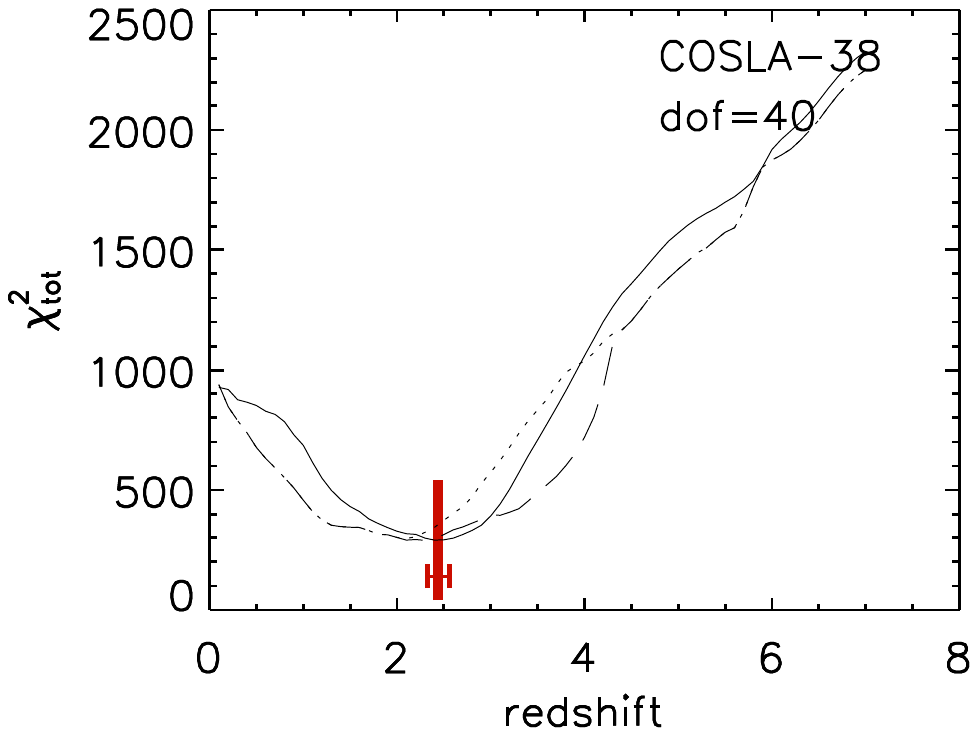}
\includegraphics[bb = 210 440 486 652,scale=0.465]{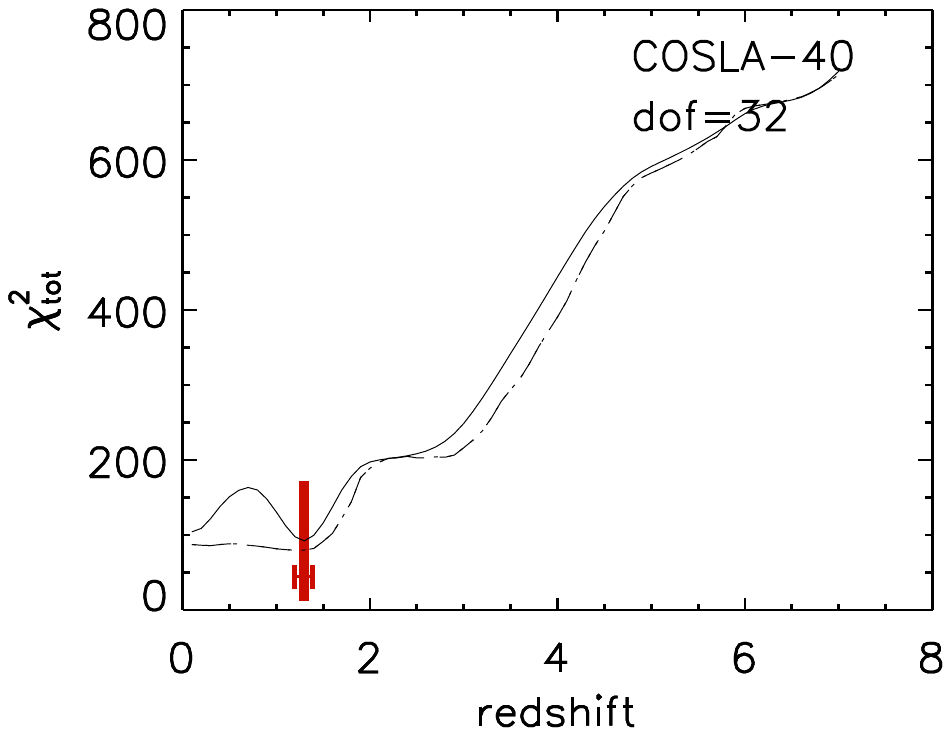}
\includegraphics[bb = 210 440 486 652,scale=0.465]{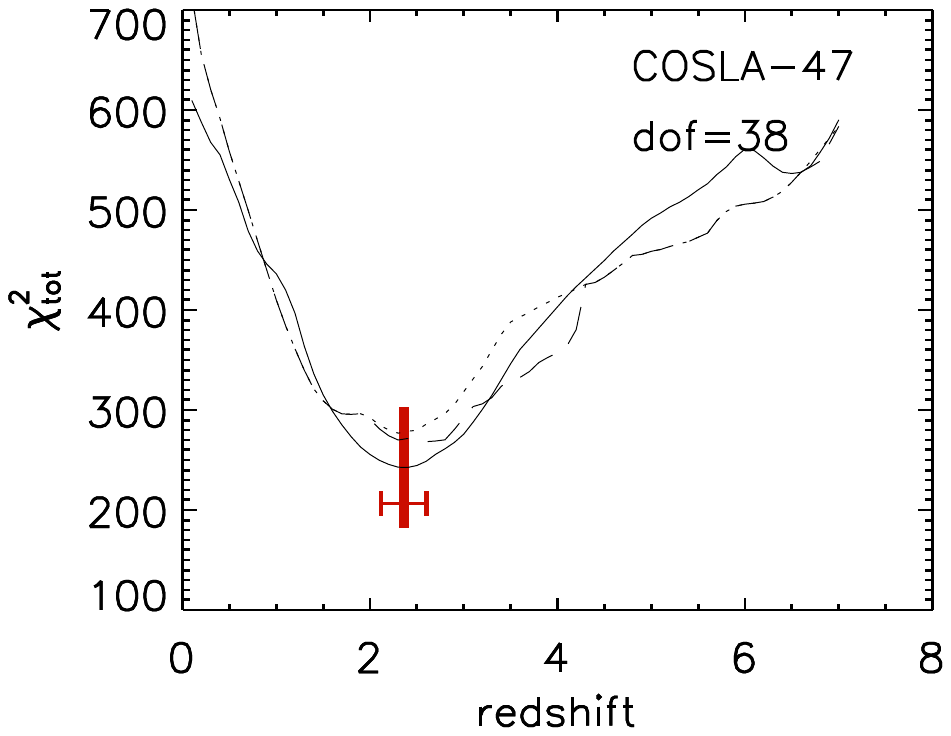}
\includegraphics[bb = 210 440 336 652,scale=0.465]{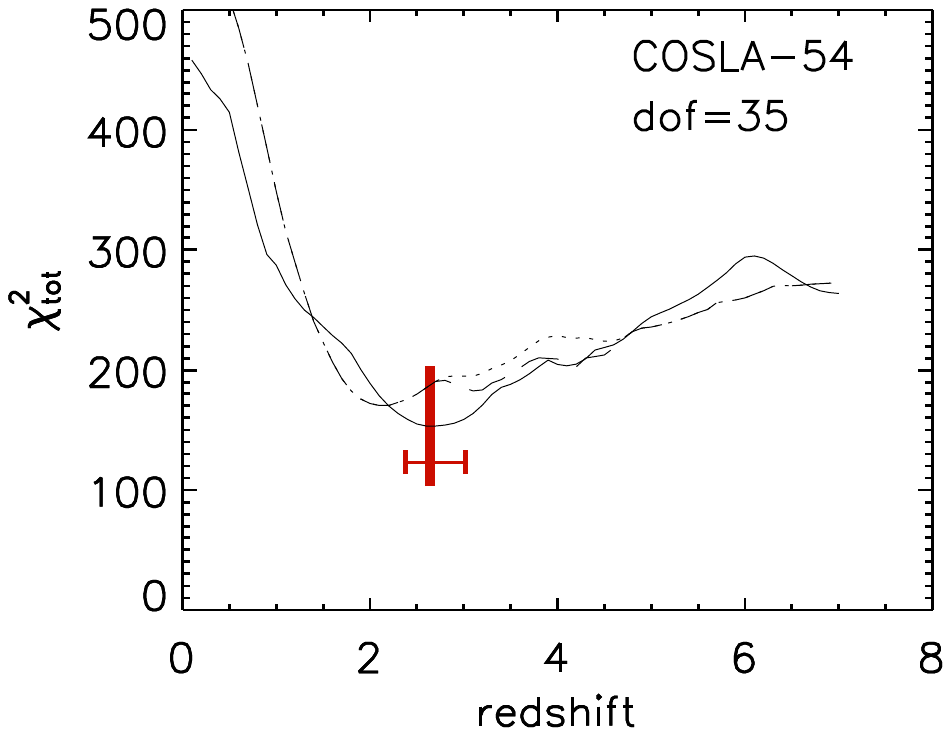}\\
\includegraphics[bb = 130 420 336 652,scale=0.465]{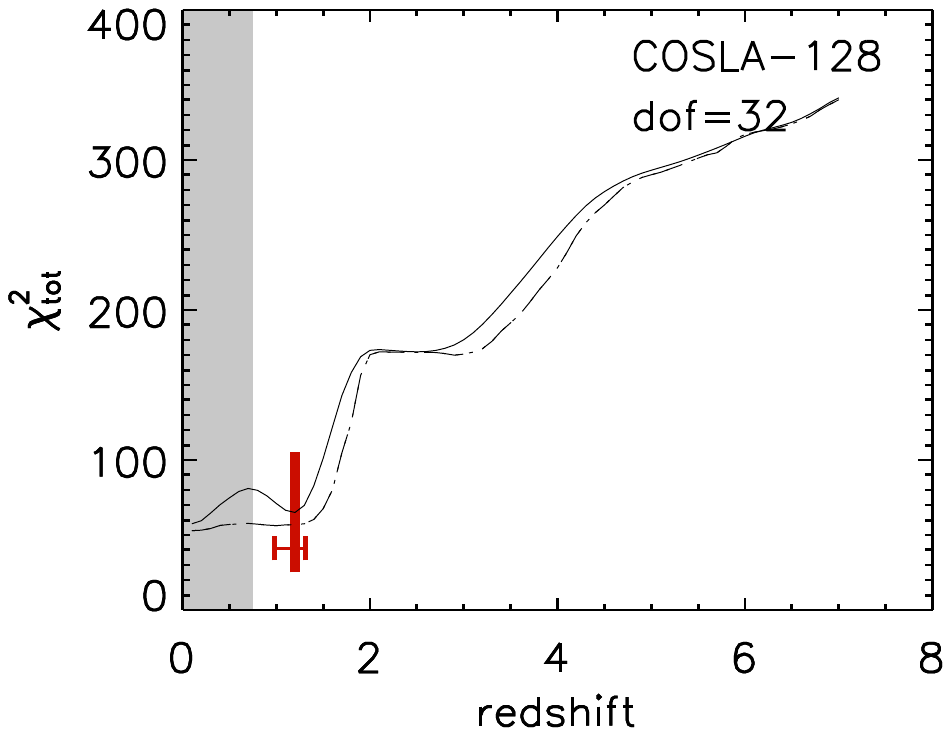}
\caption{Same as \f{fig:photz} , but for our LABOCA/PdBI COSMOS SMGs without spectroscopic redshifts. }
      \label{fig:photz2}
\end{figure*}

Using the same approach as described above we compute photometric redshifts for all SMGs in the COSMOS field with multi-wavelength counterparts determined via mm-interferometry mapping and without spectroscopic redshifts. We present their photometric redshift total $\chi^2$ distributions (prior to any systematic correction)  in Figs.\ \ref{fig:photz1} and \ref{fig:photz2}, and tabulate  their photometric redshifts (not corrected for the systematic offset)  in  Table~\ref{tab:interf} and Table~\ref{tab:pdbi}.

\subsection{AGN considerations}

As photometric redshifts are typically computed using libraries for
the stellar light only, it may be argued that substantial AGN
contribution to the UV-MIR SED for some SMGs may affect our
photometric redshift estimate. Note however that only bright Type 1 (broad line) AGN need special treatment for photometric redshift estimates (see Salvato et al.\ 2010). For low-luminosity
(Seyfert, Type 2) AGN, with SEDs dominated by the stellar light of a galaxy (e.g.\ Kauffmann et al.\ 2003) usual photometric redshift computations, as the one presented here, are expected to yield satisfactory results.

To address the AGN issue in our SMG sample we
have utilized the X-ray data from the Chandra-COSMOS survey (Elvis et al.
2009), which provide the most direct way to identify AGN associated with the SMGs in our
1.1mm- and 870$\mu$m-selected samples. 

Only COSLA-161, for which we
find a good agreement between its photometric and spectroscopic
redshifts, is found to be associated with X-ray emission (note however
that given the X-ray 0.1-10~keV rest-frame luminosity of
$(6.2\pm2.7)\times10^{40}$ ergs s$^{-1}$ at the source's low spectroscopic redshift, it is not clear whether the
source of X-rays is star-formation or emission from the nucleus; see
Appendix for more details). In order to put further constraints on the
AGN properties of our SMGs, we derive the average X-ray flux in the
0.5--2~keV band using all COSMOS SMGs with interferometric positions.  This
is done in such a way that for each SMG we extract the X-ray counts
from the 0.5--2~keV band image within a circular aperture of $1.5"$ in
radius, and then convert this to an average X-ray flux. 

For this stacking analysis we only used the so called best PSF Chandra
mosaic (Elvis et al. 2009), that has a continuous coverage of the
central 0.5\,deg$^2$ of COSMOS at 50ks depth, in order to be able to
use a small extraction region and therefore reduce contamination.  The
background counts were estimated using the stowed Chandra background
data after normalizing the background image to the average background
rate in a source-free zone. After background subtraction we find a
marginal detection at a $1.5\sigma$ level in the stack with
$F_\mathrm{0.5-2keV}=(0.9\pm0.6)\times10^{-17}$~erg~s$^{-1}$~cm$^{-2}$.
For SMGs at redshifts $z=$2, 3, and 4, and assuming a power-law X-ray
spectrum with photon index 1.8 (typical for AGN), the obtained average
flux converts to average bolometric X-ray luminosities (rest-frame
0.1-10~keV) of $(7.5\pm5)\times10^{41}$, $(1.9\pm1.3)\times10^{42}$,
and $(3.7\pm2.5)\times10^{42}$~erg~s$^{-1}$ (given the marginal
detection, these values should be considered as upper limits). 

The inferred X-ray luminosities are typical for normal galaxies rather
than strong AGN ($L_X>10^{42}$~erg~s$^{-1}$; e.g.\ Brusa et al.\ 2007).  This rules out a
major AGN contribution within our SMG sample (consistent with previous studies of SMGs; Alexander et al.\ 2005; Menendez-Delmestre et al.\ 2009), and thus also a
significant influence of AGN on the accuracy of our photometric
redshift estimates.  Furthermore, as demonstrated by the source Vd-17871 (see previous Section and Karim et al., in prep.)
  buried AGN only obvious in the IR SED do not appear to affect the
  method. This is consistent with the results from Wardlow et al.\ (2011) who have found that the accuracy of photometric redshifts is not affected for SMGs showing an IR (8~$\mu$m) excess likely due to an AGN component. 

\section{Redshift distribution of SMGs in the COSMOS field}
\label{sec:zdistrib}

In this section we present the redshift distributions for our 1.1mm- and 870$\mu$m-selected samples.
To derive the redshift distributions, we take spectroscopic redshifts if available, and otherwise photometric redshifts based on Michalowski et al.\ (2010) spectral templates, and corrected for the systematic offset as discussed in the previous Section (see Table~\ref{tab:statsamples}). 

\subsection{Redshift distribution of AzTEC/JCMT SMGs with mm-interferometric positions}

Our \mmsample \ contains 17 SMGs\footnote{when AzTEC-11 is treated as two separate sources; see Appendix~B for details.}  with accurate positions from 890~$\mu$m
interferometric observations at intermediate resolution ($\sim2"$)
with the SMA (Younger et al.\ 2007; 2009). 
Spectroscopic
redshifts, based on optical (DEIMOS) and/or CO (CARMA/PdBI)
spectroscopic observations, are available for 7 out of the 17
AzTEC/JCMT/SMA SMGs (see Table~\ref{tab:interf}). For 7 of the remaining
sources we use photometric redshifts, derived as described in
\s{sec:redsft} . Three sources (AzTEC-11S, AzTEC-13 and AzTEC-14E) cannot be
associated with multi-wavelength counterparts in our deep COSMOS
images. Thus, for these we use the mm-to-radio flux ratio based
redshifts, often utilized for the derivation of distances to SMGs
(Carilli \& Yun 1999, 2000). Consistent with the faintness at optical,
IR, and radio wavelengths the mm-to-radio flux based redshifts suggest
$z\gtrsim3$ (see Table~\ref{tab:statsamples}) for all three sources when the PdBI 1.3~mm fluxes and an Arp~220
template are used (following Aravena et al.\ 2010a). The redshifts for
AzTEC/JCMT COSMOS SMGs are summarized in Table~\ref{tab:statsamples}.

The redshift distribution for the 17 AzTEC/JCMT SMGs mapped by SMA is shown in the left panel of  \f{fig:photzdistrib} . Given that for three sources we only have lower redshift limits, we compute the mean redshift using the statistical package ASURV which relies on survival analysis, and takes upper/lower limits properly into account (assuming that sources with limiting values follow the same distribution as the ones well constrained).
 We infer a mean redshift of $3.06\pm0.37$ for the AzTEC/JCMT/SMA sources. We note that treating the source AzTEC-11  as a single source yields a mean redshift of $3.00\pm0.38$.

\subsection{Redshift distribution of  LABOCA-COSMOS SMGs with mm-interferometric positions}

Unlike the  \mmsample , the $870\mu$m-selected one is not strictly limited in flux or signal-to-noise. 
 This is because a fraction of $\mathrm{S/N_{870\mu m}}\gtrsim3.8$ LABOCA-COSMOS SMGs have not been detected by our PdBI observations at 1.3~mm. The least biased sample that can be constructed from these detections is a sample of the LABOCA-COSMOS sources detected with interferometers at mm-wavelengths, but without prior assumptions, such as e.g. multi-wavelength counterpart associations (as described in \s{sec:obs} , we required that our $\mathrm{S/N_{1.3mm}}\leq4.5$ PdBI sources had to be confirmed by independent multi-wavelength detections).  This yields 16 sources (9 PdBI detected with S/N$>4.5$, and 7 detected with SMA or CARMA\footnote{Out of the total of 8 LABOCA-COSMOS sources detected by CARMA or SMA (see Table~\ref{tab:interf}), only 7 were detected without priors (AzTEC/C1, $z_\mathrm{phot}=5.6\pm1.2$ was detected at $3.2\sigma$ with the CARMA interferometer, and verified by a coincident $4.4\sigma$ radio source; \smo \ et al.\ 2012).}) in our least-biased \submmsample , listed in Table~\ref{tab:statsamples}.  It is hard to assess the completeness of this sample. However, if our detection rate of the LABOCA sources can be considered random i.e.\ devoid of any redshift-biases, and the properties of the non-detected LABOCA sources are similar to those of the detected ones, then one can assume that this subsample reflects the distribution of the parent LABOCA SMG sample within the same flux limits. Furthermore, note that the 9 LABOCA/PdBI sources detected with S/N$_\mathrm{1.3mm}>4.5$ within our least-biased \submmsample \ can be regarded as a 1.3~mm flux-limited sample ($F_\mathrm{1.3mm}\gtrsim2.1$~mJy; given that we reached an rms of $\sim0.46$~mJy in our PdBI maps). Hence, if the redshift distribution of the least-biased \submmsample \ is consistent with that of the PdBI sub-sample, we can assume that this reflects the distribution of SMGs with $F_\mathrm{1.3mm}\gtrsim2.1$~mJy.

Five out of the 16 SMGs in our least-biased \submmsample \footnote{COSLA-2, COSLA-3, COSLA-4, COSLA-60, COSLA-73} have spectroscopic redshifts. For another eight we use photometric redshifts as derived in \s{sec:redsft} , and for 3 others  we use the mm-to-radio flux based redshifts. Their redshift distribution, shown in the middle panel of \f{fig:photzdistrib} , is similar to that of the PdBI subsample.  Using the ASURV statistical package we infer a mean redshift of $2.59\pm0.36$, and $2.29\pm0.48$ for the least-biased \submmsample , and the PdBI $\mathrm{S/N_{1.3mm}}>4.5$ subsample, respectively.

Given the faintness of the counterpart of COSLA-17-S its photometric redshift ($0.7^{+0.21}_{-0.22}$) is rather poorly constrained, and discrepant compared to the mm-to-radio based one ($\gtrsim4$). Thus it is possible that this source is at high redshift and further mm-observations of COSLA-17-N are required to affirm the reality of this source (see Appendix~A for details).  Nonetheless, excluding COSLA-17 N and S from the sample, we obtain consistent mean redshifts for the least-biased \submmsample \ ($2.65\pm0.38$) and the PdBI $\mathrm{S/N}>4.5$ subsample within ($2.34\pm0.55$). 

We show the redshift distribution of the joint \mmsample\ and least-biased \submmsample \  in the right panel of \f{fig:photzdistrib} . A mean redshift of $2.80\pm0.28$\footnote{A consistent mean redshift ($\bar{z}=2.76\pm0.28$) is found if AzTEC-11 is treated as a single source.} is found. A comparison with results from literature is given in \f{fig:photzdistribcompare} \ and discussed in detail in \s{sec:discussion} .

\begin{figure*}
\includegraphics[bb = 80 400 426 792,scale=0.45]{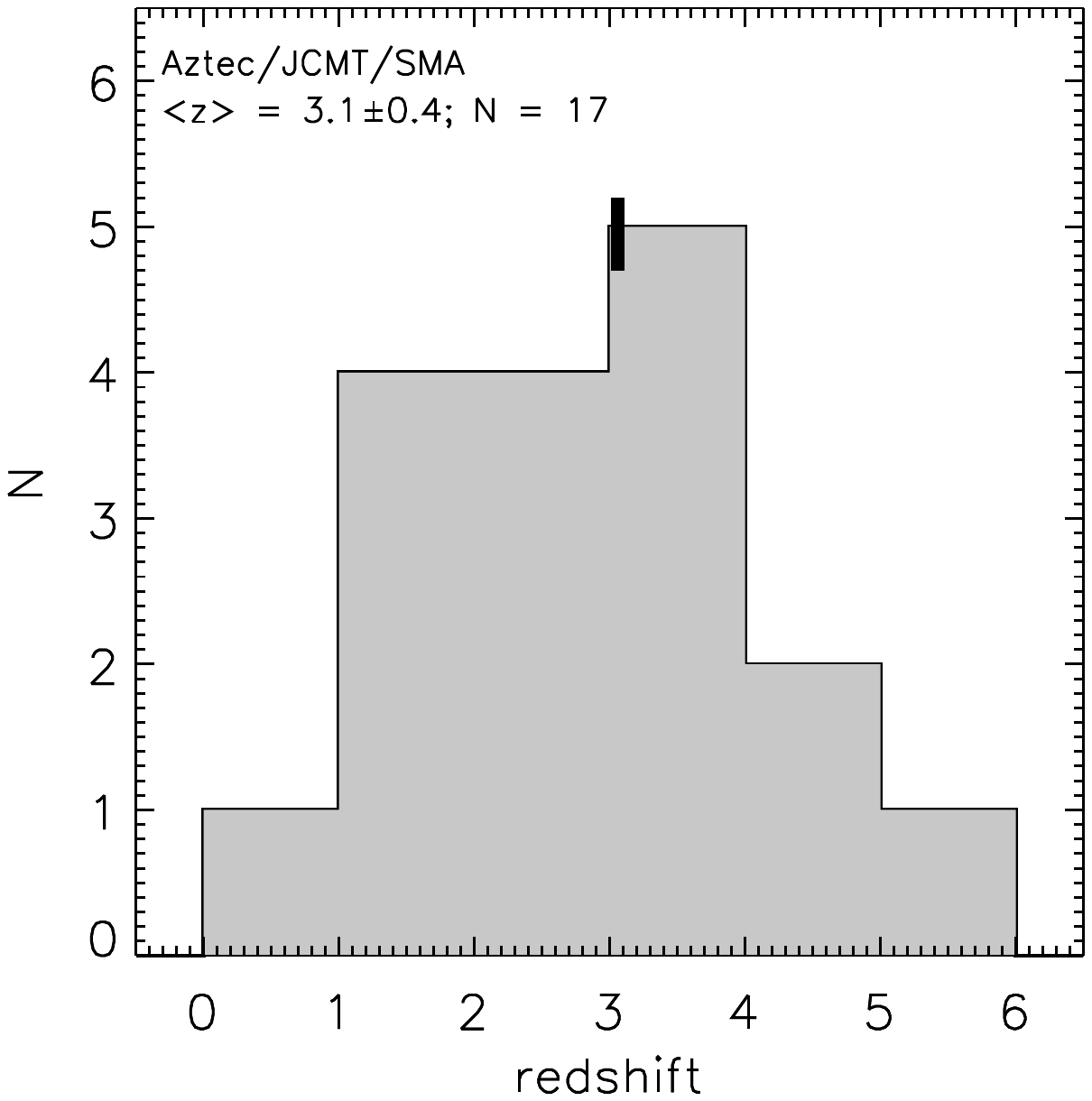}
\includegraphics[bb = 40 400 426 792,scale=0.45]{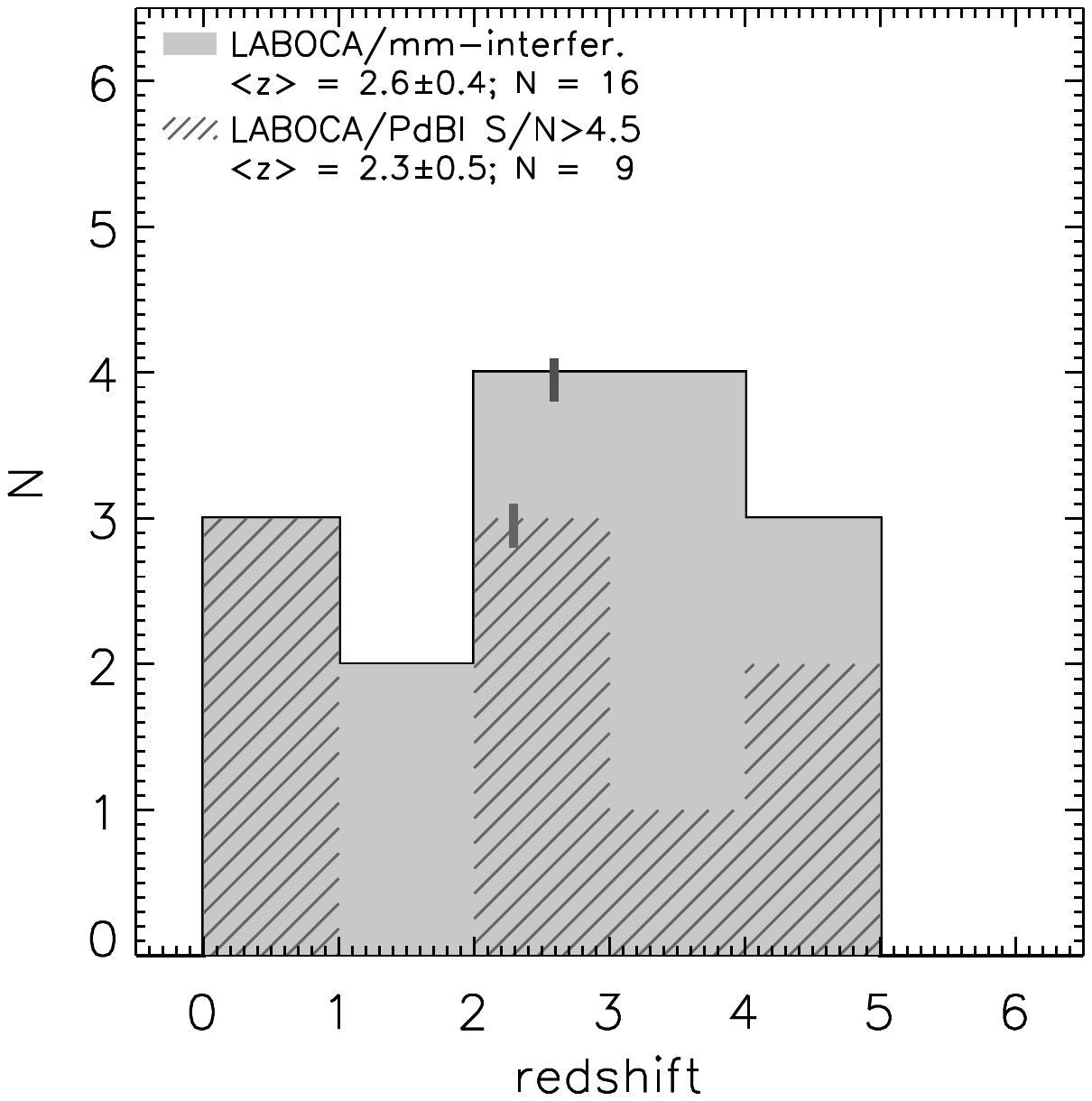}
\includegraphics[bb = 40 400 426 792,scale=0.45]{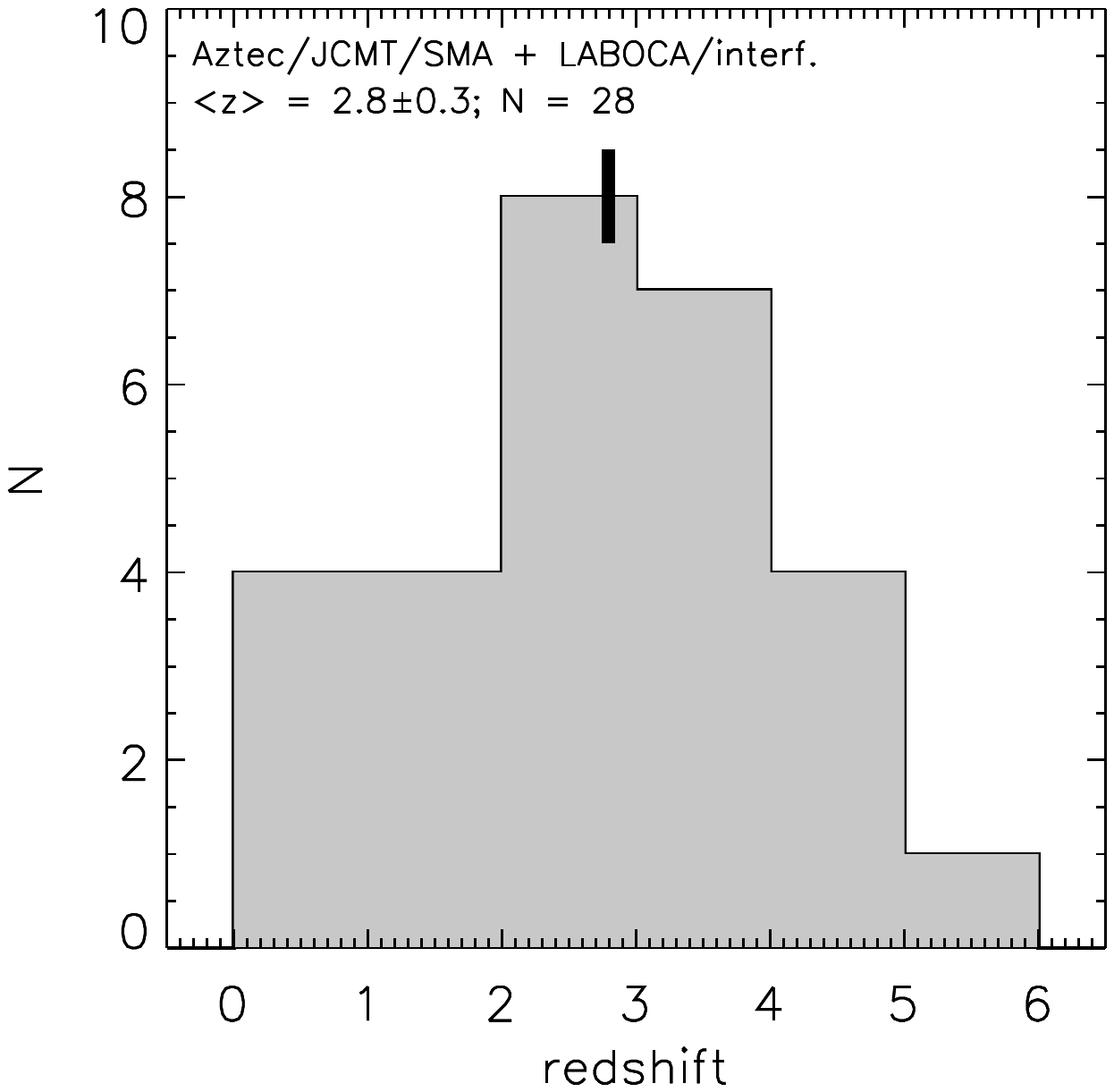}
\caption{ Redshift distribution of our \mmsample \ (left panel), \submmsample \ (middle panel), and the two samples combined (with sources present in both samples counted only once; right panel). In the middle panel we also show the redshift distribution of our $\mathrm{S/N_{1.1mm}>4.5}$ PdBI-detected LABOCA sample (hatched histogram).  Mean redshift values, and corresponding errors obtained using the statistical package ASURV, as well as the number of sources in each sample are indicated in each panel. Mean redshifts for every sample distribution are also indicated by the thick vertical lines. }
      \label{fig:photzdistrib}
\end{figure*}

\section{Discussion}
\label{sec:discussion}

\subsection{The redshift distribution of SMGs}

In \f{fig:photzdistribcompare} \ we compare the (normalized) redshift
distribution of the AzTEC/JCMT/SMA (left panel) and LABOCA/interferometric (middle panel)
COSMOS samples, and their joint distribution (right panel), with redshift distributions of SMGs derived for other
surveys (Chapman et al.\ 2005; Banerji et al.\ 2011; Wardlow et al.\ 2011; Yun et al.\
2012). The redshift distribution derived by Chapman et al.\ (2005) is
based on a sample of 76 SMGs drawn from various SCUBA 850~$\mu$m
surveys with counterparts identified via radio sources present within
the SCUBA beam, and observed with Keck~I to obtain optical spectroscopic redshifts
 (see their Tab.~2). To account for the redshift desert at $z= 1.2-1.7$ in the Chapman et al.\ sample we supplement it with 19 SMGs with DEIMOS spectra drawn from Banerji et al.\ (2011; see their Tab.~2). We combine these two data sets normalizing each by the observed area (556~arcmin$^2$ for the Banerji et al.\ , and 721~arcmin$^2$ for the Chapman et al.\ samples;  Chapman, priv.\ com.).
 
 The distribution published by Wardlow et al.\ (2011)
is based on 74 SMGs drawn from the LESS survey at 870~$\mu$m that
could be assigned robust counterparts based on the P-statistic (using
radio, 24~$\mu$m and IRAC data; Biggs et al.\ 2011).  Wardlow et al.\ derived photometric redshifts for these galaxies (see their Tab.~2) accurate to $\sigma_{\Delta z/(1+z)}=0.037$. Using the
P-statistic to associate counterparts to SMGs (although using a
slightly modified method to that utilized by Biggs et al.\ 2011) Yun
et al.\ (2012) identified $44$ (robust and tentative) counterparts to SMGs detected with
AzTEC at 1.1~mm in the GOODS-S field. For 16 sources in this sample a
spectroscopic redshift is used, for 21 a photometric redshift was
inferred by Yun et al., and for 7 only a mm-to-radio based redshift
could be derived (see their Tab.~3).

From \f{fig:photzdistribcompare} \ it is immediately obvious that the
redshift distribution of the COSMOS SMGs is much broader 
compared to that derived from previous surveys, in which the SMG
counterparts were identified statistically within the large bolometer
beam. In particular, significant high-redshift ($z\gtrsim4$) and
low-redshift ($z<2$) ends are present. 
 In the \submmsample \ we find five\footnote{COSLA-3, COSLA-4, COSLA-6S, COSLA-17S, COSLA-128} out of 16 SMGs (i.e.\ $\sim30\%$) at $z<1.5$. While the redshifts of two of these are spectroscopically confirmed, the photometric redshifts for the other three show possible secondary (higher) redshift solutions, which are more consistent with their mm-to-radio flux based redshifts. In our \mmsample \ we find four\footnote{AzTEC-2, AzTEC-6, AzTEC-9, AzTEC-14W} out of 17 ($23.5\%$) SMGs at $z<1.5$. The redshifts for three of these are spectroscopically confirmed. Thus, in total we find roughly 20-30\% of SMGs at $z<1.5$ (see left and middle panels in \f{fig:photzdistribcompare} ).  Such low redshift SMGs, present in the combined Chapman et al. (2005) and Banerji et al.\ (2011) sample but interestingly missed in the 1.1~mm AzTEC-GOODS-S and 870~$\mu$m-LESS samples, are expected in models of the evolution of infrared galaxies (see e.g.\ Fig. 7 in B\'{e}thermin et al.\ 2011), and they could be explained by cold dust temperatures in these sources (e.g.\ Greve et al.\ 2006; Banerji et al.\ 2011). A more detailed analysis of the physical properties of these SMGs will be presented in an upcoming publication.

We find significantly more SMGs at the high-redshift end ($z\gtrsim4$) in both our 1.1mm- and least-biased-870$\mu$m-selected samples, compared to the other
surveys. As discussed in detail by Chapman et al.\ (2005) and Wardlow
et al.\ (2011) this is likely due to the low-redshift bias of
statistical counterpart assignment methods.  Using statistical means
to overcome this bias Wardlow et al.\ (2011) estimate that  $\sim30\%$ (and at most $\sim45\%$)
of all SMGs in their sample are at redshifts $z\gtrsim3$. Our combined
AzTEC/JCMT/SMA and LABOCA/interferometric COSMOS data yield that
$\sim50\%$ of the COSMOS SMGs with interferometrically identified
counterparts are at $z\gtrsim3$. Exploring these two samples
separately, we find that $\sim50\%$ (i.e.\ 9/17) of the AzTEC/JCMT/SMA SMGs, and
$\sim40\%$ (i.e.\ 6/16) of the LABOCA/interferometric SMGs  have $z\gtrsim3$.

It is possible that the discrepancies between the $z\gtrsim3$ SMG
fractions in these different samples are due to their different
average flux densities.
Namely, past studies have suggested the
existence of a correlation between SMG brightness and redshift, in
such a way that the brightest SMGs lie at the highest redshift (e.g.
Ivison et al.\ 2002; Pope et al.\ 2005; Biggs et al.\ 2011). The LESS
survey source flux limit is $F_\mathrm{870\mu m}>4.4$~mJy (Biggs et
al.\ 2011). Assuming a power-law of 3 this translates into a limit of
2.2~mJy at 1.1~mm, and 1.3~mJy at 1.3~mm. The AzTEC/JCMT/SMA COSMOS
source flux limit is $F_\mathrm{1.1mm}>4.2$~mJy, thus about 2 times
higher, and the LABOCA/PdBI limit is $\gtrsim2.1$~mJy, thus a factor
of 1.6 higher compared to the LESS sample. Indeed, we find
 higher mean redshifts ($\bar{z}=3.1\pm0.4$ for the \mmsample , and $\bar{z}=2.6\pm0.4$ for the \submmsample ) and thus also
a higher fraction of high-redshift sources compared to the results
based on the LESS survey ($\bar{z}=2.5\pm0.3$). This is consistent with the
suggestions from past studies that {\em on average} brighter SMGs are
at higher redshifts. On the other hand, it has also been suggested that mm-selected samples lie on average at higher redshifts, compared to sub-mm-selected samples (e.g.\ Yun et al.\ 2012). Although our results are also consistent with this hypothesis, a more conclusive answer, disentangling these degeneracies, will have to await for deeper mm- and sub-mm selected samples with interferometric counterparts and accurately determined redshifts.

\begin{figure*}
\includegraphics[bb = 80 360 426 792,scale=0.45]{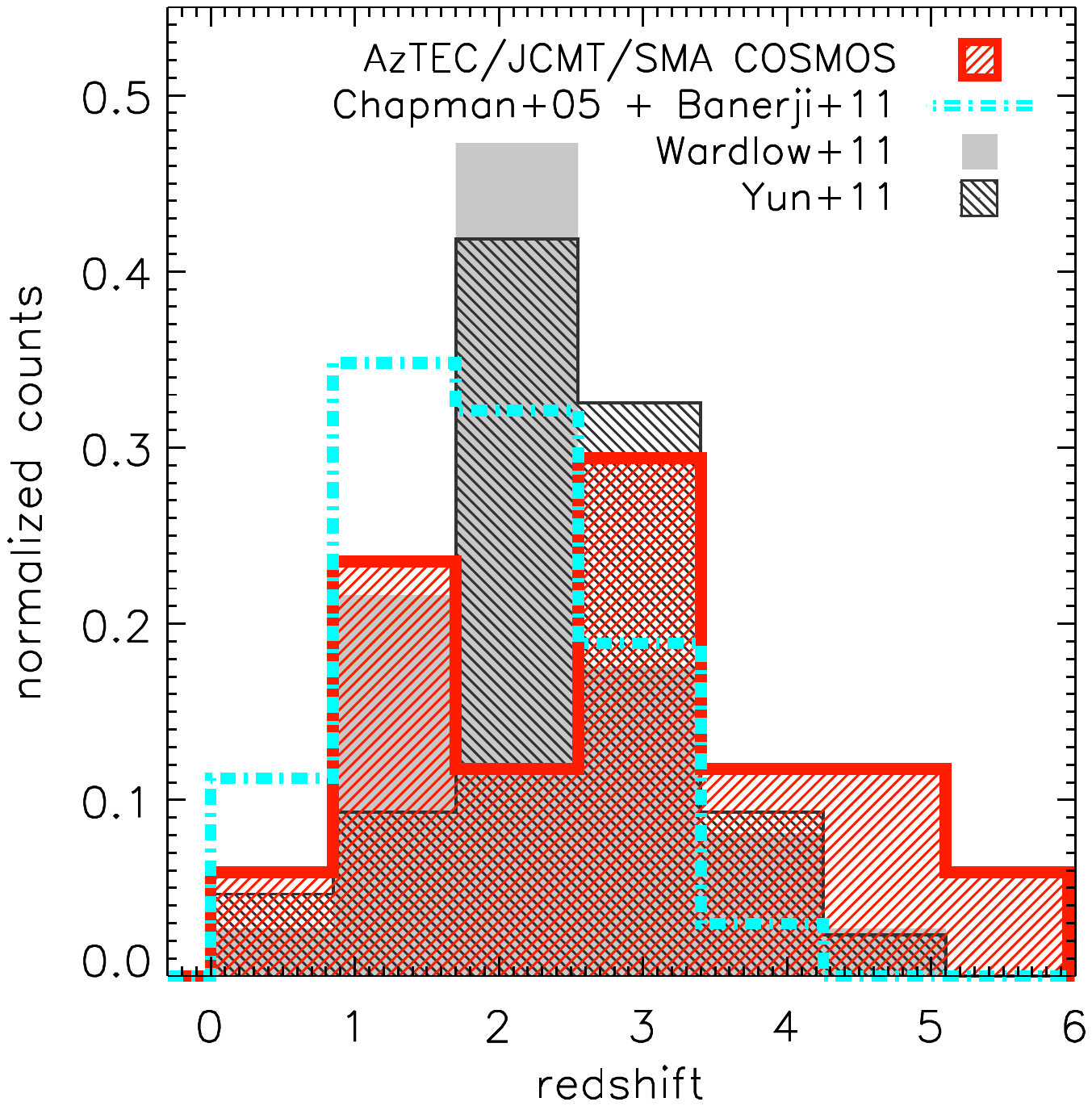}
\includegraphics[bb = 40 360 426 792,scale=0.45]{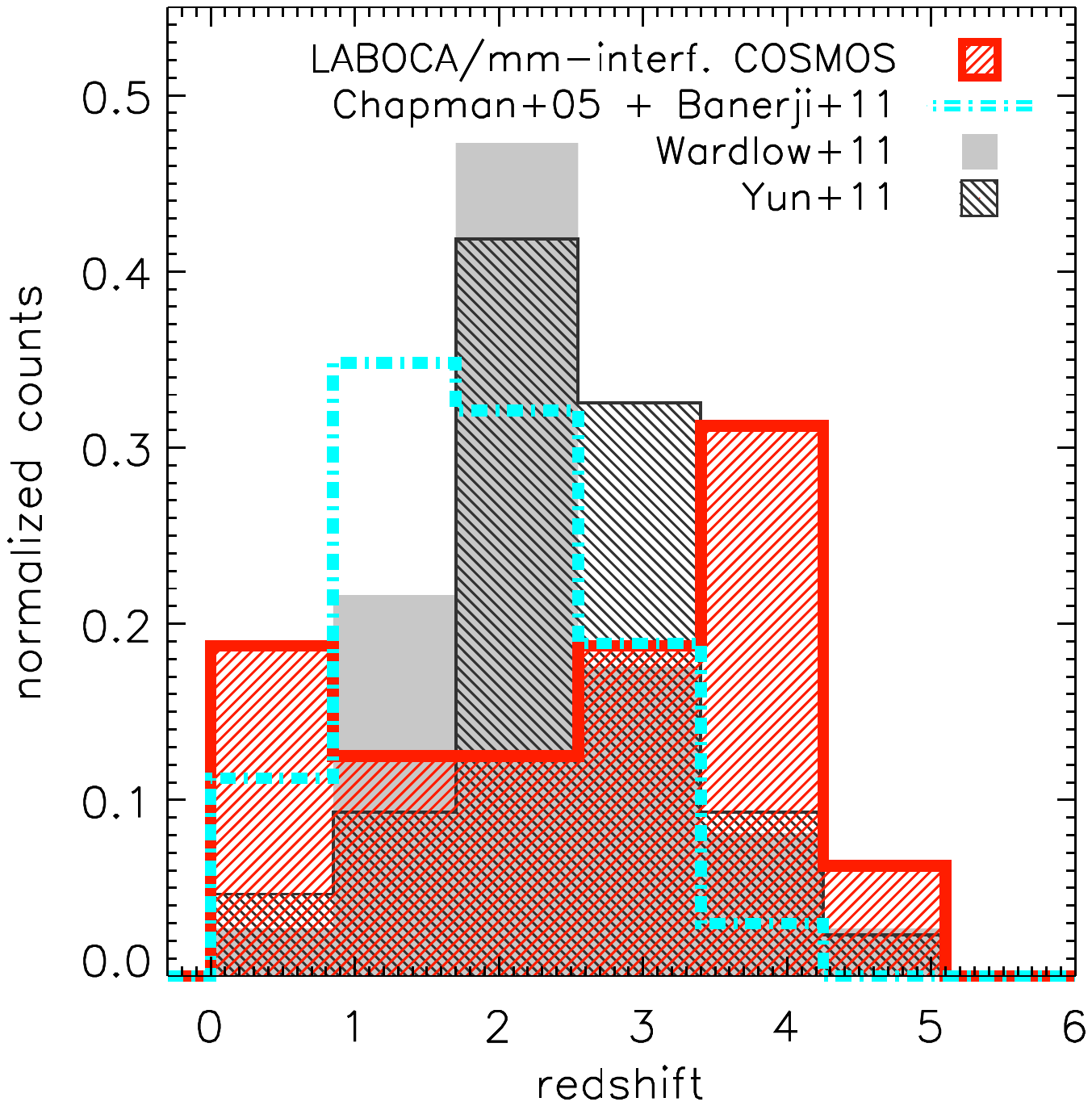}
\includegraphics[bb = 40 360 426 792,scale=0.45]{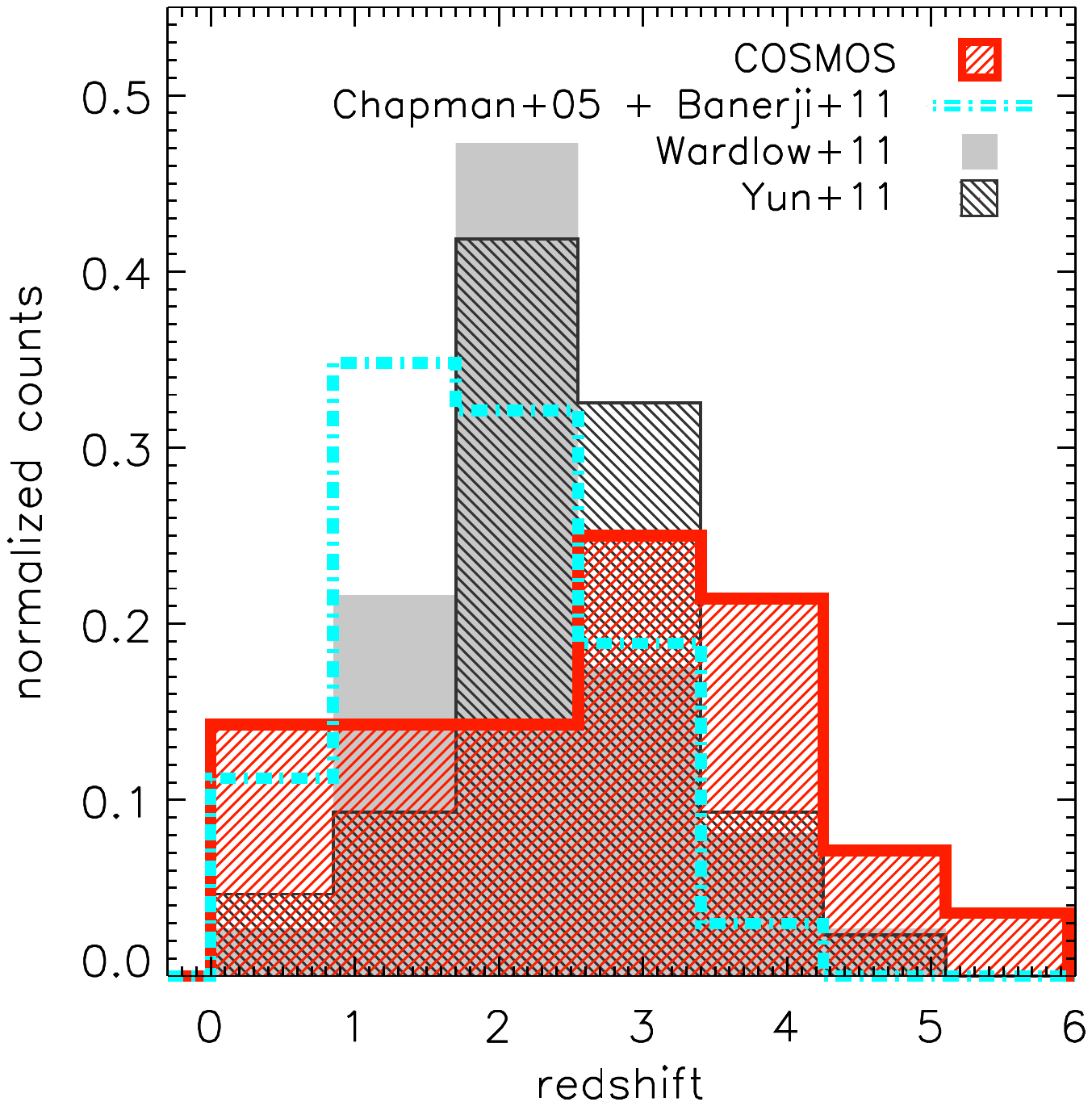}
\caption{ Normalized redshift distributions for SMGs drawn from various studies in the literature, indicated in the panel.  }
      \label{fig:photzdistribcompare}
\end{figure*}

\subsection{High redshift SMGs}
In our 1.1mm- , and 870$\mu$m-selected samples we find 9 (3 of which have radio counterparts) and 8 (4 of which have radio counterparts) $z\gtrsim3$ SMGs. We find 5-8\footnote{AzTEC-1, AzTEC3, AzTEC-4, AzTEC-5, J1000+0234, and possibly AzTEC-11S, AzTEC-13, and AzTEC-14E for which only lower redshift limits are available; see Table~\ref{tab:interf} and Table~\ref{tab:statsamples}.} SMGs at
$z\gtrsim4$ in our \mmsample , and and 3-4\footnote{COSLA-6-1, COSLA-23-N, COSLA-60, and possibly COSLA-158 with only a lower-redshift limit; see Table~\ref{tab:statsamples}.} SMGs at
$z\gtrsim4$ in our \submmsample .  This corresponds to $\sim30-50\%$ of the \mmsample , and $\sim20\%$ of the \submmsample . As our \submmsample , %selected from the 0.7\sqdeg\ LABOCA-COSMOS field, 
is not complete, we can infer only a
lower limit for the $z\gtrsim4$ SMG surface density of $\geq
3/0.7\approx4$~deg$^{-2}$. The \mmsample \ is however nearly complete at the
given 1.1~mm flux limit of $F_\mathrm{1.1mm}>4.2$~mJy, and drawn from
a uniform area of 0.15\sqdeg .  Four SMGs (AzTEC-1, 3, 4, and 5) in the \mmsample \ are found
to be at $z\gtrsim4$ (three of those have spectroscopic redshifts; see Table~\ref{tab:statsamples}).
J1000+0234, with a 1.1~mm flux of $4.8\pm1.5$~mJy (i.e.\ $\sim3\sigma$ and thus
not included in our \mmsample ), is also spectroscopically confirmed
to be at $z>4$ (Capak et al.\ 2008, Schinnerer et al.\ 2008). Furthermore, only lower redshift limits are available for 
AzTEC-11S, 13 and 14E. Thus, these three SMGs may possibly also lie at $z\gtrsim4$. 
Hence,
these 5-8 $z\gtrsim4$ SMGs with $F_\mathrm{1.1mm}>4.2$~mJy in the 0.15\sqdeg \ field yield a surface
density in the range of $\sim34\pm14$~deg$^{-1}$ to $\sim54\pm18$~deg$^{-1}$  (Poisson errors).
 Both values are substantially higher than what is expected in cosmological
models (Baugh et al.\ 2005; Swinbank et al.\ 2008; see also Coppin et
al.\ 2009, 2010), even if the AzTEC/JCMT-COSMOS field were affected by cosmic variance of to a factor of 3 overdensity, as suggested by Austermann et al.\ (2009). 

Based on the galaxy formation model of Baugh et al.\ (2005; top-heavy
IMF; $\Lambda$ cold dark matter cosmology) a surface density of
$\sim7$~deg$^{-1}$ for $z>4$ SMGs with 850~$\mu$m fluxes brighter than
5~mJy  is expected (see also Swinbank et al.\
2008; Coppin et al.\ 2009).  As the AzTEC/JCMT/SMA 4.2~mJy flux limit
at 1.1~mm translates to about a factor of two higher flux (i.e.\ 9.6~mJy) at 850~$\mu$Jy, the models would predict an even lower surface density at this flux threshold. 

In the GOODS-N field to date four $z>4$ SMGs were found  (Daddi et al.\ 2009a, b; Carilli et al.\ 2011). Given the $10\times16.5$~arcmin$^2$ area  (but with a highly non uniform rms in the SCUBA map with average $1\sigma= 3.4$~mJy; Pope et al.\ 2005) this implies a surface density of $\gtrsim87$~deg$^{-2}$, an order of magnitude higher than predicted by the models. The GOODS-N $z>4$ SMGs are however associated with a protocluster at $z\sim4.05$ which increases the surface density value. The COSMOS $z>4$ SMGs were selected from a larger field, and although an overall overdensity of bright  SMGs was found in the AzTEC/JCMT-COSMOS field (Austermann et al.\ 2009), the $z>4$ SMGs do not seem to be associated with each other (Capak et al.\ 2009, 2010; Schinnerer et al.\ 2009; Riechers et al.\ 2010; \smo\ et al.\ 2011). We still find significantly more $z>4$ SMGs than current models predict.
If the AzTEC/JCMT COSMOS SMGs are representative of the overall SMG population ($F_\mathrm{1.1mm}>4.2$~mJy), then our results imply that
current semi-analytic models underpredict the number of high-redshift
starbursts.

\section{Summary}
\label{sec:summary}

We presented PdBI continuum observations at 1.3~mm with $\sim1.5"$
angular resolution and an rms noise level of $\sim0.46$~mJy/beam towards 28
SMGs selected from the (single-dish) LABOCA-COSMOS survey of $27"$
angular resolution. Nine SMGs remain undetected, while the remainder
yields 9 highly significant ($\mathrm{S/N}>4.5$) and 17 tentative ($3<\mathrm{S/N}\leq4.5$ with multi-wavelength source association required) detections.
Combining these with other single-dish identified SMGs detected via intermediate
($\lesssim2"$) angular resolution mm-mapping in the COSMOS field
we present the largest sample of this kind to-date, containing 50 sources.
Based on 16 interferometrically confirmed SMGs with spectroscopic redshifts, 
we show that photometric redshifts derived from optical to MIR photometry 
are as accurate for SMGs as for other galaxy populations.
We derived photometric redshifts for those SMGs in our sample which lack
spectroscopic redshifts.

We distinguish two statistical samples within the total sample of 50 COSMOS
SMGs detected at $\lesssim2"$ angular resolution at mm-wavelengths:
i) a \mmsample , forming a significance-
($\mathrm{S/N_{1.1mm}}>4.5$) and flux- limited
($F_\mathrm{1.1mm}>4.2$~mJy) sample containing 17 SMGs with interferometric positions drawn from the
AzTEC/JCMT 0.15\sqdeg \ COSMOS survey, and ii) a \submmsample , containing 27 single-dish SMGs drawn
from the LABOCA 0.7\sqdeg \ COSMOS survey and detected with various
(CARMA, SMA, PdBI) mm-interferometers at intermediate angular
resolution. 

Within our samples we find that $\gtrsim15\%$, and up to $\sim40\%$ of single-dish identified SMGs
tend to separate into multiple components when observed at
intermediate angular resolution.

The common P-statistics counterpart identification correctly associates
counterparts to $\sim50\%$ of the parent single-dish SMG samples analyzed here.

We derive the redshift distribution of the SMGs with secure counterparts
identified via intermediate $\lesssim2"$ resolution mm-observations, and compare this
distribution to previous estimates that were based on {\em statistically} identified 
counterparts.
We find a broader redshift distribution
with a higher abundance of low- and high-redshift SMGs. The 
mean redshift is higher than in previous estimates.
This may add
evidence to previous claims that brighter and/or mm-selected SMGs are located at higher redshifts. 

We derive a surface density of
$z\gtrsim4$ SMGs ($F_\mathrm{1.1mm}\gtrsim4.2$~mJy) of $\sim34-54$~deg$^{-1}$, which is significantly higher than
what has been predicted by current galaxy formation models. 

\section*{Acknowledgments}

We thank the anonymous referee for helpful comments on the manuscript.
Based on data products from observations made with ESO Telescopes at the La Silla Paranal Observatory under ESO programme ID 179.A-2005 and on data products produced by TERAPIX and the Cambridge Astronomy Survey Unit on behalf of the UltraVISTA consortium.
The research leading to these results has received funding from the European Community's Seventh Framework Programme (/FP7/2007-2013/) under grant agreement No 229517 and through the DFG-SFB 956 and the DFG Priority Program 1573.  The National Radio Astronomy Observatory is a facility of the National Science Foundation operated under cooperative agreement by Associated Universities, Inc.

{}

\appendix

\section{Notes on individual LABOCA-COSMOS targets observed with the PdBI}
\label{sec:individuals}

\begin{figure*}
\hspace{-0.3mm}\includegraphics[scale=0.65]{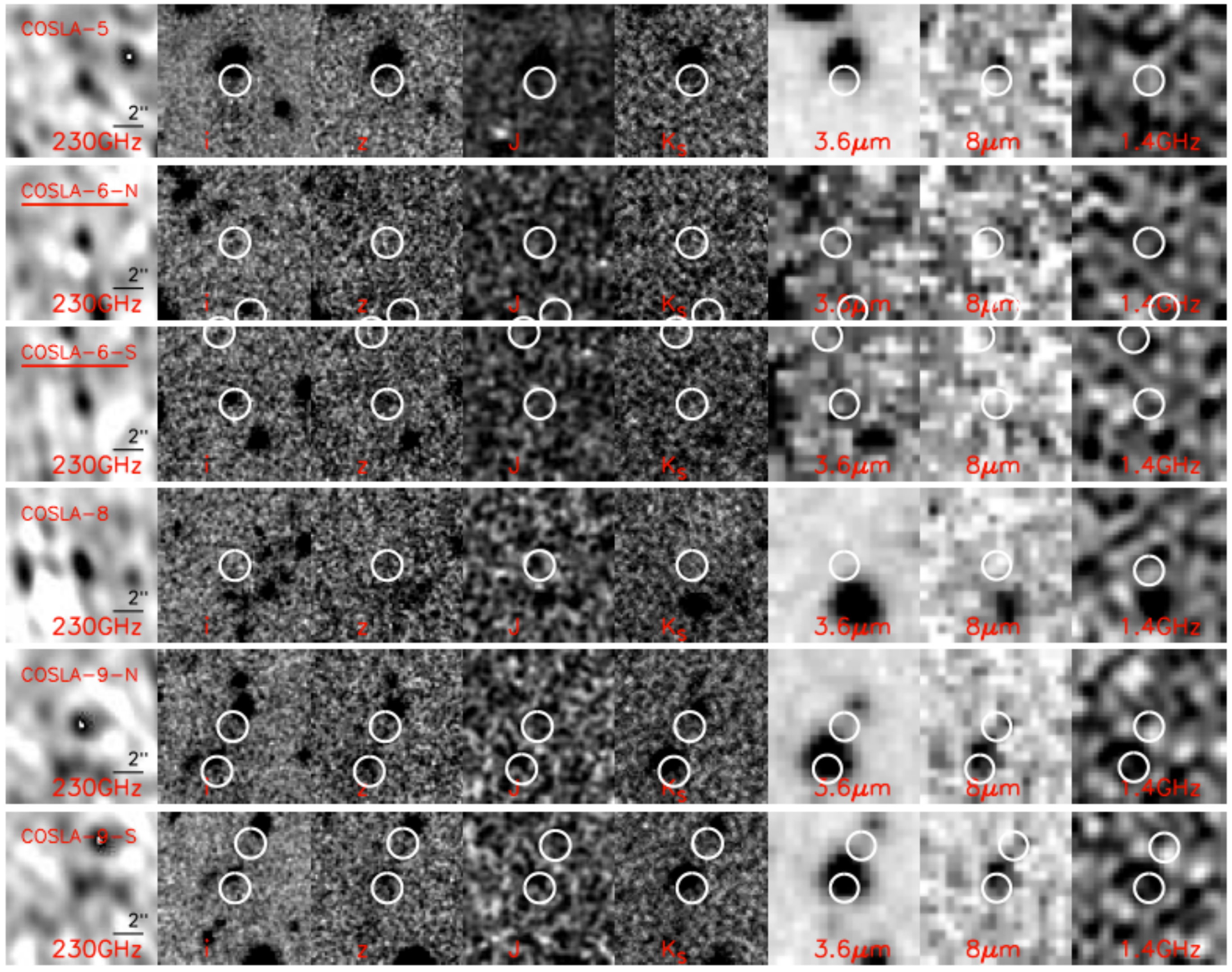}
\vspace{-0.5mm}\includegraphics[scale=0.65]{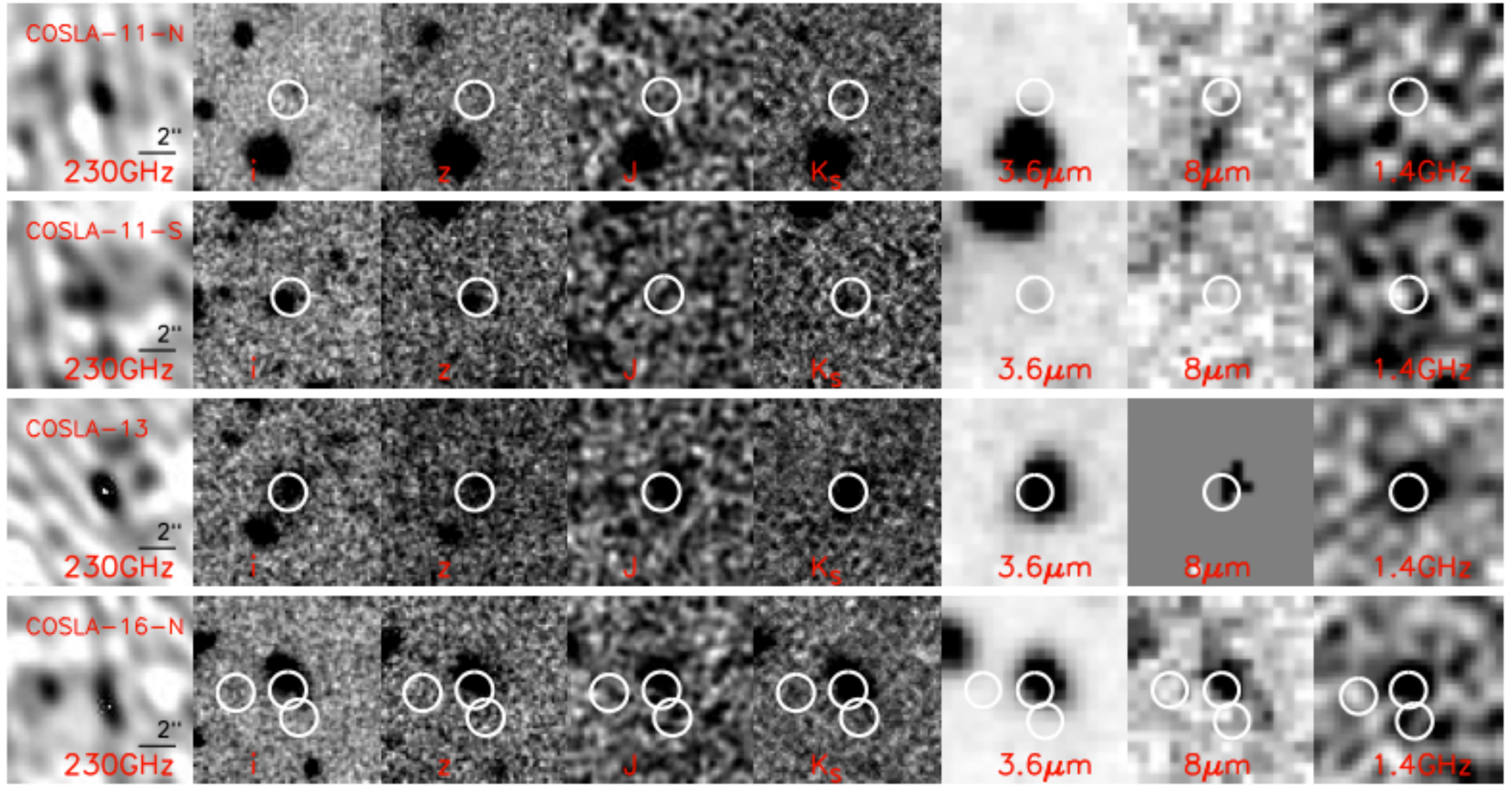}
\caption{ Optical to radio stamps for the LABOCA-COSMOS SMGs detected by PdBI. Names of sources with $\mathrm{S/N}>4.5$ in PdBI maps are underlined.  \label{fig:pdbistamps1}}
\end{figure*}

\begin{figure*}
\hspace{-0.3mm}\includegraphics[scale=0.65]{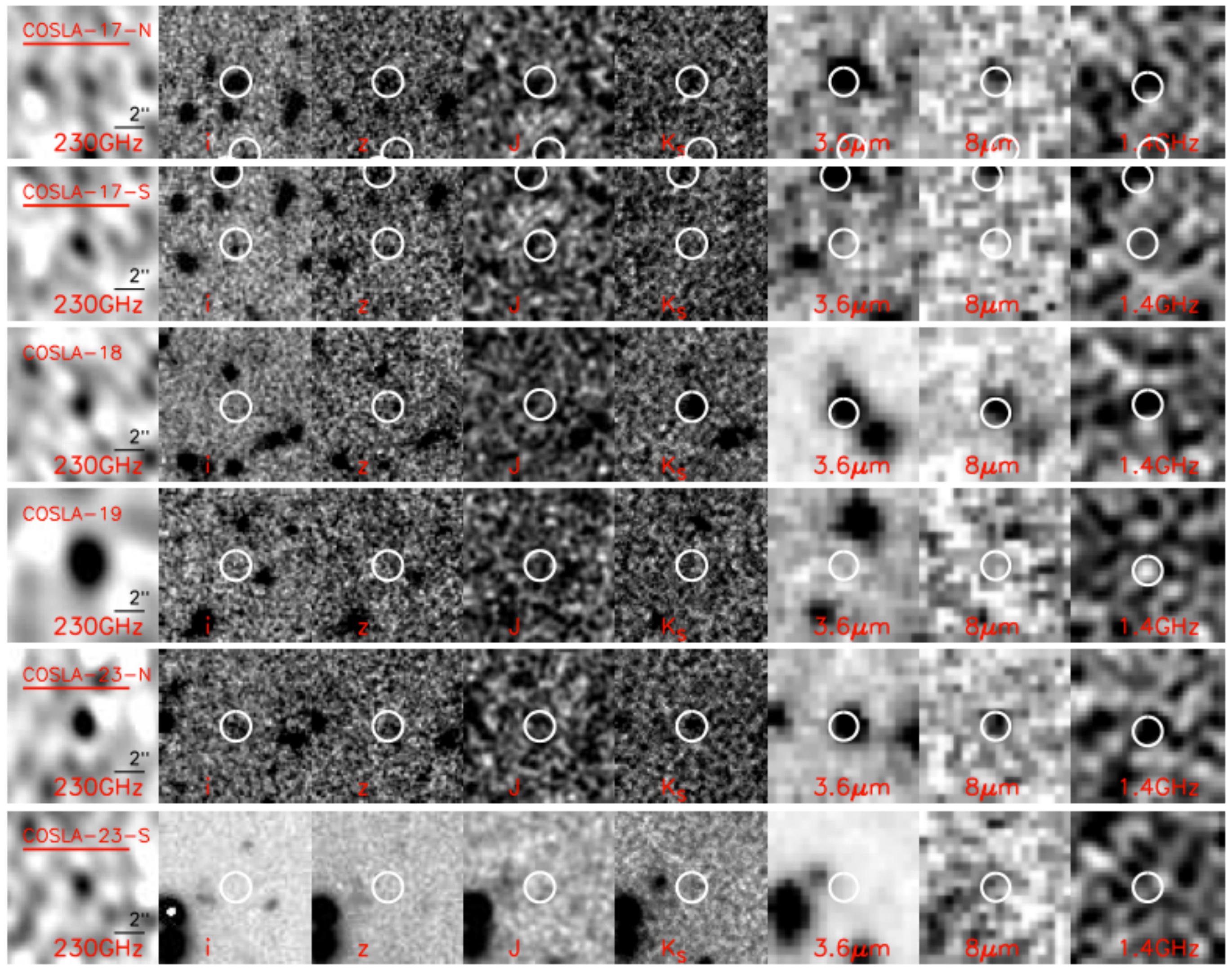}
\vspace{-1.5mm}\includegraphics[scale=0.65]{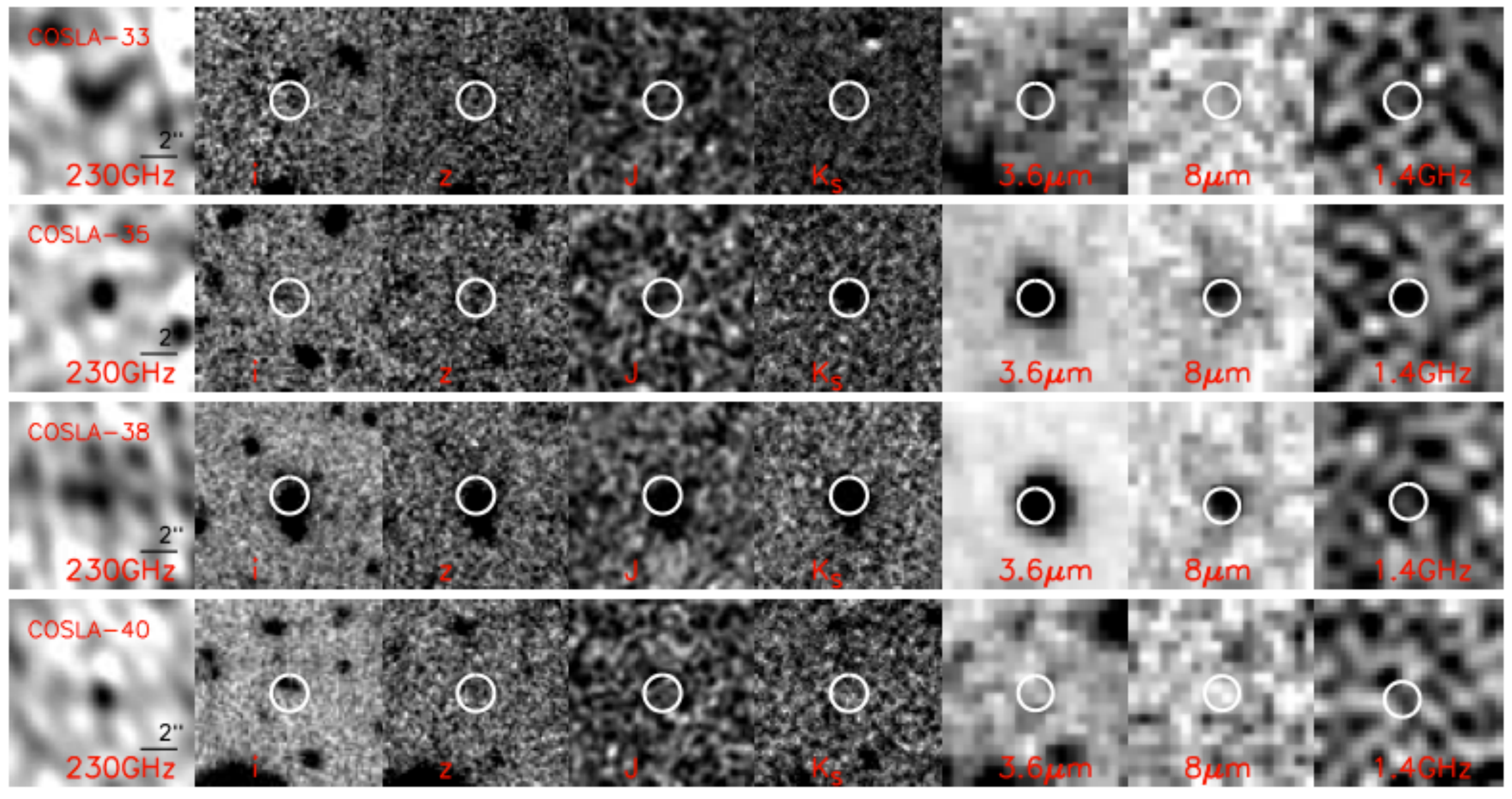}
\caption{ Continuation of \f{fig:pdbistamps1} \label{fig:pdbistamps2}}
\end{figure*}

\begin{figure*}
\vspace{1mm}
\vspace{-1.5mm}\includegraphics[scale=0.65]{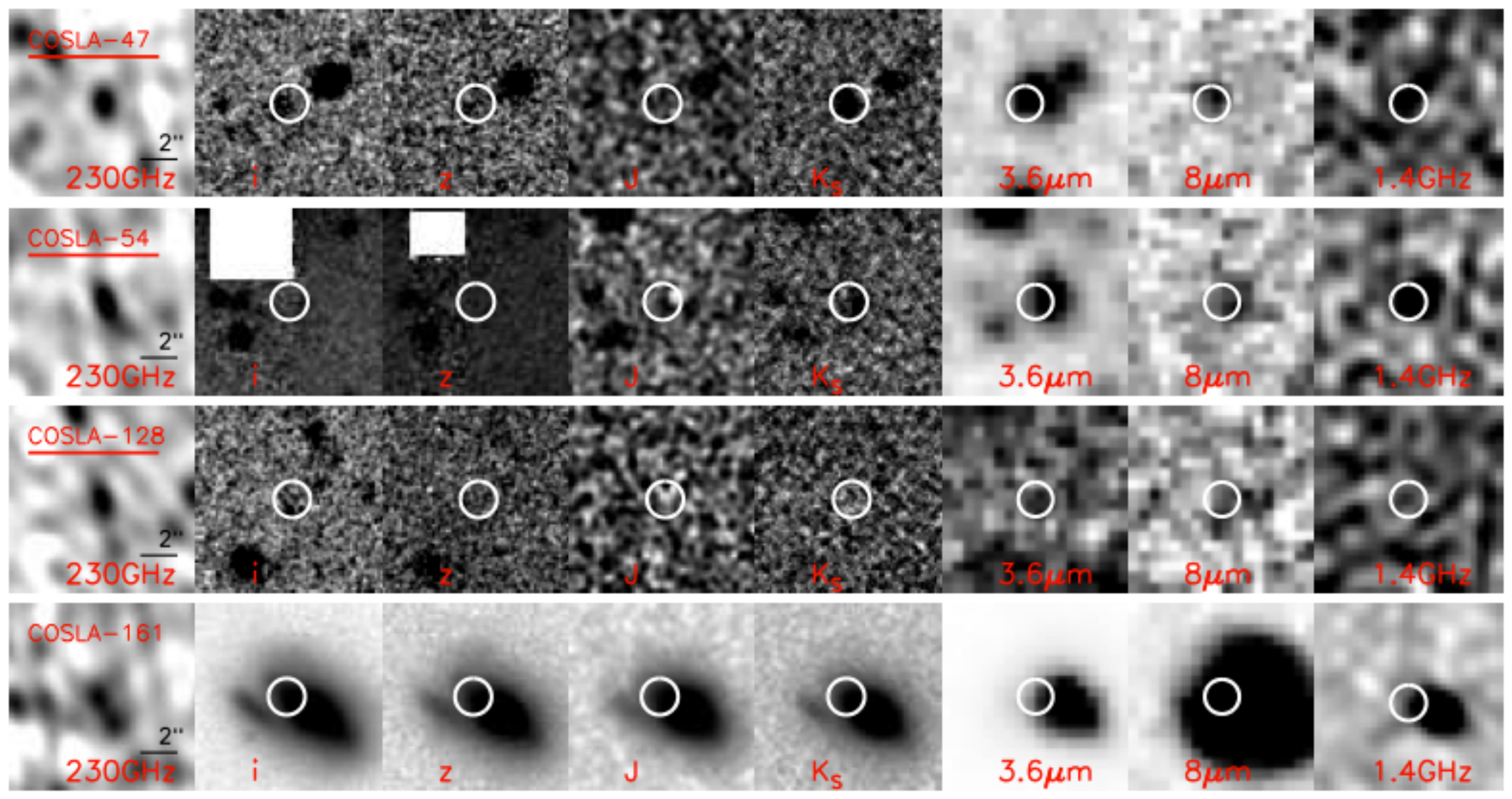}
\caption{ Continuation of \f{fig:pdbistamps1}  \label{fig:pdbistamps3}}
\end{figure*}

%PdBi maps

\begin{figure*}\vspace{-4mm}
\includegraphics[scale=0.65]{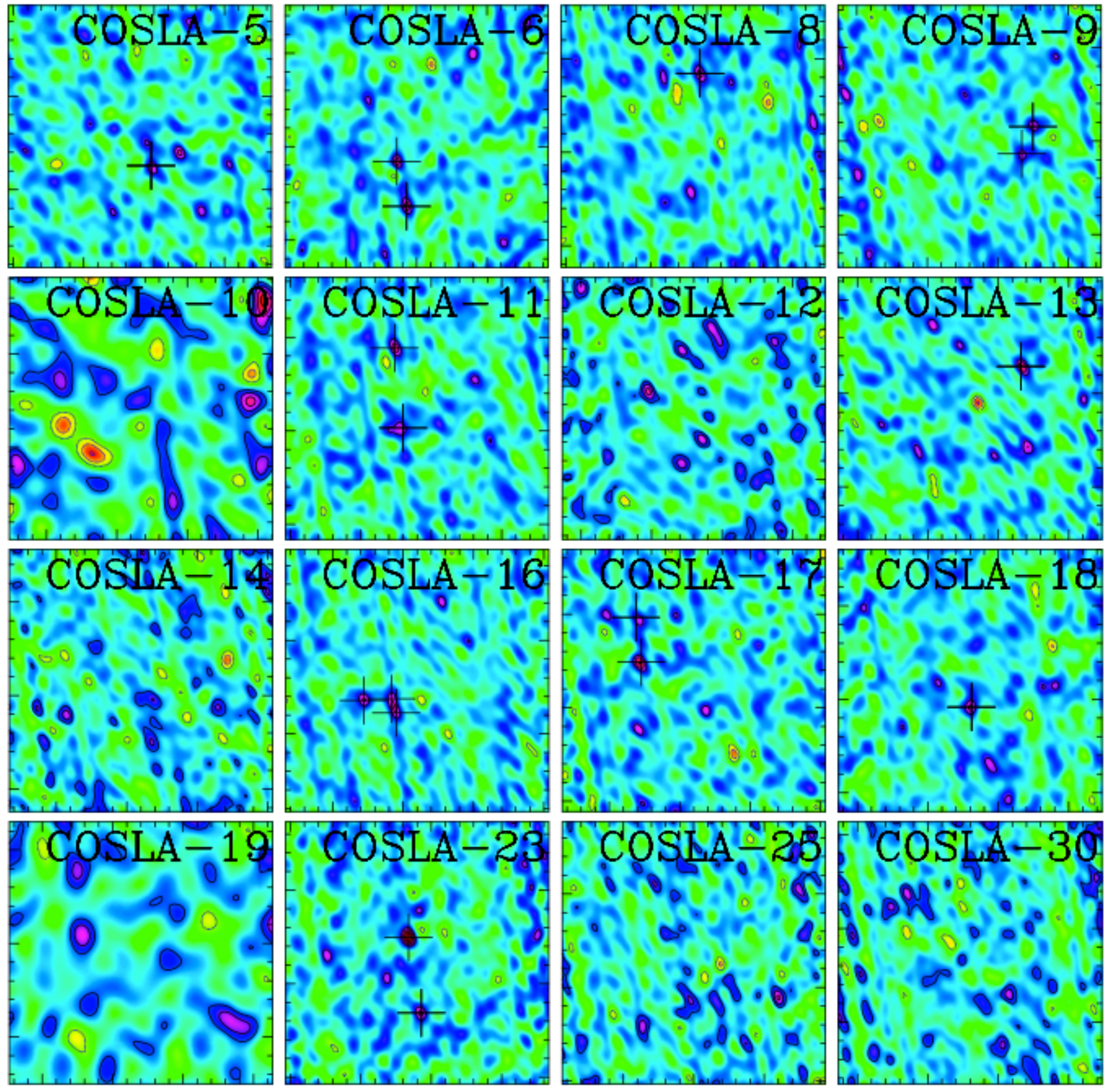}
\includegraphics[scale=0.65]{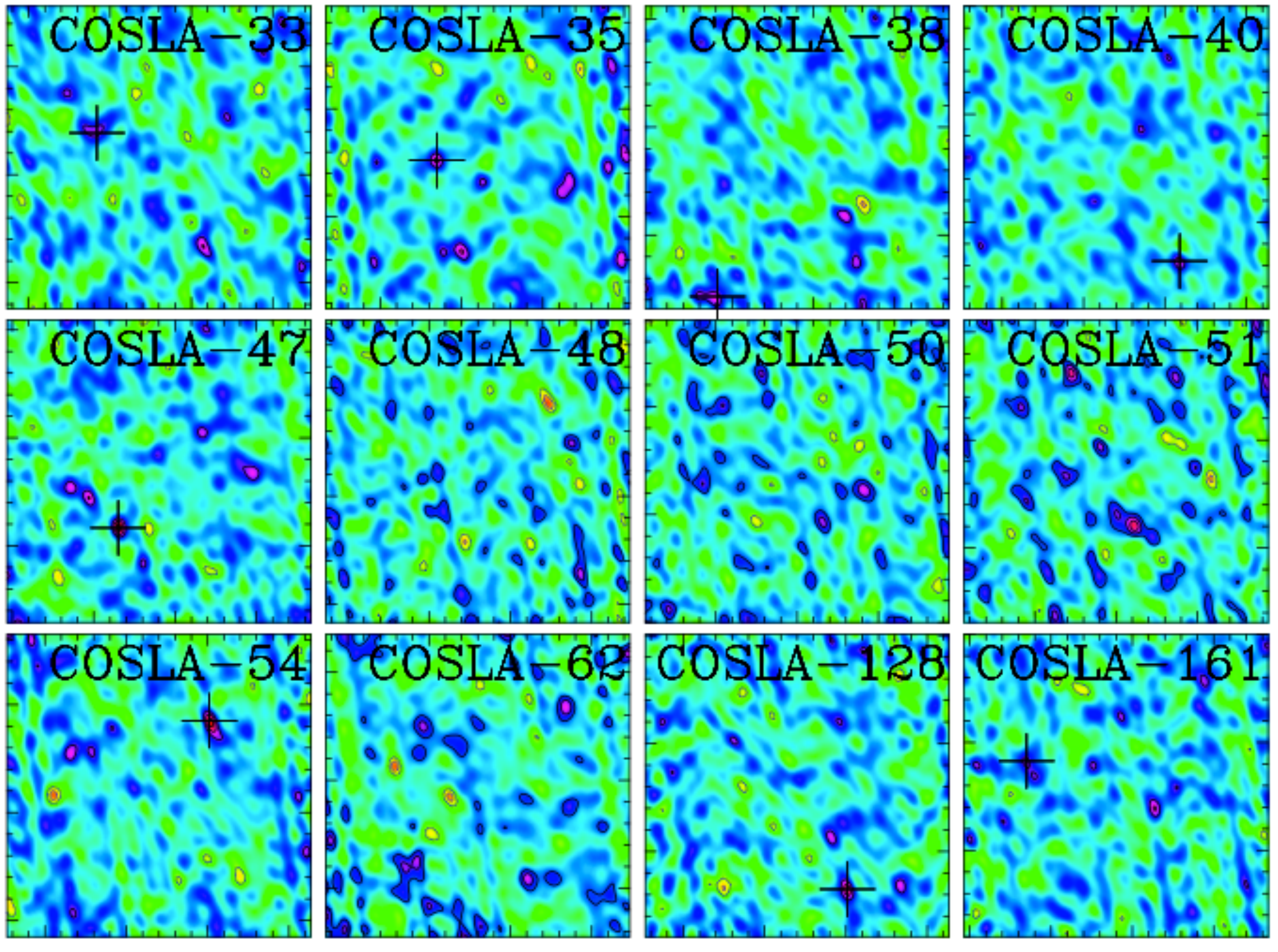}
\caption{ Cleaned PdBI maps  (color scale), $30"$ on the side, with $\pm2\sigma,3\sigma,..$ contours overlaid. Detections (identified in the {\em dirty} maps, see text for details) are marked by crosses. \label{fig:pdbistamps}}
\end{figure*}

Here we present detailed notes on individual LABOCA-COSMOS (i.e.\ COSLA) SMGs observed with PdBI at 1.3~mm and $\sim1.5"$ resolution. We use a spectral index of 3, i.e.\ assuming $S_\nu\propto\nu^{2+\beta}$, where $S_\nu$ is the flux density at frequency $\nu$ and $\beta=1$ the dust emissivity index, to convert fluxes from/to various (sub-)mm wavelengths (if not stated otherwise). 

\vspace{3mm}
\noindent\underline{\bf \large COSLA-5}
\vspace{1mm}

COSLA-5 is a $\mathrm{S/N}=4.1$ detection located at $\alpha=$~10  00  59.521, $\delta=$~+02  17  02.57.  The PdBI-source is found at a separation of $3.4"$ from the LABOCA source center. The 1.3~mm flux density of the source is $2.04\pm0.49$~mJy. Scaling this to the MAMBO (1.2~mm) and LABOCA (870~$\mu$m) wavelengths we find $2.6\pm0.6$ and $6.9\pm1.7$~mJy, respectively. This is slightly lower than the extracted (and deboosted) MAMBO and LABOCA fluxes ($4.78\pm1$, and $12.5\pm2.6$~mJy) and may indicate the presence of another mm-source, not detected in our PdBI map. Based on the rms reached in the PdBI observations, we can put a 3$\sigma$ upper limit to this  potential second source of 1.4~mJy.

The PdBI peak is $1.3"$ away from a source that is independently detected in the optical ($i^+=22.5$), UltraVista, and IRAC bands. The photometric redshift of this source is well constrained,  $z_\mathrm{phot}=0.85^{+0.07}_{-0.06}$. Interestingly, at a separation of $1.1"$ towards the SW of the PdBI detection we find a faint source present only in the Ks band images, but not included in the catalogs (it is present in both the WIRCam and UltraVista images). 

The mm-source is not associated with a radio detection suggesting a mm-to-radio flux based redshift  for COSLA-5 of $z_\mathrm{mm/radio}\gtrsim3.8$. Here we take the optical/IRAC/UltraVista source as the counterpart noting that given the $\sim4\sigma$ significance of the PdBI peak, and a separation of $\gtrsim1"$ from multi-wavelength sources, further follow-up is required to confirm this source and its redshift.

\vspace{3mm}
\noindent\underline{\bf \large COSLA-6}
\vspace{1mm}

Two significant ($\mathrm{S/N}>4.5$) sources are detected within the COSLA-6 PdBI map.  

COSLA-6-N (S/N=5.4, $\alpha=$10  01  23.64,  $\delta=$+02  26  08.42) has a  230~GHz (1.3~mm) flux density of $2.7\pm0.5$~mJy. No IRAC/UltraVista source is found nearby. The closest source is a faint optical (no UltraVista/IRAC/radio) source $2.0''$ away. Given the high significance of the mm-detection the mm-positional accuracy is $\sim0.3"$. Thus, it seems unlikely that this optical source is the counterpart of the mm-detection.  COSLA-6-N is however coincident with a $2.1\sigma$ peak in the radio map ($F_\mathrm{1.4GHz}=19.5\pm9.4~\mu$Jy). Based on this, we infer a mm-to-radio flux based redshift of $z_\mathrm{mm-radio}=4.01^{+1.51}_{-0.83}$.

COSLA-6-S ($\mathrm{S/N}=4.75$, $\alpha=$10  01  23.57,  $\delta=$+02  26  03.62) has a 1.3~mm flux density of $3.1\pm0.6$~mJy. It might be associated with a source detected in the optical (separation$=0.5"$; $i^+=26.15$), but not in near- or mid-IR. It is coincident with a $3.3\sigma$ peak in the radio map ($F_\mathrm{1.4GHz}=33.3\pm10.1~\mu$Jy). Based on the multi-wavelength photometry, we infer a photometric redshift of $z_\mathrm{phot}=0.48^{+ 0.19}_{-0.22}$ for this source. A second potential redshift solution (although not as likely as the first one) exists at $z\sim4$, and it is supported by the mm-to-radio flux based redshift, $z_\mathrm{mm/radio}=3.44^{+ 0.83}_{-0.58}$ . 

The combined 1.3~mm fluxes of COSLA-6-N and COSLA-6-S, scaled using a spectral index of 3, yield an expected flux density of $19.4\pm2.7$~mJy at 870~$\mu$m. This is in very good agreement with the deboosted LABOCA 870~$\mu$m flux of  $16.0 \pm 3.4$. 

\vspace{3mm}
\noindent\underline{\bf \large COSLA-8}
\vspace{1mm}

COSLA-8 is detected at $\mathrm{S/N}=4.2$, and
located at $\alpha=$~10~00~25.55, $\delta=$~+02~15~08.44. Its 1.3~mm
flux density is $F_\mathrm{1.3mm}=2.65\pm0.62$~mJy. Using a spectral
index of 3 this extrapolates to an 870~$\mu$m flux density of
$8.9\pm2.1$~mJy, in good agreement with the LABOCA deboosted flux
of $6.9\pm1.6$~mJy. The source is located in a crowded region. A
radio, IRAC, UltraVista detection is present at a separation of
$\sim3''$, however the closest source to the mm-source (separation=1.0")
is detected only in the optical ($i^+=27.4$). A $3.3\sigma$ peak is found at the mm-position in the radio map ($F_\mathrm{1.4GHz}=26.2\pm8.0~\mu$Jy). The most probable photometric redshift for this source is $z\mathrm{phot}=1.83^{+ 0.4}_{-1.31}$, however with a rather flat $\chi^2$ distribution as reflected in the uncertainties. 

\vspace{3mm}
\noindent\underline{\bf \large COSLA-9}
\vspace{1mm}

We identify two $3.2\sigma$ peaks at $\alpha=$~10~00~13.83, $\delta=$~+01~56~38.64 (COSLA-9-S; $5.8"$ away from the LABOCA source center), and $\alpha=$~10~00~13.75, $\delta=$~+01~56~41.54 (COSLA-9-N; $7"$ away from the LABOCA source center). COSLA-9-S can be matched to an optical/IRAC/UltraVista source ($i^+=24.8$, separation=0.8"), while the closest source to COSLA-9-N is detected only in the optical ($i+=26.1$; separation=0.4") however it is only $1.3"$ away from an optical/UltraVista/IRAC source. 

The primary beam corrected 1.3~mm flux densities of COSLA-9-N and COSLA-9-S are $1.69\pm0.47$~mJy, and $1.87\pm0.58$~mJy, respectively. Added together, and scaled to the LABOCA ($13.2\pm2.1$~mJy) and AzTEC ($6.3\pm1.0$~mJy) frequencies yields a good match to the deboosted LABOCA ($14.4\pm3.3$~mJy) and AzTEC ($8.7\pm1.1$~mJy) fluxes. Given the low significance of the sources, further follow-up is required to confirm their reality.

\vspace{3mm}
\noindent\underline{\bf \large COSLA-10}
\vspace{1mm}

No significant source is present in the 1.3~mm map. The statistical counterpart association (see \s{sec:pstat} \ and \f{fig:statstampscoslanodet} \ for details) suggests three separate potential (tentative) counterparts to this LABOCA source. The sum of the extracted 1.3~mm fluxes (taken as maximum flux within a circular area of $1"$ in radius centered at the statistical counterpart), corrected for the primary beam response, is 2.43~mJy. This yields a flux of 8.1~mJy when scaled to 870~$\mu$m, in very good agreement with the LABOCA flux of $7.3\pm1.7$~mJy (see \f{fig:flux} ). This suggests that the LABOCA source may be fainter at 1.3~mm than can be detected given our PdBI sensitivity and that it possibly breaks up into multiple components when observed at $1.5"$ resolution.

\vspace{3mm}
\noindent\underline{\bf \large COSLA-11}
\vspace{1mm}

The brightest peaks in the PdBI map are at $3.5\sigma$ (COSLA-11-N) and $3\sigma$ (COSLA-11-S). 

COSLA-11-N is located at $\alpha=$~10  01  14.260, $\delta=$~01  48  18.86, and  it can be associated with a faint optical detection $0.64"$ away ($i^+=27.75$). Its 1.3~mm flux density is $F_\mathrm{1.3mm}= 2.15\pm0.62$~mJy, and the source is not detected in the radio map. Based on the multi-wavelength photometry of the counterpart of COSLA-11-N, we find a photometric redshift of $z_\mathrm{phot}=0.75^{+0.23}_{-0.25}$. The mm-to-radio based redshift however suggests $z_\mathrm{mm/radio}\gtrsim3.6$.

COSLA-11-S ($\alpha=$~10 01 14.200, $\delta=$~ +01 48 10.31)  is only $2.6"$ away from the center of the LABOCA source, and it coincides (separation=$0.5"$) with independent optical and UltraVista  H-band, and UltraVista J-band detections.  Although fairly low S/N, the UltraVista detection increases the probability that it is a real source.  Its 1.3~mm flux density is $F_\mathrm{1.3mm}= 1.43 \pm0.48$~mJy.  Our photometric redshift estimate yields two, almost equally probable, redshifts at $z\sim0.2$ and $z\sim3$. Given that the SMG is not detected in the radio map,  the higher redshift solution, also consistent with the mm-to-radio flux ratio based redshift ($z_\mathrm{mm/radio}\gtrsim3$), is more likely. We thus adopt the higher redshift solution for this source yielding $z_\mathrm{photo}=3.00^{+0.14}_{-0.07}$ (where the errors reflect the 99\% confidence interval derived using only $z>1.5$ $\chi^2$ values).

The combined flux densities of the two detections, scaled to 870~$\mu$m, yield a flux density of $12.0 \pm 2.6$~mJy at this wavelength. This is in good agreement with the LABOCA 870~$\mu$m flux density of $19.4\pm4.5$~mJy, and thus further affirms the reality of the sources.

\vspace{3mm}
\noindent\underline{\bf \large COSLA-12}
\vspace{1mm}

No significant source is present in the 1.3~mm map. The statistical counterpart association (see \s{sec:pstat} \ and \f{fig:statstampscoslanodet} \ for details) suggests two separate potential (one robust and one tentative) counterparts to this LABOCA source. The sum of the extracted 1.3~mm fluxes (taken as the maximum flux within a circular area of $1"$ in radius centered at the statistical counterpart), corrected for the primary beam response, is 2.68~mJy. Scaled to 870~$\mu$m this implies a flux of 8.9~mJy which is lower than the  LABOCA flux of $18.3\pm4.2$~mJy  (see \f{fig:flux} ). This suggests that the LABOCA source may be fainter at 1.3~mm than can be detected given our PdBI sensitivity and/or it breaks up into multiple components when observed at $1.5"$ resolution.

\vspace{3mm}
\noindent\underline{\bf \large COSLA-13}
\vspace{1mm}

COSLA-13 is detected at $\mathrm{S/N}=3.9$ ($\alpha=$~10 00   31.840, $\delta=$~+02  12  42.81).  Its 1.3~mm flux density is $F_\mathrm{1.3mm}= 1.37\pm0.61$~mJy. This SMG is detected within the LABOCA (870~$\mu$m), AzTEC (1.1~mm), and MAMBO (1.2~mm) surveys. Scaling the PdBI 1.3~mm flux to 870~$\mu$m, 1.1~mm, 1.2~mm (using a slope of 4.6 which corresponds to the mean slope between the AzTEC/LABOCA and AzTEC/MAMBO detected fluxes) we find flux densities of 8.8 $\pm$ 3.9~mJy (870~$\mu$m), 2.9 $\pm$     1.3~mJy  (1.1~mm), and 2.0 $\pm$    0.9~mJy (1.2~mm). These are consistent with the AzTEC/LABOCA fluxes, and slightly lower than the MAMBO flux (note that we find consistent results when using a slope with a spectral index of 3).

The PdBI mm-source is coincident with an optical/UltraVista/IRAC/radio source (separation=$0.55"$) with an optical spectrum at redshift $z_\mathrm{spec}=2.175$.

\vspace{3mm}
\noindent\underline{\bf \large COSLA-14}
\vspace{1mm}

No significant source is present in the 1.3~mm map. The statistical counterpart association (see \s{sec:pstat} \ and \f{fig:statstampscoslanodet} \ for details) suggests one robust statistical counterpart   to this LABOCA source. The extracted 1.3~mm flux (taken as maximum flux within a circular area of $1"$ in radius centered at the statistical counterpart), corrected for the primary beam response, is 1.64~mJy.  The scaled 870~$\mu$m flux of 5.5~mJy is  fairly consistent with the LABOCA flux of $9.0\pm2.1$~mJy when scaled to 870~$\mu$m (see \f{fig:flux} ). This suggests that the LABOCA source may be fainter at 1.3~mm than can be detected given our PdBI sensitivity.

\vspace{3mm}
\noindent\underline{\bf \large COSLA-16}
\vspace{1mm}

A significant extended source is found $\sim3.5"$ away from the LABOCA source center. It is best fit by a double-Gaussian (using the AIPS task jmfit and fixing the width of the Gaussians), yielding two sources located at $\alpha=$~10  00  51.5854, $\delta=$~+02  33  33.5648 (COSLA-16-N) and $\alpha=$~10  00  51.5541, $\delta=$~+02  33  32.0948 (COSLA-16-S). The 2-Gaussian fit yields 1.3 mm flux densities of $F_\mathrm{1.3~mm}=1.39\pm0.32$~mJy (COSLA-16-N) and $F_\mathrm{1.3~mm}=1.19\pm0.33$~mJy (COSLA-16-S).

COSLA-16-N can be associated  with an optical/UltraVista/IRAC/radio source (separation $=0.79"$). 
COSLA-16-S is not associated with a separate source in the multi-wavelength catalogs. The radio emission associated with the position of COSLA-16-N is significant ($F_\mathrm{1.4GHz}=95.6\pm10.1~\mu$Jy), while a $3.3\sigma$ peak, that seems to be the extension of the significant radio source, is associated with COSLA-16-S ($F_\mathrm{1.4GHz}=33.3\pm10.1~\mu$Jy). The multi-wavelength photometry of COSLA-16-N implies a photometric redshift of $z_\mathrm{phot}=2.16^{+0.12}_{-0.25}$. The mm-to-radio based redshift inferred for COSLA-16-S is $z_\mathrm{mm/radio}=2.40^{0.62}_{-0.51}$, suggesting it is associated with COSLA-16-N.

A third $3.9\sigma$ peak (COSLA-16-E) with 1.3~mm flux density of $F_\mathrm{1.3~mm}=2.26\pm0.58$~mJy is found $6"$ east of the LABOCA source center and it is coincident with a faint optical source (separation $=0.41"$, $i^+=29.20$). Our photometric redshift computation yields s redshift of $z_\mathrm{phot}=1.25^{+3.03}_{-1.15}$, however (as also reflected in the error) the $\chi^2$ distribution is fairly flat below $z\lesssim4$ thus making all redshifts below $z\sim4$ almost equally probable. The mm-to-radio flux based redsfhit suggests $z_\mathrm{mm/radio}\geq 3.7$.

The combined 1.3~mm flux density of the 3 identified sources, scaled to 870~$\mu$m yields $16.3\pm 2.5$~mJy, in very good agreement with the deboosted LABOCA flux ($14.0\pm3.6$~mJy).

\vspace{3mm}
\noindent\underline{\bf \large COSLA-17}
\vspace{1mm}

Two significant $\mathrm{S/N}>4.5$ detections are found within the PdBI map. 

COSLA-17-S is detected at high significance ($\mathrm{S/N}=5.3$; $\alpha=$~10  01  36.772, $\delta=$~+02  11  04.87). Its 1.3~mm flux density is $F_\mathrm{1.3~mm}= 3.0\pm0.6$~mJy, and it can be associated with a faint source $0.23"$ away with $m_\mathrm{NB816}=26.2$.
No IR or radio source is associated with this detection. We find a photometric redshift for this source of $z_\mathrm{phot}=0.7^{+0.21}_{-0.22}$, while the mm-to-radio based redshift suggests $z_\mathrm{mm/radio}\gtrsim4$. 

COSLA-17-N is found at S/N=4.6 ($\alpha=$~10  01  36.811, $\delta=$~+02  11  09.66) with a 1.3~mm flux density of $F_\mathrm{1.3mm}=3.55\pm    0.677$~mJy. It is perfectly coincident with an optical/UltraVista/IRAC/radio source (separation=0.09"), however it is within the sidelobe region of the brighter COSLA-17-S source. Hence further follow-up with more complete uv-coverage is required to affirm this source. We find a photometric redshift of $z_\mathrm{phot}=3.37^{+0.14}_{-0.22}$, consistent with the mm-to-radio based redshift of $z_\mathrm{mm/radio}=3.27^{+0.60}_{-0.49}$.

This SMG is detected by both LABOCA and AzTEC/ASTE surveys, and the flux ratio using these two surveys suggests a spectral index of 2.08. Using this value to scale the combined 1.3~mm PdBI fluxes to 870~$\mu$m (1.1~mm) we find a flux density of $15.2\pm2.2$~mJy ($9.2\pm1.4$~mJy), consistent with the deboosted LABOCA (AzTEC/ASTE) flux of $12.5\pm 3.2$~mJy ($7.5^{+1.0}_{ -1.1}$~mJy).

 \vspace{3mm}
\noindent\underline{\bf \large COSLA-18}
\vspace{1mm}

COSLA-18 is detected at  $\mathrm{S/N}=4.5$ ($\alpha=$~10  00  43.19, $\delta=$~+02  05  19.17). Its 1.3~mm flux density is $F_\mathrm{1.3mm}=2.15 \pm   0.48$~mJy. Scaled to 870~$\mu$m (1.1~mm) this yields a flux density of $7.2\pm1.6$~mJy ($3.5\pm0.8$~mJy), consistent with the deboosted LABOCA (AzTEC/ASTE) fluxes of $10.0\pm 2.6$~mJy ($3.8^{+1.1}_{ -1.2}$~mJy). The source is coincident with an optical/UltraVista/IRAC/radio source (separation $=0.67"$; $i^+=28.96$). Our photometric redshift computation yields 2 almost equally probable photometric redshifts at $z\sim3$ and $z\sim5$. The significant radio detection ($F_\mathrm{1.4GHz}=60.1\pm8.9~\mu$Jy) would argue in favor of the lower redshift solution, consistent with the mm-to-radio flux based redshift of $z_\mathrm{mm/radio}=2.40^{+0.35}_{-0.34}$. Thus, here we adopt the low-redshift solution for this source, yielding $z_\mathrm{phot}=2.90^{+0.31}_{-0.43}$, noting that a second solution of $z_\mathrm{phot}=4.92^{+0.38}_{-0.34}
$ is possible.

 \vspace{3mm}
\noindent\underline{\bf \large COSLA-19}
\vspace{1mm}

Cosla-19 is detected at $\mathrm{S/N}=4.1$ with a 1.3~mm flux density of $3.17\pm0.76$~mJy. Scaling this flux to 1.2~mm, 1.1~mm, and 870~$\mu$Jy yields fluxes of $4.1\pm1.0$~mJy, $5.1\pm1.2$~mJy, $10.7\pm2.6$~mJy, respectively, consistent with the deduced MAMBO, AzTEC, and LABOCA fluxes of $5.55\pm0.9$~mJy, $5.3^{+1.1}_{-1.2}$~mJy, and $7.4\pm1.8$~mJy, respectively. The closest multi-wavelength source to Cosla-19 is an optical/UltraVista source $2.0"$ away ($i^+=25.66$).  Such a separation makes it unlikely that this source is the counterpart of the mm-detection although given the mm-resolution and significance a $\sim0.8"$ positional uncertainty is expected. A $2\sigma=16.1~\mu$Jy radio peak is associated with the PdBI mm peak yielding a mm-to-radio-flux ratio based redshift for Cosla-19 of $z_\mathrm{mm/radio}=3.98^{+1.62}_{-0.90}$.

\vspace{3mm}
\noindent\underline{\bf \large COSLA-23}
\vspace{1mm}

Within the COSLA-23 LABOCA beam two significant ($\mathrm{S/N}>5$) sources are found in the PdBI 1.3~mm map. 
COSLA-23-N is detected at $\mathrm{S/N}=7.3$ at $\alpha=$~10  00  10.161, $\delta=$~+02  13  34.95. It is coincident with an optical/UltraVista/IRAC/radio source (separation$=0.44"$; $i^+=26.3$). COSLA-23-S is detected at $\mathrm{S/N}=6.2$ at $\alpha=$~10  00  10.070, $\delta=$~+02  13  26.87. It can be matched to an optical/IRAC source (separation$=0.87"$, $i^+=28.49$), but it is not detected in the radio map. Based on the multi-wavelength photometry of the counterparts we find photometric redshifts of $z_\mathrm{phot}=4.00^{+0.67}_{-0.90}$ (COSLA-23-N) and $z_\mathrm{phot}=2.58^{+1.52}_{-2.48}$ (COSLA-23-S). 

The 1.3~mm flux densities for COSLA-23-N and COSLA-23-S are 3.42 $\pm$ 0.47~mJy, and 3.70 $\pm$    0.60~mJy, respectively. Only COSLA-23-N is within the MAMBO $11"$ beam, and the scaled 1.3~mm flux (4.4$\pm$ 0.6~mJy) agrees well with the COSBO-2 flux (5.77$\pm$0.9~mJy).

\vspace{3mm}
\noindent\underline{\bf \large COSLA-25}
\vspace{1mm}

No significant source is present in the 1.3~mm map.  No statistical counterpart could be associated with this SMG. Therefore, it remains unclear whether the LABOCA source is spurious, breaks up into multiple components at $1.5"$ angular resolution, or simply is below the PdBI detection limit.

\vspace{3mm}
\noindent\underline{\bf \large COSLA-30}
\vspace{1mm}

No significant source is present in the 1.3~mm map.  No statistical counterpart could be associated with this SMG. Therefore, it remains unclear whether the LABOCA source is spurious, breaks up into multiple components at $1.5"$ angular resolution, or simply is below the PdBI detection limit.

\vspace{3mm}
\noindent\underline{\bf \large COSLA-33}
\vspace{1mm}

The most prominent feature in the PdBI map within the LABOCA beam is a
 $3.1\sigma$ peak  $6.0"$ away from the LABOCA source center that can be associated with an optical/UltraVista/IRAC source ($i^+=25.2$, separation=$0.95"$). Its 1.3~mm flux density is $1.78\pm0.58$~mJy which scales to $6.02\pm1.95$~mJy at 870~$\mu$m, in good agreement with the deboosted LABOCA flux ($6.8\pm1.1$~mJy). Given the low significance of the 1.3~mm source further follow-up is required to affirm its reality.

\vspace{3mm}
\noindent\underline{\bf \large COSLA-35}
\vspace{1mm}

COSLA-35 is detected at a signal-to-noise of $\mathrm{S/N}=4.2$ ($\alpha=$~10  00  23.65, $\delta=$~+02  21  55.22).  Its 1.3~mm flux density is $F_\mathrm{1.3mm}=2.15 \pm   0.51$~mJy.  
This flux scaled to 870~$\mu$m (1.1~mm) yields a flux density of $7.3\pm1.7$~mJy ($3.5\pm0.8$~mJy), consistent with the observed LABOCA (AzTEC/ASTE) fluxes of $8.2\pm 1.1$~mJy ($5.1^{+1.2}_{ -1.1}$~mJy). 
The mm-detection is 
coincident with an optical/UltraVista/IRAC/radio source
(separation$=0.55"$; $i^+=27.24$). We find a photometric redshift of $z_\mathrm{phot}=1.91^{+1.75}_{-0.64}$.

\vspace{3mm}
\noindent\underline{\bf \large COSLA-38}
\vspace{1mm}

COSLA-38 is detected at $\mathrm{S/N}=4.4$ at  $\alpha=$~10  00  12.59, $\delta=$~+02  14  44.31, $14.8"$ away from the LABOCA source position (thus essentially outside the LABOCA beam; FWHM=$27"$). It is however only $0.67"$ away from the MAMBO source Cosbo-19, and coincident with a radio/UltraVista/IRAC/optical
source (separation=0.23", $i^+=24.08$). We infer a 1.3~mm flux density of $F_\mathrm{1.3mm}=8.19 \pm 1.85$~mJy, which should however be treated with caution as the correction for the primary beam response at that distance from the PdBI phase center applied to the flux is about a factor of 4. We find a photometric redshift of $z_\mathrm{phot}=2.44^{+0.12}_{-0.11}$ for this SMG.

\vspace{3mm}
\noindent\underline{\bf \large COSLA-40}
\vspace{1mm}

COSLA-40 is detected at 
$\mathrm{S/N}=3.4$ ($\alpha=$~09 59 25.91, $\delta=$~+02 19 56.40), $11.3"$ away from the LABOCA source center. Its 1.3~mm flux density is $F_\mathrm{1.3mm}=3.41\pm1.02$~mJy. Scaling
this flux to 870~$\mu$m yields a flux density of $11.5\pm3.4$~mJy, in
very good agreement with the deboosted LABOCA flux
($F_\mathrm{870\mu m}=11.1\pm3.4$~mJy). The source is coincident with an
optical source (separation=0.51"; $i^+=25.52$), but not detected in the radio. Given the expected
flux density, and the coincidence of the source with an optical
detection we assume this source to be real. We find a photometric redshift of $z_\mathrm{phot}=1.30^{+0.09}_{-0.11}$, but we note that $\chi^2$ dips are also found at lower and higher redshift values, and that the mm-to-radio flux based redshift suggests $z_\mathrm{mm/radio}\gtrsim4.5$.

\vspace{3mm}
\noindent\underline{\bf \large COSLA-47}
\vspace{1mm}

COSLA-47 is detected at $\alpha=$~10  00  33.350, $\delta=$~+02  26  01.66  and $\mathrm{S/N}=5.3$, $6.4"$ away from the LABOCA source center. Its 1.3~mm flux density is 3.11 $\pm$  0.59~mJy, and consistent with the LABOCA/AzTEC fluxes, when scaled to these frequencies. 
The PdBI source is coincident (separation=0.48") with a source independently detected at optical, IR, and radio wavelengths. We find a well constrained photometric redshift of $z_\mathrm{phot}=2.36^{+0.24}_{-0.24}$.

Within the LABOCA beam several more $\mathrm{S/N}>4$ peaks can be associated with optical/UltraVista/IRAC sources. They are however within sidelobe contaminated regions. This LABOCA SMG may be a blend of several sources, but further follow-up 
is required to confirm this.

\vspace{3mm}
\noindent\underline{\bf \large COSLA-48}
\vspace{1mm}

No significant source is present in the 1.3~mm map. The statistical counterpart association (see \s{sec:pstat} \ and \f{fig:statstampscoslanodet} \ for details) suggests two potential tentative counterparts to this LABOCA source. The sum of the extracted 1.3~mm fluxes (taken as maximum flux within a circular area of $1"$ in radius centered at the statistical counterpart), corrected for the primary beam response, is 1.56~mJy.  This flux, scaled to 870~$\mu$m (5.2~mJy) is in very good agreement with the LABOCA flux of $6.1\pm1.7$~mJy  (see \f{fig:flux} ). This suggests that the LABOCA source may be fainter at 1.3~mm than can be detected given our PdBI sensitivity and it breaks up into multiple components when observed at $1.5"$ resolution.

\vspace{3mm}
\noindent\underline{\bf \large COSLA-50}
\vspace{1mm}

No significant source is present in the 1.3 mm map. The statistical counterpart association (see \s{sec:pstat} \ and \f{fig:statstampscoslanodet} \ for details) suggests two potential (robust and tentative) counterparts  to this LABOCA source. The sum of the extracted 1.3~mm fluxes (taken as maximum flux within a circular area of $1"$ in radius centered at the statistical counterpart), corrected for the primary beam response, is 2.61~mJy.  When scaled to 870~$\mu$m this flux ($8.7$~mJy) is fairly consistent with the LABOCA flux of $5.6\pm1.6$~mJy  (see \f{fig:flux} ). This suggests that the LABOCA source may be fainter at 1.3~mm than can be detected given our PdBI sensitivity and it breaks up into multiple components when observed at $1.5"$ resolution.

\vspace{3mm}
\noindent\underline{\bf \large COSLA-51}
\vspace{1mm}

No significant source is present in the 1.3~mm map. The statistical counterpart association (see \s{sec:pstat} \ and \f{fig:statstampscoslanodet} \ for details) suggests one robust potential counterpart  to this LABOCA source. The extracted 1.3~mm flux (taken as maximum flux within a circular area of $1"$ in radius centered at the statistical counterpart), corrected for the primary beam response, is 1.27~mJy, consistent with the LABOCA flux of $6.2\pm1.7$~mJy when scaled to 870~$\mu$m (4.5~mJy; see \f{fig:flux} ). This suggests that the LABOCA source is fainter at 1.3~mm than can be detected given our PdBI sensitivity and/or it breaks up into multiple components when observed at $1.5"$ resolution.

\vspace{3mm}
\noindent\underline{\bf \large COSLA-54}
\vspace{1mm}

COSLA-54 is detected at $\mathrm{S/N}=5.0$
($\alpha=$09 58 37.99, $\delta=$~+02 14 08.52), $7.6"$ away from the LABOCA source center.  Its 1.3~mm flux
density is $F_\mathrm{1.3mm}=3.26\pm0.65$~mJy. Scaling this flux to
870~$\mu$m (1.1~mm) yields a flux density of $11.0\pm2.2$~mJy
($5.3\pm1.1$~mJy), in agreement with the deboosted LABOCA (AzTEC/ASTE)
flux of $11.6\pm4.1$~mJy ($8.7^{+1.3}_{-1.4}$~mJy). The mm-detection
can be associated with an optical/IRAC/radio source (separation$=0.75''$;
$i^+$=25.21). We find a photometric redshift of $z_\mathrm{phot}=2.64^{+0.38}_{-0.26}$.

\vspace{3mm}
\noindent\underline{\bf \large COSLA-62}
\vspace{1mm}

No significant source is present in the 1.3~mm map. No statistical counterpart could be associated with this SMG. Therefore, it remains unclear whether the LABOCA source is spurious, breaks up into multiple components at $1.5"$ angular resolution, or simply is below the PdBI detection limit.

\vspace{3mm}
\noindent\underline{\bf \large COSLA-128}
\vspace{1mm}

COSLA-128 is detected at 
$\mathrm{S/N}=4.8$ ($\alpha=$~10 01 37.99, $\delta=$~+02 23
26.50). Its 1.3~mm flux density is
$F_\mathrm{1.3mm}=4.50\pm0.94$~mJy. Scaling this flux to 870~$\mu$m
(1.1~mm) yields a flux density of $15.2\pm3.2$~mJy ($7.3\pm1.5$~mJy),
in agreement with the LABOCA (AzTEC/ASTE) flux of $11.0\pm3.5$~mJy
($4.4\pm1.1$~mJy). The source is coincident with an optical detection
(no MIR/radio; separation$=0.55"$; $i^+=26.57$). We find a photometric redshift of $z_\mathrm{phot}=0.10^{+ 0.19}_{-0.00}$, with secondary and tertiary possible solutions at $z\sim1.2$, and $z\sim3$.

\vspace{3mm}
\noindent\underline{\bf \large COSLA-161}
\vspace{1mm}

 COSLA-161 is detected at $\mathrm{S/N}=3.5$  ($\alpha=$~10  00  16.150, $\delta=$~+02  12  38.27).  Its 1.3~mm flux density is $F_\mathrm{1.3mm}=2.54\pm0.674$~mJy. Scaling this flux to 870~$\mu$m, 1.1~mm, and 1.2~mm, using a spectral index of 3, yields $10.1\pm4.8$, $4.1\pm1.2$, and $3.3\pm0.9$~mJy, respectively. This is in very good agreement with the deboosted LABOCA ($10.1\pm4.8$mJy), AzTEC/ASTE ($3.2\pm1.1$~mJy), and MAMBO fluxes ($1.4\pm0.9$~mJy). 
 
The SMG is coincident with an optical/IR/radio source with an available (VIMOS/IMACS) spectrum at 
$z_\mathrm{spec}=0.187$. The source is also detected by Chandra in the X-rays, and we find a 0.5-2~keV band flux of $1.9\pm0.8$~erg~s$^{-1}$~cm$^{-2}$. At a redshift of 0.187 this corresponds to a bolometric X-ray luminosity (0.1-10~keV) of $(6.2\pm2.8)\times10^{40}$~erg~s$^{-1}$ (assuming a power law spectral shape with photon index 1.8). Given this X-ray luminosity it is not clear whether it arises from star-formation processes or a low-power AGN. 

It is interesting that a second radio source is present within the LABOCA beam
($z_\mathrm{spec}=2.947$), and is not associated with
mm-emission (however there is a $2.3\sigma$ peak at its position in the PdBI map).

%\pagebreak
\begin{table*}
\caption{Photometry table for our LABOCA SMGs with PdBI detections (magnitudes listed are total AB magnitudes corrected for galactic extinction)}
\label{tab:phot}
{\scriptsize
\begin{sideways}
\begin{tabular}{llccccccccccc}
\hline
    Source       &  ID & $r^+$ & $i^+$ & $z^+$ & J & H & Ks &
    m$_{3.6\mathrm{\mu m}}$ & m$_{4.5\mathrm{\mu m}}$ & m$_{5.8\mathrm{\mu m}}$ & m$_{8.0\mathrm{\mu m}}$ & $F_\mathrm{20cm}$ [$\mu$Jy]\\
    \hline 
COSLA-5 &     970338 & 23.25 $\pm$  0.03 & 22.58 $\pm$  0.02 & 22.23 $\pm$  0.02 & 21.69 $\pm$  0.01 & 21.50 $\pm$  0.02 & 21.02 $\pm$  0.01 & 20.98 $\pm$  0.01 & 21.23 $\pm$  0.03 & 21.00 $\pm$  0.11 & 21.25 $\pm$  0.39 & 17.3$\pm$8.6\\ 
COSLA-6-N &  -- &  -- & -- & -- & -- & -- & -- & -- & -- & -- & -- &   19.4$\pm$9.4\\ 
COSLA-6-S &    1201029 & 26.32 $\pm$  0.19 & 26.24 $\pm$  0.23 & 26.04 $\pm$  0.24 & -- & -- & -- & -- & -- & -- & -- & 18.6$\pm$9.3\\ 
COSLA-8 &     999816 & 28.06 $\pm$  0.36 & 27.45 $\pm$  0.25 & $>$ 26.00 & -- & -- & -- & -- & -- & -- & -- & 26.2$\pm$8.0\\ 
COSLA-9N &     572563 & 27.18 $\pm$  0.28 & 26.23 $\pm$  0.17 & 25.92 $\pm$  0.17 & 99.00 $\pm$ 25.40 & 24.73 $\pm$  0.39 & 24.21 $\pm$  0.26 & --& -- & -- & -- & 16.2   $\pm$  8.1\\
COSLA-9S &     571877 & 25.56 $\pm$  0.13 & 24.84 $\pm$  0.09 & 24.26 $\pm$  0.08 & 22.71 $\pm$  0.13 & 21.96 $\pm$  0.06 & 21.06 $\pm$  0.03 & 20.49 $\pm$  0.01 & 20.23 $\pm$  0.01 & 20.16 $\pm$  0.05 & 20.25 $\pm$  0.16  & 25.7  $\pm$  8.4\\
COSLA-11-N &     542814 & 28.18 $\pm$  0.50 & 27.85 $\pm$  0.46 & 26.96 $\pm$  0.28 & -- & -- & -- & -- & -- & -- & -- & 20.1$\pm$10.1\\ 
COSLA-11-S &     543122 & 24.61 $\pm$  0.06 & 24.48 $\pm$  0.06 & 24.69 $\pm$  0.08 & 24.95 $\pm$  0.15 & 24.03 $\pm$  0.11 & 23.90 $\pm$  0.12 & -- & -- & -- & -- & 20.1$\pm$10.1\\ 
COSLA-13 &    1006169 & 25.14 $\pm$  0.10 & 24.57 $\pm$  0.07 & 24.23 $\pm$  0.08 & 22.83 $\pm$  0.03 & 21.97 $\pm$  0.02 & 21.22 $\pm$  0.01 & 20.29 $\pm$  0.01 & 19.89 $\pm$  0.01 & 19.36 $\pm$  0.02 & 19.19 $\pm$  0.06 & 11.9$\pm$7.4\\ 
COSLA-16-E &    1452379 & 29.54 $\pm$  0.58 & 29.29 $\pm$  0.48 & 28.23 $\pm$  0.25 & -- & -- & -- & -- & -- & -- & -- & 20.3$\pm$10.2\\ 
COSLA-16-N &    1452311 & 23.69 $\pm$  0.04 & 23.29 $\pm$  0.03 & 22.75 $\pm$  0.03 & 21.94 $\pm$  0.01 & 21.30 $\pm$  0.01 & 20.63 $\pm$  0.01 & 20.09 $\pm$  0.01 & 19.84 $\pm$  0.01 & 19.76 $\pm$  0.04 & 20.00 $\pm$  0.13 & 95.6$\pm$10.1\\ 
COSLA-16S  & -- &  -- & -- & -- & -- & -- & -- & -- & -- & -- & -- & 33.3 $\pm$ 10.1 \\
COSLA-17-S &  959541 & 28.26 $\pm$  0.85 & $>$ 28.10 & $>$ 26.00 & $>$ 25.40 & $>$ 25.40 & $>$ 25.00 & -- & -- & -- & -- & 19.4$\pm$9.7\\ 
COSLA-17-N &     959312 & 24.94 $\pm$  0.08 & 24.71 $\pm$  0.08 & 24.70 $\pm$  0.08 & 24.33 $\pm$  0.09 & 23.98 $\pm$  0.11 & 23.01 $\pm$  0.05 & 22.43 $\pm$  0.04 & 21.93 $\pm$  0.05 & 21.84 $\pm$  0.23 & 21.30 $\pm$  0.40 & 43.7$\pm$.2\\ 
COSLA-18 &     771619 & 29.59 $\pm$  0.58 & 29.04 $\pm$  0.49 & 27.70 $\pm$  0.24 & -- & -- & -- & 20.79 $\pm$  0.01 & 20.38 $\pm$  0.01 & 20.04 $\pm$  0.05 & 20.04 $\pm$  0.11 & 60.1$\pm$8.9\\ 
COSLA-19 & -- &  -- & -- & -- & -- & -- & -- & -- & -- & -- & -- &    16.1 $\pm$ 8.0\\
COSLA-23-N &    1029389 & 28.40 $\pm$  1.33 & 26.41 $\pm$  0.30 & 26.15 $\pm$  0.31 & -- & -- & 23.02 $\pm$  0.05 & 22.17 $\pm$  0.03 & 21.66 $\pm$  0.04 & 21.64 $\pm$  0.21 & 21.01 $\pm$  0.31 & 53.8 $\pm$8.8\\ 
COSLA-23-S &    1029701 & $>$ 28.40 & 28.58 $\pm$  0.37 & 30.27 $\pm$  2.13 & -- & -- & -- & -- & -- & -- & -- & 15.9 $\pm$8.0\\ 
COSLA-33 &  1267015 & 25.35 $\pm$  0.11 & 25.28 $\pm$  0.14 & 25.19 $\pm$  0.15 & 24.88 $\pm$  0.41 & 24.30 $\pm$  0.41 & 24.15 $\pm$  0.38 & 23.75 $\pm$  0.16 & 23.95 $\pm$  0.35 & 22.42 $\pm$  0.47 & 99.00 $\pm$ 22.96 & 17.8 $\pm$ 8.9\\
COSLA-35 &    1235579 & 26.58 $\pm$  0.34 & 27.32 $\pm$  0.87 & 25.86 $\pm$  0.33 & -- & -- & 21.85 $\pm$  0.03 & 21.07 $\pm$  0.01 & 20.62 $\pm$  0.02 & 20.38 $\pm$  0.06 & 20.15 $\pm$  0.15 & 47.2 $\pm$8.9\\ 
COSLA-38 &    1026384 & 24.48 $\pm$  0.06 & 24.17 $\pm$  0.05 & 23.68 $\pm$  0.04 & 22.69 $\pm$  0.02 & 21.52 $\pm$  0.01 & 20.86 $\pm$  0.01 & 20.05 $\pm$  0.01 & 19.82 $\pm$  0.01 & 19.66 $\pm$  0.04 & 19.97 $\pm$  0.12 & 25.8 $\pm$8.3\\ 
COSLA-40 &    1289293 & 25.74 $\pm$  0.14 & 25.61 $\pm$  0.15 & 25.97 $\pm$  0.25 & -- & -- & -- & -- & -- & -- & -- & 18.8 $\pm$9.4\\ 
COSLA-47 &    1225491 & 25.69 $\pm$  0.12 & 25.83 $\pm$  0.17 & 25.36 $\pm$  0.14 & -- & 23.25 $\pm$  0.06 & 22.24 $\pm$  0.03 & 21.18 $\pm$  0.02 & 20.79 $\pm$  0.02 & 20.45 $\pm$  0.08 & 20.53 $\pm$  0.20 & 40.2 $\pm$9.3\\ 
COSLA-54 &    1078145 & 26.78 $\pm$  0.31 & -- & 28.96 $\pm$  2.88 & -- & 24.11 $\pm$  0.11 & 22.66 $\pm$  0.03 & 21.24 $\pm$  0.02 & 20.87 $\pm$  0.02 & 20.39 $\pm$  0.06 & 20.46 $\pm$  0.16 & 121.7 $\pm$10.7\\ 
COSLA-128 &    1183799 & 26.20 $\pm$  0.26 & 26.66 $\pm$  0.51 & 26.64 $\pm$  0.60 & -- & -- & -- & -- & -- & -- & -- & 18.6 $\pm$9.3\\ 
COSLA-161 &    1006116 & 18.74 $\pm$  0.00 & 18.27 $\pm$  0.00 & 17.67 $\pm$  0.00 & 17.23 $\pm$  0.00 & 16.76 $\pm$  0.00 & 16.37 $\pm$  0.00 & 17.33 $\pm$  0.00 & 17.40 $\pm$  0.00 & 17.45 $\pm$  0.01 & 15.66 $\pm$  0.00  & 217.4$\pm$8.7\\ 
\hline
\end{tabular}
\end{sideways}
}
\end{table*}

\section{Notes on the 1.1~mm-selected sample }
\label{sec:aztec}

Our \mmsample \ is based on the SMA follow-up  of 15 brightest SMGs drawn from the 1.1~mm AzTEC/JCMT-COSMOS survey at $18"$ angular resolution (AzTEC-1 to AzTEC-15; see \t{tab:interf} ; Younger et al.\ 2007, 2009). Detailed notes on individual targets are given in Younger et al.\ (2007, 2009). Here we have extracted the multi-wavelength photometry, tabulated in \t{tab:aztecphot} , for the counterparts of these SMGs using the deep COSMOS multi-wavelength catalog, with UltraVista data added. 
The photometry in the IRAC bands had to be deblended for AzTEC-8 (see \f{fig:aztec8} ), and that for AzTEC-10 had to specifically be extracted as this source was not present in the catalog (see Younger et al.\ 2009). The photometry extraction and deblending were performed following \ the procedure described in detail by \smo\ et al.\ (2012). Furthermore, AzTEC-11 is a peculiar source that required particular attention. Younger et al.\ (2009) find that the SMA detection is best fit by a double Gaussian, suggesting a multiple component (N \& S) source, labeled AzTEC-11-N and AzTEC-11-S.\footnote{Note that the N \& S labels are inverted (see Tab.~1 in Younger et al.\ 2009). } They present three positions for this SMG: i) AzTEC-11 when the SMA detection is fit using a single-Gaussian, and ii) AzTEC-11-N and AzTEC-11-S  when the SMA detection is fit using a double Gaussian. AzTEC-11 is coincident with an optical/MIR/radio source with a spectroscopic redshift ($z_\mathrm{spec}=1.599$). AzTEC-11-S (which is actually the northern component of the source) cannot be matched to a multi-wavelength counterpart in the deep COSMOS maps. Thus, given the rms in the 20~cm VLA-COSMOS survey we estimate a mm-to-radio based redshift of $z_\mathrm{mm/radio}>2.58$. AzTEC-11-N (which is actually the southern component of the source) has an independent UltraVista and IRAC counterpart. To extract its photometry we have deblended the counterpart of AzTEC-11-N by subtracting a 2D-Gaussian from the maps at the position of the counterpart of AzTEC-11,
In \f{fig:aztec11} \ we show the deblended maps for AzTEC-11-N. We find a photometric redshift of $z_\mathrm{phot}=1.51^{+0.41}_{-0.92}$ for this component.
%\pagebreak

%\pagebreak
\begin{table*}
\caption{Photometry table for our \mmsample \ (magnitudes listed are total AB magnitudes corrected for galactic extinction)}
\label{tab:aztecphot}
{\scriptsize
\begin{sideways}
\begin{tabular}{llccccccccccc}
\hline
    Source       &  ID & $r^+$ & $i^+$ & $z^+$ & J & H & Ks &
    m$_{3.6\mathrm{\mu m}}$ & m$_{4.5\mathrm{\mu m}}$ & m$_{5.8\mathrm{\mu m}}$ & m$_{8.0\mathrm{\mu m}}$ & $F_\mathrm{20cm}$ [$\mu$Jy]\\
    \hline 
AzTEC-1 &    1485894 & 26.17 $\pm$  0.16 & 25.20 $\pm$  0.10 & 24.94 $\pm$  0.09 & 25.16 $\pm$  0.17 & 24.58 $\pm$  0.19 & 23.46 $\pm$  0.08 & 22.44 $\pm$  0.05 & 22.27 $\pm$  0.07 & 21.10 $\pm$  0.14 & 21.04 $\pm$  0.32  & 41.6$\pm$11.1\\ 
AzTEC-2$^*$ & -- &  -- & -- & -- & -- & -- & -- & -- & -- & -- & -- & 47.3 $\pm$ 10.9 \\
AzTEC-3 &    1447531 & 26.26 $\pm$  0.19 & 25.01 $\pm$  0.09 & 24.39 $\pm$  0.07 & 23.75 $\pm$  0.07 & 23.88 $\pm$  0.11 & 23.73 $\pm$  0.11 & 22.67 $\pm$  0.05 & 22.03 $\pm$  0.06 & 22.45 $\pm$  0.43 & 21.35 $\pm$  0.41 & 24.3$\pm$9.8\\ 
AzTEC-4 &    1507528 & 28.14 $\pm$  0.78 & 26.45 $\pm$  0.21 & 26.68 $\pm$  0.32 & -- & -- & 24.07 $\pm$  0.10 & 22.17 $\pm$  0.04 & 21.88 $\pm$  0.04 & 21.45 $\pm$  0.19 & 20.80 $\pm$  0.22 & 28.0$\pm$11.4\\ 
AzTEC-5 &    1455197 & 25.90 $\pm$  0.20 & 25.25 $\pm$  0.15 & 24.97 $\pm$  0.16 & -- & -- & 22.90 $\pm$  0.07 & 21.69 $\pm$  0.02 & 21.27 $\pm$  0.03 & 20.94 $\pm$  0.10 & 20.48 $\pm$  0.18 & 92.5$\pm$10.8\\ 
AzTEC-6 &    1708424 & 25.62 $\pm$  0.11 & 25.16 $\pm$  0.10 & 24.54 $\pm$  0.07 & 25.57 $\pm$  0.24 & -- & -- & -- & -- & -- & -- & 29.0$\pm$11.6\\ 
AzTEC-7 &    1899647 & 24.26 $\pm$  0.07 & 23.84 $\pm$  0.06 & 23.33 $\pm$  0.05 & 21.83 $\pm$  0.03 & 21.15 $\pm$  0.03 & 20.55 $\pm$  0.02 & 19.79 $\pm$  0.01 & 19.52 $\pm$  0.01 & 19.31 $\pm$  0.03 & 19.51 $\pm$  0.09 & 103.4$\pm$18.7\\ 
AzTEC-8 &    1473458 & 26.25 $\pm$  0.13 & 25.90 $\pm$  0.14 & 26.00 $\pm$  0.17 & -- & -- & 23.67 $\pm$  0.07 & 21.73 $\pm$  0.03 & 21.22 $\pm$  0.03 & 20.46 $\pm$  0.07 & 20.06 $\pm$  0.14 & 88.8$\pm$10.6\\ 
AzTEC-9 &    1271178 & 26.29 $\pm$  0.20 & 25.34 $\pm$  0.12 & 25.01 $\pm$  0.11 & -- & -- & -- & 22.92 $\pm$  0.07 & 22.65 $\pm$  0.10 & $>$23.82 & 22.59 $\pm$  1.27 & 53.5$\pm$9.52\\ 
    AzTEC-10 & -- & -- & -- & -- & -- & -- & 23.47$\pm$0.10 & 21.76$\pm$0.048 &  21.21$\pm$0.038 & 20.80$\pm$0.063 & 20.81$\pm$0.15 & 26.5$\pm$11.6\\
AzTEC-11 &    1704741 & 23.70 $\pm$  0.04 & 23.22 $\pm$  0.03 & 22.87 $\pm$  0.03 & 22.10 $\pm$  0.01 & 21.62 $\pm$  0.01 & 21.28 $\pm$  0.01 & 20.22 $\pm$  0.01 & 19.84 $\pm$  0.01 & 19.64 $\pm$  0.04 & 19.85 $\pm$  0.11 & 139.4$\pm$12.7\\ 
    AzTEC-11N & -- &      -- & -- & -- & 22.97 $\pm$ 0.20 &    22.45$\pm$0.20  & 21.78$\pm$0.20 & 21.25$\pm$0.30 & 20.96$\pm$0.11  &  20.86$\pm$0.11 & 20.88$\pm$0.18 & 120.8$\pm$12.6\\
AzTEC-11S& -- &  -- & -- & -- & -- & -- & -- & -- & -- & -- & -- & 25.9 $\pm$ 13.0 \\
AzTEC-12 &    1671195 & 24.63 $\pm$  0.07 & 24.19 $\pm$  0.06 & 23.90 $\pm$  0.06 & 23.07 $\pm$  0.04 & 21.66 $\pm$  0.02 & 21.24 $\pm$  0.02 & 20.27 $\pm$  0.01 & 19.95 $\pm$  0.01 & 19.47 $\pm$  0.03 & 19.52 $\pm$  0.08 & 104.2$\pm$13.8\\ 
AzTEC-13& -- &  -- & -- & -- & -- & -- & -- & -- & -- & -- & -- & 20.2 $\pm$ 10.1\\
AzTEC-14W &    1484311 & 26.09 $\pm$  0.21 & 25.70 $\pm$  0.19 & 26.64 $\pm$  0.53 & -- & -- & -- & -- & -- & -- & -- & 18.9$\pm$9.4\\ 
AzTEC-14E& -- &  -- & -- & -- & -- & -- & -- & -- & -- & -- & -- & 19.4 $\pm$ 9.7\\
AzTEC-15 &    1473978 & 27.53 $\pm$  0.32 & 27.36 $\pm$  0.36 & 26.53 $\pm$  0.23 & -- & 24.43 $\pm$  0.09 & 23.66 $\pm$  0.05 & 21.77 $\pm$  0.03 & 21.20 $\pm$  0.03 & 21.05 $\pm$  0.12 & 20.33 $\pm$  0.17 & 19.5$\pm$9.7\\ 
\hline
\end{tabular}
\end{sideways}
}$^*$AzTEC-2, at a spectroscopic redshift of 1.125 (Balokovi\'{c} et al., in prep), is heavily blended by a bright, extended foreground galaxy.
\end{table*}

\pagebreak

\begin{figure*}
\includegraphics[scale=0.65]{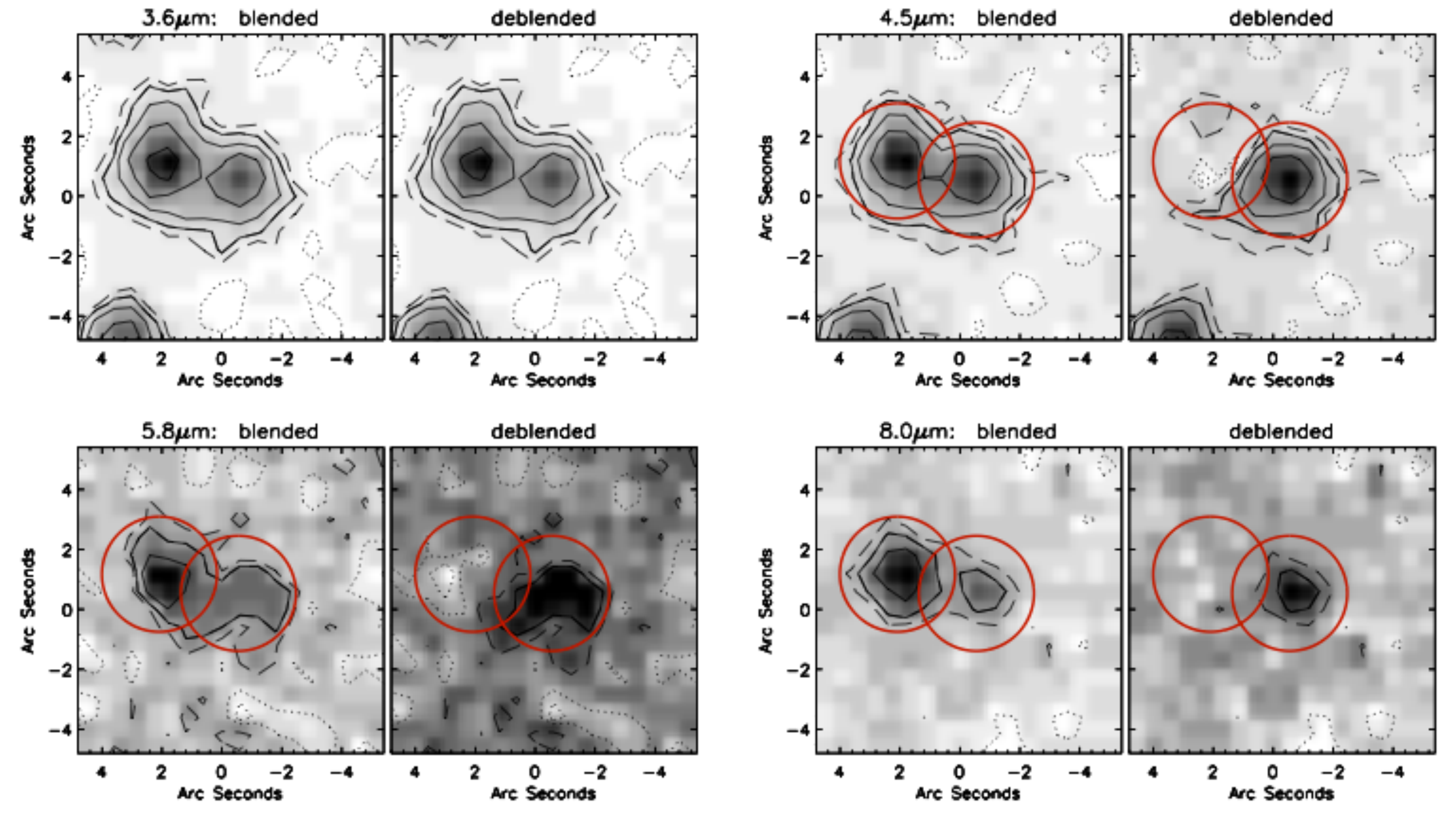}
\caption{ Deblending of AzTEC-8 in Spitzer/IRAC bands.}
      \label{fig:aztec8}
\end{figure*}

\begin{figure*}
\includegraphics[scale=0.65]{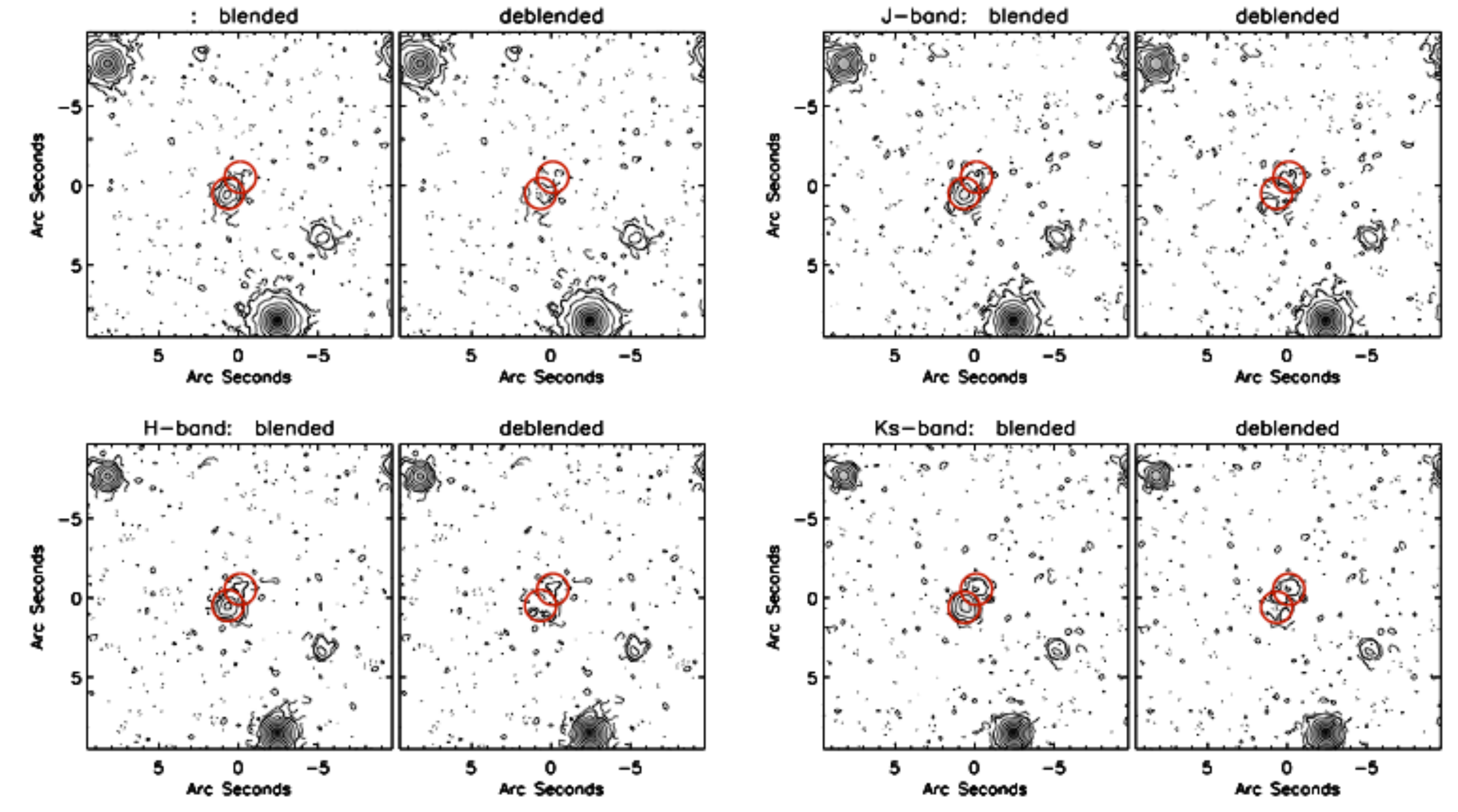}
\includegraphics[scale=0.65]{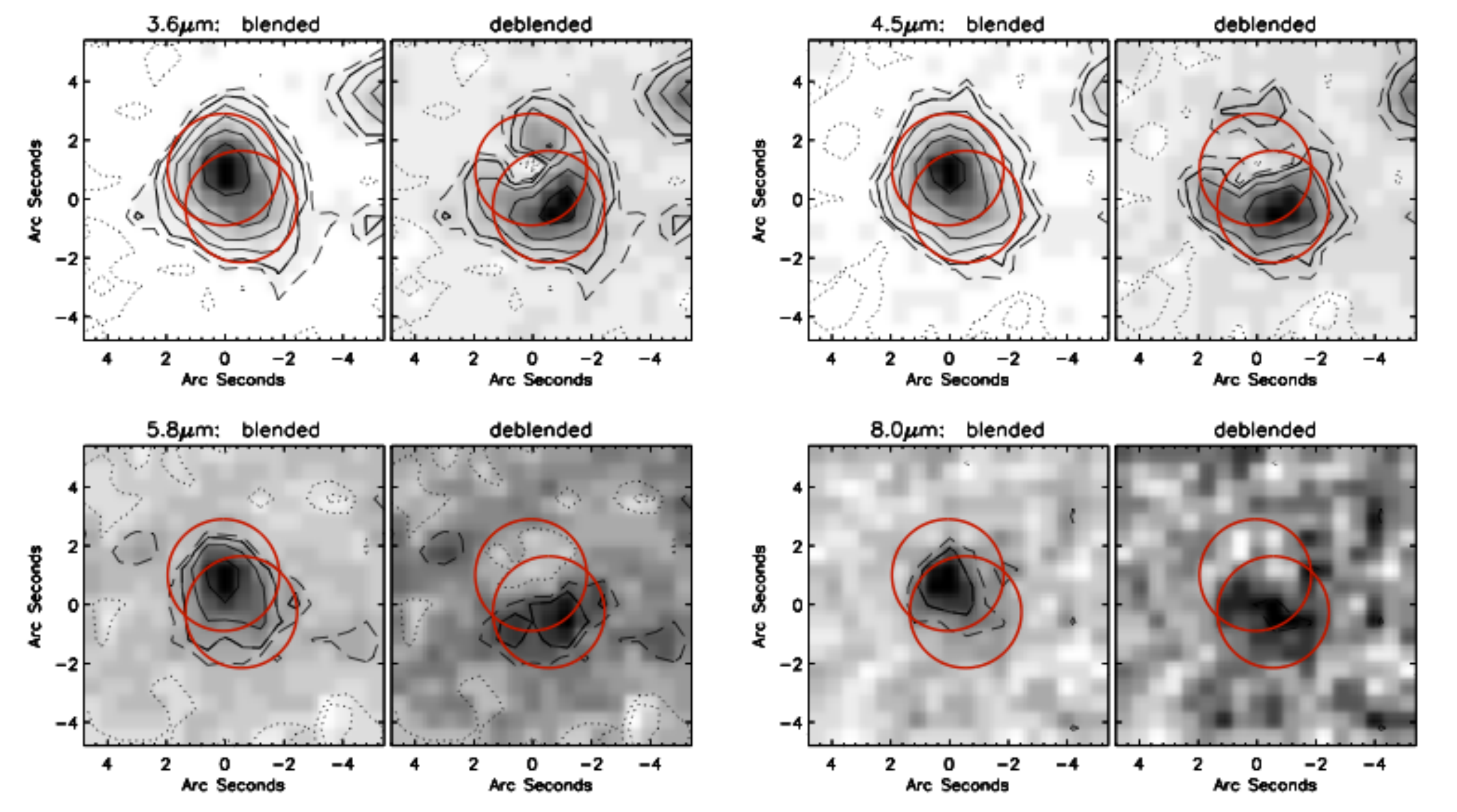}\caption{ Deblending of AzTEC-11-N in UltraVista YJHK and Spitzer/IRAC bands.}
      \label{fig:aztec11}
\end{figure*}

\end{document}